\documentclass{jpp}
\usepackage{graphicx}
\usepackage{epstopdf, epsfig}
\usepackage{hyperref}
\usepackage{bm}
\usepackage{url}
\usepackage{amsmath}
\usepackage{verbatim}
\usepackage{hyphenat}
\usepackage[mathscr]{euscript}

\usepackage{enumerate}
\usepackage{setspace}
\usepackage{color}
\usepackage{natbib}
\usepackage[bf, footnotesize]{caption2}

\newcommand{\derv}[1]{\frac{\partial}{\partial #1}}

\newcommand{\deriv}[2]{\frac{\partial #1}{\partial #2}}

\newcommand{\F}{{\mathcal F}}
\newcommand{\G}{{\mathcal G}}
\newcommand{\Ham}{{\mathcal H}}
\newcommand{\C}{{\mathcal C}}

\newcommand{\Grad}{\text{grad}}
\newcommand{\Div}{\text{div}}
\newcommand{\Curl}{\text{curl}}
\newcommand{\hook}{\rfloor}

\newcommand{\X}{\mathrm{X}}
\newcommand{\pr}{\text{pr}}
\newcommand{\lieder}[1]{{\mathcal L}_{#1}}

\newcommand{\beqn}{\begin{equation}}
\newcommand{\eeqn}{\end{equation}}
\newcommand{\beqnar}{\begin{eqnarray}}
\newcommand{\eeqnar}{\end{eqnarray}}

\newtheorem{theorem}{Theorem}[section]
\newtheorem{proposition}[theorem]{Proposition}

\clearpage

\shorttitle{Action Principles for CGL Plasmas}
\shortauthor{G.M. Webb, S.C. Anco, S.V. Meleshko, and G.P. Zank} 

\title{Action Principles and Conservation Laws for Chew-Goldberger-Low Anisotropic Plasmas}

\author{G. M. Webb\aff{1}
  \corresp{\email{gmw0002@uah.edu}},
  S.  C. Anco\aff{2},
  S. V. Meleshko\aff{3},
  \& \ G. P. Zank\aff{1,4}}

\affiliation{\aff{1}Center for Space Plasma and Aeronomic Research, the University of Alabama in Huntsville, Huntsville AL 35805, USA.
\aff{2}Department of Mathematics \& Statistics, Brock University, St. Catharines, ON L2S 3A1 Canada.
\aff{3} School of Mathematics, Institute of Science, Suranaree University of Technology, Nakhon Ratchasima, 30000, Thailand.
\aff{4} Department of Space Science, The University of Alabama in Huntsville, 
Huntsville AL35899, USA.
Submitted\today}
\begin{document}
\maketitle

\begin{abstract}
The ideal CGL plasma equations, including the double adiabatic 
conservation laws for the parallel ($p_\parallel$) and perpendicular pressure ($p_\perp$), 
are investigated using a Lagrangian variational principle. 
An Euler-Poincar\'e variational principle is developed
and the non-canonical Poisson bracket is obtained, 
in which the non-canonical variables consist of 
the mass flux ${\bf M}$, the density $\rho$,  the entropy variable 
$\sigma=\rho S$
% three entropy variables, $\sigma=\rho S$, $\sigma_\parallel=\rho S_\parallel$, $\sigma_\perp=\rho S_\perp$ 
%($S_\parallel$ and $S_\perp$ are the two scalar entropy invariants), 
and the magnetic induction ${\bf B}$. 
Conservation laws of the CGL plasma equations are derived via Noether's theorem. 
The Galilean group leads to conservation of energy, momentum, center of mass, and 
angular momentum. 
Cross helicity conservation arises from a fluid relabeling symmetry, 
and is local or nonlocal depending on whether the gradient of $S$ 
% $S_\parallel$, $S_\perp$ and $S$ 
is perpendicular to ${\bf B}$ or otherwise. 
The point Lie symmetries of the CGL system are shown to comprise the 
Galilean transformations and scalings. 
\end{abstract}

\section{Introduction}
In magnetohydrodynamics (MHD), 
the ideal Chew-Goldberger-Low (CGL) equations (\cite{Chew56})
describe plasmas in which 
there is not enough scattering of the particles to have an isotropic pressure
and only lowest order gyro-radius effects terms are taken 
into account in the particle transport 
(i.e. finite Larmor-radius terms are neglected). 
These equations can be viewed as 
the small gyro-radius limit of the Vlasov fluid moment equations,  
where the pressure tensor has the anisotropic (gyrotropic) form 
${\sf p}=p_\perp {\sf I} + (p_\parallel -p_\perp){\bm\tau}{\bm\tau}$.
Here ${\bm\tau}=|{\bf B}|^{-1}{\bf B}$ is the unit vector 
along the magnetic field
and ${\sf I}$ is the identity tensor. 
The pressure component $p_\parallel-p_\perp$ controls the anisotropy, i.e. 
the pressure tensor is isotropic if $p_\parallel=p_\perp$. 

Higher order gyro-radius effects and gyro-viscosity 
lead to the extended anisotropic MHD equations 
(see e.g. \cite{Macmahon65},  \cite{Ramos05a,Ramos05b}; 
\cite{Sulem15} and \cite{Hunana19a}). \cite{Braginskii65}, \cite{Schnack09}; 
and \cite{Devlen10} use an 
isotropic pressure decomposition, which is not an anisotropic
CGL decomposition. \cite{Hunana22} considered various generalizations
of  the \cite{Braginskii65} model using the Landau collisional operator 
for the case of multi-species plasmas. 
A recent Hamiltonian version of extended gyro-viscous MHD
has been developed by \cite{Lingam20}. 

In the present work, 
we develop a variational formulation of 
the ideal CGL plasma equations
starting from the Lagrangian formulation of (\cite{Newcomb62}). 
Our main goals are to derive and discuss:
\\
(1) an Euler-Poincare action principle;
\\
(2) a non-canonical Poisson bracket and its Casimirs;
\\
(3)  conservation laws from application of Noether's theorem to the Galilean group of Lie point symmetries;
\\
(4) a cross-helicity conservation law from particle relabeling symmetries
and its nonlocal nature when the entropy gradients are non-orthogonal to the magnetic field. 

Previous work on the CGL equations can be found in 
\cite{Shrauner67}; \cite{Hazeltine13}; 
\cite{Holm86}; \cite{Ramos05a,Ramos05b};  \cite{Hunana19a,Hunana19b}; 
\cite{Du20}. 

\cite{CheviakovBog04} and \cite{CheviakovAnco08} derived exact, 
anisotropic MHD equilibria solutions of the CGL equations, 
with a modified equation of state
for incompressible fluid flows. \cite{CheviakovBog04} obtained an infinite 
group of Lie symmetries of the anisotropic plasma equilibium equations for steady flow configurations. An infinite dimensional family of transformations 
between the isotropic MHD equilibrium equations and solutions of the 
anisotropic CGL equations were obtained. \cite{Ilgisonis96} studies the 
stability of steady CGL plasma equilibria in a tokamak for a generalized 
Grad-Shafranov 
equation, taking into account the fluid relabelling symmetry, 
and covers previous stability criteria for ideal CGL plasma and ideal MHD
cases.   

Our methods are adapted from MHD and ideal fluid mechanics. 
It is well known how to use a Lagrangian map to relate Eulerian and Lagrangian fluid quantities (\cite{Newcomb62}),
and this underpins the Euler-Poincar\'e action principle 
which is a Lagrangian counterpart of a Hamiltonian formulation 
(\cite{Holm98}; \cite{Webb18} and \cite{Webb19}).  
This action principle is based on the Lagrangian map 
and utilizes an associated Lie algebra. 

Recent work by \cite{Dewar20} and \cite{Dewar22} developed variational 
principles to describe time-~dependent relaxed MHD using a 
global cross helicity constraint with a phase space Lagrangian action 
principle (PSL) as opposed to a  configuration space Lagrangian (CSL) action. 
The theory has been used to describe   
multi-region, extended MHD (RxMHD) in fusion plasma devices.
 In ideal MHD current sheets can develop. One 
of the main ideas in this work is to provide a global fitting together 
of ideal MHD sub-regions, subject to global cross helicity and magnetic 
helicity constraints by using Lagrange multipliers.  This is a rather 
complicated theory that lies beyond the scope of the present paper.    

Noether's theorem has been used to obtain conservation laws for ideal MHD 
(\cite{Webb19}; see also \cite{Padhye96a, Padhye96b}; \cite{Padhye98}).    
A different approach to deriving conservation laws is Lie dragging of differential 
forms, vector fields and tensors, as developed by 
(\cite{Moiseev82}; \cite{Sagdeev90};  
\cite{Tur93};  \cite{Webb14a,Webb14b}; 
\cite{Gilbert20}; \cite{Besse17}; \cite{Anco20}).

%***** work 
%of Bogoyanlenskij, Cheviakov and Anco on the steady-state solutions and symmetries
%of the anisotropic MHD equations. 

Section~\ref{sec:cgleqns} summarizes the ideal CGL plasma equations. 
The anisotropic pressure components $p_\parallel$ and $p_\perp$
satisfy double adiabatic conservation laws. 
The total energy equation for the system arises from 
combining the internal energy equation, the kinetic energy equation, 
and the electromagnetic energy equation (i.e. Poynting's theorem). 
The cross helicity and Galilean conservation laws are obtained for 
barotropic and non-barotropic gas equations of state. 
The magnetic helicity transport equation and the conservation of magnetic helicity 
is described. 
The thermodynamics of CGL plasmas are discussed. 
%The first law is derived from the pressure tensor equation generalized 
%to include dissipative effects. 
This leads to an internal energy density of the form $e=e(\rho,S,B)$ for the plasma 
(e.g. \cite{Holm86}; \cite{Hazeltine13}), 
where $B=|{\bf B}|$ is the magnitude of the magnetic field. 
It applies to the situation when reversible energy changes occur, 
with the temperature of the plasma being given by $T=e_S$, 
and it yields an equation of state which incorporates the double adiabatic conservation laws. 
A comparison is given with a more restrictive different approach that applies the ideal gas law to the plasma (e.g. \cite{Du20}), 
with an equation of state being derived as a consequence. 

Section~\ref{sec:Lagrangianmap} 
describes the Lagrangian map between 
the Eulerian fluid particle position ${\bf x}$ 
and the Lagrangian particle position ${\bf x}_0$.
This map is obtained from integrating the system of ordinary differential equations
$d{\bf x}/dt={\bf u}({\bf x},t)$ with the fluid velocity ${\bf u}({\bf x},t)$ 
assumed to be a known function of ${\bf x}$ and $t$ and ${\bf x}={\bf x}_0$ at time $t=0$. 
The Lagrangian map ${\bf x}({\bf x}_0,t)$ is used to write the fluid equations as a variational principle (\cite{Newcomb62}; \cite{Holm98}). 
In this description the canonical coordinates are ${\bf q}={\bf x}({\bf x}_0,t)$, 
and the canonical momenta are defined by the Legendre transformation 
${\bf p}=\partial L_0/\partial \dot{\bf x}$ where $L_0$ is the Lagrangian density 
in a frame moving with the fluid. 

Section~\ref{sec:EulerPoincare} 
develops the Euler-Poincar\'e action principle for the CGL plasma equations
by using the general method of \cite{Holm98} and the results of \cite{Newcomb62}.

In Section~\ref{sec:Poissonbracket}, 
the Poisson bracket $\{F,G\}$ of functionals $F$ and $G$ is described. 
{We obtain the noncanonical Poisson bracket of \cite{Holm86} but give more
details}. 
By converting the variational derivatives for functionals from the canonical coordinates
$({\bf q},{\bf p})$ to physically motivated non-canonical coordinates, 
the Poisson bracket is converted to its non-canonical form in the new variables. 
The symplectic (i.e. Hamiltonian) form of the non-canonical Poisson bracket is determined,
which is useful in proving the Jacobi identity and for writing down the Casimir 
determining equations. 
The Casimirs of the Poisson bracket commute with the Hamiltonian functional 
characterizing the system and are obtained using the approach 
of \cite{Hameiri04} 
(see also \cite{Marsden94}, \cite{Morrison82}, \cite{Holm85},  
\cite{Padhye96a,Padhye96b} and \cite{Padhye98}).  

Section 6 
describes Noether's theorem for the CGL variational principle
using the Lagrangian variables. 
Noether's theorem produces conservation laws in terms of these variables 
from variational symmetries. The Eulerian form of the conservation laws is obtained 
through the Lagrangian map by use of a result of \cite{Padhye98}. 
The variational symmetries include the Galilean group: in particular, 
(\romannumeral1)\ time translation invariance yields energy conservation;
(\romannumeral2)\ space translation invariance yields momentum conservation;
(\romannumeral3)\ invariance under  Galilean boosts yields center of mass conservation;
(\romannumeral4)\ invariance of under rotations yields angular momentum conservation.

In addition, fluid relabeling symmetries are shown to be variational symmetries which yield the generalized cross helicity conservation law for CGL plasmas. 
The conditions under which this conservation law is local or nonlocal are delineated.
A local conservation law is shown to arise when the parallel and perpendicular entropies,
$S_\parallel$ and $S_\perp$, have zero gradient along the magnetic field.
Alternatively, for an internal energy density $e=e(\rho,S,B)$, 
a local conservation law occurs when ${\bf B}{\bf\cdot}\nabla S=0$. 
When these conditions fail to hold in the plasma, 
the cross helicity conservation law is nonlocal and depends on the integration of 
the temperatures back along the Lagrangian fluid particle paths. 

The present paper can be thought of as an extension of 
the work of \cite{Holm86}. We restrict our analysis to non-relativistic flows,
whereas \cite{Holm86} study both relativistic and non-relativistic flow versions of 
the CGL equations. Our analysis is more complete than \cite{Holm86} 
in that: (\romannumeral1)\ our analysis takes into account 
more recent developments in anisotropic moment equations for plasmas 
with an anisotropic pressure tensor (e.g. \cite{Hazeltine13} 
and \cite{Hunana19a}). In particular, the formulations of the equation 
of state and the first law of thermodynamics used by \cite{Holm86} 
were not clear because of the brevity 
of the exposition. (\romannumeral2)  We show that a slightly more 
general, non-separable equation of state can be used than that of 
 \cite{Hazeltine13} (see \ref{eq:eos} et seq.); 
(\romannumeral3)\ We provide a more complete treatment of the conservation 
laws for the CGL system, and identify the symmetries of the Lagrangian 
which give rise to the conservation  laws in Noether's theorem. Our analysis 
shows that the cross helicity conservation law is a consequence 
of a fluid relabelling symmetry, which is in general 
a nonlocal conservation law unless ${\bf B}$ is normal to the 
entropy gradients of $S_\parallel$, $S_\perp$ and $S$, in which case 
the conservation law is local. (\romannumeral4)\ We describe 
the CGL Poisson bracket of  
 \cite{Holm86}  which uses the noncanonical 
variables $\rho$, ${\bf M}=\rho {\bf u}$, $\sigma=\rho S$ and ${\bf B}$.
The entropy variable used in the non-canonical Poisson bracket is 
$\sigma=\rho S$ where $S$ is the entropy. The entropy $S$ can in turn be 
decomposed into the form $S=S_\parallel+S_\perp$. 
 The non-canonical Poisson bracket 
derivation in Appendix D is more rigorous than that used by 
most authors as it takes into account the variations in the basis vectors 
used to define vectors and tensors. However these changes 
do not modify the net Poisson bracket, because they only lead at most to 
changes in the boundary terms in the bracket which are discarded (see 
e.g. \cite{Holm83c} for the more rigorous general derivation 
of the non-canonical Poisson bracket in MHD). (\romannumeral5)\ 
We provide a modern version of Noether's theorem which only uses 
the evolutionary symmetry operator. We describe  
the classical form of Noether's theorem  obtained by \cite{Bluman89},
 in Appendix H. In Section 6.1 we derive CGL conservation laws 
using the more modern form of Noether's theorem.  
 Section 6.2 gives the classical 
version of Noether's theorem to derive CGL plasma conservation laws.

Section~\ref{sec:conclude}
concludes with a summary and discussion.

The various technical derivations used in the main results, 
and the Lie point symmetries of the CGL system, 
are summarized in Appendices~A to~G.

%section 2
\section{CGL Equations and Conservation Laws}\label{sec:cgleqns}
In this section, we first summarize the basic CGL plasma model equations,
and then we discuss the thermodynamics of CGL plasmas, 
followed by the key conservation laws in Eulerian form:  
total energy; cross helicity and nonlocal conserved cross helicity; 
magnetic helicity. 
These conservation laws, in later sections, will be related to the variational symmetries of the action principle for the CGL equations and expressed in Lagrangian form. 
Some general remarks about Eulerian conservation law equations, 
which will be used in the discussions, are summarized in Section~\ref{sec:ConsLaws}.

\subsection{CGL Model}
The physical variables which describe CGL plasmas are 
the fluid velocity ${\bf u}$, fluid density $\rho$, magnetic field induction ${\bf B}$, 
the anisotropic pressure components $p_\parallel$ and $p_\perp$, 
and the entropy $S$. 
The ideal CGL plasma equations are similar to the MHD equations and consist of:
%(e.g. \cite{Hunana19a,Hunana19b})
the mass continuity equation
\begin{equation}
\deriv{\rho}{t}+\nabla{\bf\cdot} (\rho{\bf u})=0; 
\label{eq:2.1}
\end{equation}
the momentum equation (in semi-conservative form) 
\begin{equation}
\derv{t}(\rho{\bf u})+\nabla{\bf\cdot}\left(
\rho {\bf u}{\bf u}+{\sf p} + {\sf M}_B \right)
=-\rho\nabla\Phi,
\label{eq:2.2}
\end{equation}
in which ${\sf p}$ is the gyrotropic pressure tensor 
(which replaces the isotropic gas pressure $p{\sf I}$ used in MHD) 
given by 
\begin{equation}
{\sf p}=p_\perp{\sf I}+(p_\parallel-p_\perp)\bm{\tau}\bm{\tau}, 
\quad 
\bm{\tau}=\frac{\bf B}{B}, 
\quad
B= |{\bm B}| , 
\label{eq:2.3}
\end{equation}
and ${\sf M}_B$ is the magnetic pressure tensor 
\begin{equation}
{\sf M}_B = \frac{1}{\mu_0}\left( \frac{B^2}{2} {\sf I} -{\bf B B} \right)
= \frac{B^2}{\mu_0}\left(\frac{1}{2} {\sf I} -{\bm\tau}{\bm\tau} \right) ;
\label{eq:Bpressure}
\end{equation} 
the entropy transport equation
\begin{equation}
\deriv{S}{t}+{\bf u}{\bf\cdot}\nabla S=0 ;
\label{eq:2.4}
\end{equation}
Faraday's equation and Gauss's law 
\begin{align}
\deriv{\bf B}{t}-\nabla\times\left({\bf u}\times {\bf B}\right) =&0,  
\label{eq:2.5}\\
\nabla{\bf\cdot B}=& 0 ; 
\label{eq:2.6}
\end{align}
along with the pressure equations
\begin{align}
\deriv{p_\parallel}{t}+\nabla{\bf\cdot}\left(p_\parallel{\bf u}\right)
+ 2{\sf p}^\parallel{\bf :}\nabla{\bf u} = & 0, 
\label{eq:2.7}\\
\deriv{p_\perp}{t}+\nabla{\bf\cdot}\left(p_\perp {\bf u}\right)
+ {\sf p}^\perp{\bf :}\nabla{\bf u} = & 0,
\label{eq:2.8}
\end{align}
where ${\sf p}^\parallel$ and ${\sf p}^\perp$ are the terms comprising the gyrotropic pressure tensor
\begin{equation}
{\sf p}^\parallel = p_\parallel \bm{\tau}\bm{\tau}, 
\quad
{\sf p}^\perp = p_\perp \left({\sf I} - \bm{\tau}\bm{\tau}\right),
\quad
{\sf p} = {\sf p}^\parallel + {\sf p}^\perp . 
\label{eq:defn.p}
\end{equation}
%2p_\parallel\left(\bm{\tau}{\bf\cdot}\nabla{\bf u}{\bf\cdot}\bm{\tau}\right)
%p_\perp \nabla{\bf\cdot}{\bf u}-p_\perp\left(\bm{\tau}{\bf\cdot}\nabla {\bf u}{\bf\cdot}\bm{\tau}\right)

Through Faraday's equation (\ref{eq:2.5}) and the mass continuity equation (\ref{eq:2.1}),  
the pressure equations (\ref{eq:2.7})-(\ref{eq:2.8}) can be expressed as the double 
adiabatic equations
\begin{equation}
\frac{d}{dt}\left(\frac{p_\parallel B^2}{\rho^3}\right)=0,
\quad 
\frac{d}{dt}\left(\frac{p_\perp}{\rho B}\right)=0, 
\label{eq:2.9} 
\end{equation}
which represent conservation of the particle magnetic moment 
and the second longitudinal adiabatic moment of the particles,
where $d/dt=\partial/\partial t+{\bf u}{\bf\cdot}\nabla$ is the Lagrangian time derivative following the flow. 

In the momentum equation (\ref{eq:2.2}), 
$\Phi$ is the potential of an external source of gravity 
(for example, in the case of the solar wind, it could represent 
the gravitational potential field of the Sun). 

If one counts the number of evolution equations 
in (\ref{eq:2.1})-(\ref{eq:2.9}) there are 10 equations 
for the 10 variables ${\bf u}$, ${\bf B}$, $\rho$, $S$, $p_\parallel$ and $p_\perp$. If one does not need to know $S$ then there are 9 equations 
for 9 unknowns.

The non-canonical Poisson bracket formulation of MHD in \cite{MorrisonGreene82}
uses Faraday's law (\ref{eq:2.5}) in form 
$\deriv{\bf B}{t}-\nabla\times\left({\bf u}\times {\bf B}\right) + {\bf u}\nabla{\bf\cdot}{\bf B}=0$ 
for the mathematical case in which $\nabla{\bf\cdot}{\bf B}\neq 0$ (see also \cite{Webb18}). 
In fact, the possibility of $\nabla{\bf\cdot}{\bf B}\neq 0$ arises in 
numerical MHD due to numerical errors in the Gauss' law (\ref{eq:2.6}). 

To determine the relationship between the entropy $S$, the internal energy density $\varepsilon$, and the pressure components $p_\parallel$ and $p_\perp$ 
in (\ref{eq:2.1})-(\ref{eq:2.9}) we first note that $\varepsilon$ 
is given by:
\begin{equation}
\varepsilon=\frac{p_\parallel+2p_\perp}{2}.
\label{defn.energydens}
\end{equation} 
A different, related approach comes from the entropy law for ideal gases. 
We will consider and compare both approaches. 

One link between the approaches is the observation that 
integration of the double adiabatic equations (\ref{eq:2.9}) yields 
\begin{equation}
p_\parallel=\exp\left({\bar S}_\parallel\right) \frac{\rho^3}{B^2}, 
\quad
p_\perp=\exp\left({\bar S}_\perp\right)\rho B, 
\label{eq:2.9a}
\end{equation}
where the quantities ${\bar S}_\parallel$ and ${\bar S}_\perp$ 
are dimensionless forms of entropy integration constants 
arising from integrating the double adiabatic equations (\ref{eq:2.9}). 
In principle, ${\bar S}_\parallel$ and ${\bar S}_\perp$ 
must be scalars advected with the flow. In particular these quantities 
could be functions of the entropy $S$ satisfying (\ref{eq:2.4}). They 
could also be functions of other advected invariants of the flow, 
such as ${\bf B}{\bf\cdot}\nabla S/\rho$ (e.g. \cite{Tur93}, \cite{Webb14a}). 
For the sake of simplicity, we assume that $S_\parallel$ and 
$S_\perp$ depend only on $S$. The overbars on ${\bar S}$, 
${\bar S}_\parallel$ and ${\bar S}_\perp$  
 denote dimensionless versions of these quantities. 

If these $S_\parallel$ and $S_\perp$  are functions solely of the entropy $S$, 
then the internal energy density $\varepsilon$ will have the 
functional form $\varepsilon(\rho,S,B)$.  
This form also arises from the first law of thermodynamics coming from 
the transport equation for $\varepsilon$ when reversible thermodynamic 
processes are considered. 
Strictly speaking, we should use normalized 
or dimensionless variables in (\ref{eq:2.9a}), i.e. we should have used
the variables:
\begin{equation}
{\bar p}_\parallel=\frac{p_\parallel}{p_{\parallel0}}, 
\quad {\bar p}_\perp=\frac{p_\perp}{p_{\perp0}}, 
\quad {\bar{\bf B}}=\frac{{\bf B}}{B_0},
\quad \bar{\rho}=\frac{\rho}{\rho_0}, \label{eq:2.9b}
\end{equation}
where the subscript zero quantities are dimensional constants. In a 
convenient abuse
of notation, we have dropped the overbar superscripts in (\ref{eq:2.9a}).

\subsection{Thermodynamic formulation}\label{sec:thermodynamics}
A physical description of $p_\parallel$ and $p_\perp$ and their relation 
to the internal energy density $\varepsilon$ was developed 
first by \cite{Holm86}, which in part uses the work of \cite{Volkov66}, 
and also later by \cite{Hazeltine13}, 
which discusses both reversible and irreversible thermodynamics. 
Here we will concentrate on the ideal reversible dynamics case.  

An analysis of reversible work done on the plasma 
reveals that the internal energy per unit mass of the CGL plasma, 
$e=\varepsilon/\rho$, 
obeys the Pfaffian equation
\begin{equation}
de=T dS+\frac{p_\parallel}{\rho^2}d\rho -\frac{p_{\Delta}}{\rho B} dB. 
\label{eq:2.16i}
\end{equation}
where $T$ is the adiabatic temperature in the plasma. 
The expression (\ref{defn.energydens}) for the internal energy density gives the relations
\begin{equation}
e=\frac{\varepsilon}{\rho}=\frac{(p_\parallel+2p_\perp)}{2\rho},
\quad 
p_{\Delta}=p_\parallel-p_\perp, 
\label{eq:2.16j}
\end{equation}
(see also \cite{Holm86})
and 
\begin{equation}
p \equiv \frac{(p_\parallel+2p_\perp)}{3} =\frac{2\varepsilon}{3}, 
\label{eq:2.16k}
\end{equation}
which is the gas pressure defined by one third of the trace of the CGL pressure tensor (\ref{eq:2.3}). 

The Pfaffian equation (\ref{eq:2.16i}) constitutes the first law of thermodynamics for a CGL plasma. 
As outlined in \cite{Hazeltine13}, 
it can be derived from the transport equation for $\varepsilon$ 
under thermodynamic processes that involve reversible work done on the plasma
described by the gyrotropic part of the Vlasov distribution function. 
The derivation can be generalized to include irreversible work comprising 
a dissipative term and a further term due to gyro-viscosity, 
which is a non-dissipative term obtained in the limit of no scattering. 
Equation (\ref{eq:2.16i}) is equivalent to the non-dissipative internal 
energy equation (\ref{eq:2.18}) in which a source term 
$\rho TdS/dt$ has been added to the right hand-side.

The above interpretation of equation (\ref{eq:2.16i}) implies 
$e=e(\rho,S,B)$ along with the thermodynamic relations 
\begin{equation}
T =e_S = \varepsilon_S/\rho,
\quad 
p_\parallel=\rho^2 e_\rho = \rho\varepsilon_\rho-\varepsilon, 
\quad 
p_{\Delta}=-\rho B e_B = -B\varepsilon_B .
\label{eq:2.16l}
\end{equation}
Thus, if $e(\rho,S,B)$ is known, i.e. an equation of state has been specified, 
then the relations (\ref{eq:2.16l}) give $p_\parallel$ and $p_\perp$ 
as functions of $\rho,S,B$. 
Consistency must hold with the double adiabatic equations (\ref{eq:2.9}), 
which constrains the possible expressions for $e(\rho,S,B)$. 

To derive the constraints, we first substitute the relations (\ref{eq:2.16l}) 
into the internal energy (\ref{eq:2.16k}), 
giving
\begin{equation}
e=\frac{1}{\rho}\left(\frac{3 p_\parallel}{2}-p_{\Delta}\right)
=\frac{3}{2}\rho e_\rho+B e_B. 
\label{eq:2.16n}
\end{equation}
This is a first order partial differential equation for $e(\rho,S,B)$ 
which is easily solved by using the method of characteristics 
(e.g. \cite{Sneddon57}): 
\begin{equation}
\frac{d\rho}{(3\rho/2)}=\frac{dB}{B}=\frac{de}{e}. 
\label{eq:2.16o}
\end{equation}
Integrating these differential equations yields the general solution 
\begin{equation}
e=\rho^{2/3} f(\xi,S), 
\quad
\xi=B\rho^{-2/3}, 
\label{eq:2.16p}
\end{equation}
where $f(\xi, S)$ is an arbitrary function of $S$ and the similarity variable $\xi$. 
Now, the relations (\ref{eq:2.16l}) imply that 
\begin{equation}
T =\rho^{2/3} f_S , 
\label{eq:2.16q}
\end{equation}
and 
\begin{equation}
p_\parallel=\frac{2}{3} \rho^{5/3} \left( f- \xi f_\xi\right), 
\quad
p_\perp =\frac{1}{3} \rho^{5/3}\left( 2 f+\xi f_\xi \right). 
\label{eq:2.16r}
\end{equation}
Substitution of expressions (\ref{eq:2.16r}) into the double adiabatic equations (\ref{eq:2.9})
followed by use of the relation $B=\rho^{2/3}\xi$ then yields
\begin{equation}
\frac{d}{dt}\left( \frac{2}{3} \xi^2f -\frac{2}{3} \xi^3 f_\xi \right) =0, 
\quad
\frac{d}{dt}\left( \frac{2}{3\xi}f+ \frac{1}{3} f_\xi \right) =0, 
\label{f.eqns}
\end{equation}
with $df/dt = f_\xi d\xi/dt$ for any process in which $S$ is adiabatic 
(cf the entropy equation (\ref{eq:2.4})). 
However, from Faraday's equation (\ref{eq:2.5}) and the mass continuity equation (\ref{eq:2.1}),  
we find that $\xi$ obeys the transport equation 
\begin{equation}
\frac{d}{dt}\xi = \left(-\frac{1}{3}\nabla{\bf\cdot}{\bf u} + {\bm \tau}{\bm \tau}{\bf :}\nabla{\bf u} \right)\xi . 
\end{equation}
Thus, $\xi$ is not an advected quantity, which implies that the adiabatic equations (\ref{f.eqns}) for $f(\xi,S)$ reduce to:
\begin{equation} 
\frac{2}{3} \xi^2f -\frac{2}{3} \xi^3 f_\xi = c_\parallel(S), 
\quad
\frac{2}{3\xi}f+ \frac{1}{3} f_\xi = c_\perp(S) .
\end{equation}
The general solution of this pair of equations is given by 
\begin{equation}
f(\xi,S) = c_\perp(S) \xi + c_\parallel(S)\frac{1}{2\xi^2} . 
\end{equation}
Consequently, expressions (\ref{eq:2.16r}) yield the equation of state 
\begin{equation}
p_\parallel= c_\parallel(S) \rho^3/B^2, 
\quad
p_\perp =c_\perp(S) \rho B, 
\label{eos}
\end{equation}
along with the relations
\begin{equation}
{\bar S}_\parallel = \ln c_\parallel(S), 
\quad
{\bar S}_\perp = \ln c_\perp(S) 
\label{eq:2.9b}
\end{equation}
from the double adiabatic integrals (\ref{eq:2.9a}). 
The corresponding internal energy (\ref{eq:2.16p}) and pressure (\ref{eq:2.16k})
have the explicit form 
\begin{equation}
e=c_\perp(S) B + c_\parallel(S)  \frac{\rho^2}{2B^2}  
=\exp({\bar S}_\perp) B + \exp({\bar S}_\parallel)  \frac{\rho^2}{2B^2}  ,
\label{eq:eos}
\end{equation}
and
\begin{equation}
p=c_\perp(S) \frac{2\rho B}{3} + c_\parallel(S)  \frac{\rho^3}{3B^2}  
=\exp({\bar S}_\perp) \frac{2\rho B}{3} + \exp({\bar S}_\parallel)  \frac{\rho^3}{3B^2}  .
\label{p.eos}
\end{equation}
Note that the expressions for $e$ and $p$ in terms of the double adiabatic integrals 
$\bar S_\parallel$ and $\bar S_\perp$ hold independently of any form of equation of state. 
Finally, either equation (\ref{eq:2.16q}) or equation (\ref{eq:2.16l}), 
both of which rely on $e=e(\rho,S,B)$, yield the plasma temperature: 
\begin{equation}
T = c_\perp'(S) B + c_\parallel'(S)  \frac{\rho^2}{2B^2} 
=\frac{d{\bar S}_\perp}{dS} \exp({\bar S}_\perp) B 
+ \frac{d{\bar S}_\parallel}{dS} \exp({\bar S}_\parallel)  \frac{\rho^2}{2B^2}  . 
\label{eq:T}
\end{equation}

\cite{Hazeltine13} arrives at a similar but less general result 
under the assumption that $f$ is separable in $S$ and $\xi$. 
This leads to $c_\parallel(S) = Q(S) d_\parallel$ and $c_\perp(S) = Q(S) d_\perp$, 
where $d_\perp$ and $d_\parallel$ are constants. 
The assumption of separability includes, for example, 
the case where the gas pressure $p$ satisfies the ideal gas law:
\begin{equation}
p=\rho R T\quad\hbox{or}\quad T=\frac{p}{\rho R}, 
\label{eq:2.16t}
\end{equation}
where $R$ is the gas constant. 
In this case, from the relation (\ref{eq:2.16k}), 
it follows that $f$ must satisfy the equation
$\deriv{f}{S}=\frac{2}{3R} f$. 
Then it further follows that $Q(S)=\exp(2S/(3R))$.

Conditions on the derivatives of the internal energy $e$ 
for thermodynamic stability are discussed in \cite{Hazeltine13}, 
but these considerations lie beyond the scope of the present work.

%gas law
\subsubsection{Ideal gas law and entropy}
The double adiabatic equations (\ref{eq:2.7}) for a CGL plasma
can be combined in the form 
\begin{equation}
\frac{d}{dt}\left(\frac{p_\parallel p_\perp^2}{\rho^5}\right)=0
\label{eq:2.10}
\end{equation}
which suggests the advected quantity 
$p_\parallel p_\perp^2/\rho^5$ may be viewed as a function of the entropy $S$. 
A specific functional relation can be motivated by considering
the MHD limit for a non-relativistic gas, where the gas entropy is given by the standard formula
\begin{equation}
S=C_v\ln\left(\frac{p}{\rho^\gamma}\right), 
\quad 
\gamma=\frac{C_p}{C_v}=\frac{5}{3}, 
\label{eq:2.12}
\end{equation}
with $C_v$ and $C_p$ being the specific heats at constant volume and pressure, respectively. 
In a CGL plasma, the gas pressure $p$ is given by expression (\ref{eq:2.16k})
in terms of the pressure components $p_\parallel$ and $p_\perp$
which are equal in the MHD limit, since the pressure tensor (\ref{eq:2.3}) 
must become isotropic. 
This results in the relation $p_\parallel=p_\perp=p$,
and as a consequence 
$p_\parallel p_\perp^2/\rho^5 = p^3/\rho^5 = \exp( 3S/C_v)$
by assuming the formula (\ref{eq:2.12}). 
Generalizing this relation away from the MHD limit then suggests the formula
\begin{equation}
S=C_v \ln \left(\frac{p_\parallel^{1/3}p_\perp^{2/3}}{\rho ^{5/3}}\right) 
\label{eq:2.11}
\end{equation}
for the entropy of a CGL plasma (see e.g. \cite{Shrauner67} and \cite{Du20}). 

Furthermore, since the double adiabatic integrals (\ref{eq:2.9a}) 
give $p_\parallel p_\perp^2/\rho^5=\exp(\bar S_\parallel +2\bar S_\perp)$
 where $\bar S_\parallel$  and $\bar S_\perp$  are each advected, 
the entropy formula (\ref{eq:2.11}) can be expressed as 
$S = (C_v/3) (\bar S_\parallel + 2\bar S_\perp)$. 
This relation now suggests that the entropy is a sum of components 
\begin{equation}
S= S_{\parallel} + S_{\perp} 
\quad\hbox{where}\quad 
S_{\parallel} \equiv C_{v\parallel}{\bar S}_\parallel
\quad\text{and}\quad
S_\perp \equiv C_{v\perp}{\bar S}_\perp, 
\label{eq:S.eqns}
\end{equation}
with 
\begin{equation}
C_{v_\parallel}\equiv\frac{1}{3} C_v, 
\quad 
C_{v_\perp}\equiv\frac{2}{3} C_v. 
\label{eq:2.16e}
\end{equation}
Correspondingly, the internal energy density (\ref{eq:2.16j}) of the plasma 
can be split into a sum of densities 
\begin{equation}
\varepsilon=\frac{p_\parallel+2 p_\perp}{2}
=\frac{p_\parallel}{(\gamma_\parallel-1)}
+\frac{p_\perp}{(\gamma_\perp-1)}
\equiv \varepsilon_\parallel+\varepsilon_\perp, 
\label{eq:2.13}
\end{equation}
with effective adiabatic indices:
\begin{equation}
\gamma_\parallel=3, \quad \gamma_\perp=2 . 
\label{eq:2.14}
\end{equation}
Further note that 
\begin{equation}
\varepsilon=\frac{p}{\gamma-1},
\quad 
p=\frac{1}{3}(p_\parallel+2 p_\perp), 
\quad 
\gamma=\frac{5}{3}
\label{eq:2.15}
\end{equation}
in terms of the gas pressure $p$. 

In addition, by applying the standard ideal gas law (\ref{eq:2.16t}) to each of the pressure components $p_\parallel$ and $p_\perp$, 
we can define corresponding temperatures: 
\begin{equation}
T_\parallel \equiv p_\parallel/(\rho R) 
\quad\text{and}\quad
T_\perp \equiv p_\perp /(\rho R) 
\label{eq:T.eqns}
\end{equation}
Then the internal energy density (\ref{eq:2.13}) can be expressed as
\begin{equation}
\varepsilon = R\rho\left(\tfrac{1}{2}T_\parallel + T_\perp\right) ,
\label{eq:2.13a}
\end{equation}
which yields the relations
\begin{equation}
C_{v_\parallel} = e_{T_\parallel} = \frac{1}{2} R, 
\quad 
C_{v_\perp} = e_{T_\perp} = R, 
\quad
C_v = \frac{3}{2} R
\label{eq:Cv.eqns}
\end{equation}
from the specific heats (\ref{eq:2.16e}). 
Finally, by substituting the double adiabatic integrals (\ref{eq:2.9a}) 
into the quantities (\ref{eq:T.eqns}) and using expressions (\ref{eq:S.eqns}) and (\ref{eq:Cv.eqns}), 
we obtain 
\begin{equation}
\deriv{T_\parallel}{S_\parallel} = 2 T_\parallel/R, 
\quad
\deriv{T_\perp}{S_\perp} = T_\perp/R,
\end{equation}
which implies the expected thermodynamic relations:
\begin{equation}
e_{S_\parallel} = T_\parallel,
\quad 
e_{S_\perp} = T_\perp. 
\end{equation}

The preceding approach to the entropy of CGL plasmas has been adopted by \cite{Du20}. 
It is equivalent to the thermodynamic approach specialized to the case of an ideal gas 
in which the gas law is assumed to hold for all three pressures $p$, $p_\parallel$, $p_\perp$. 

 The net upshot of the above discussion is that we take $e=e(\rho,S,B)$ 
as the equation of state for the gas in the rest of the paper. 
Note however, that $S_\parallel=S_\parallel (S)$ and 
$S_\perp(S)=S-S_\parallel (S)$ are functions of $S$. 
$e=e(\rho,S,B)$ is the form of the equation of state used by \cite{Holm86}. 
%However, we prefer to use the variables $\sigma=\rho S$, 
%$\sigma_\parallel=\rho S_\parallel$ and 
%$\sigma_\perp=\rho S_\perp$ as explicit variables in the CGL Poisson 
%bracket in Section 5, as it shows that the physics will change
%depending on how $S_\parallel$ and $S_\perp$ depend on $S$. 
The CGL Poisson bracket of \cite{Holm86} 
 uses the entropy variable $\sigma=\rho S$   
to describe the complicated thermodynamics of the CGL plasma.

\subsection{Total Mass and Energy Conservation Laws}\label{sec:totalenergy}
The mass continuity equation (\ref{eq:2.1}) can be expressed in the familiar co-moving form 
\begin{equation}
\deriv{\rho}{t}+{\bf u}{\bf\cdot}\nabla \rho = -\rho \nabla{\bf\cdot}{\bf u} . 
\end{equation}
On a volume $V(t)$ moving with the fluid, 
the corresponding mass integral is conserved:
\begin{equation}
\frac{d}{dt}\int_{V(t)} \rho \,d^3x =0 . 
\label{eq:2.17}
\end{equation}

A co-moving equation for the internal energy density (\ref{defn.energydens}) 
is obtained by combining the parallel and perpendicular pressure equations (\ref{eq:2.7})-(\ref{eq:2.8}), which yields 
\begin{equation}
\deriv{\varepsilon}{t}+\nabla{\bf\cdot}(\varepsilon {\bf u})
+(p_\parallel-p_\perp)\bm{\tau}\bm{\tau}{\bf :}\nabla{\bf u}
+p_\perp\nabla{\bf\cdot}{\bf u}=0. 
\label{eq:2.16}
\end{equation}
This equation can be written in terms of the gyrotropic pressure tensor ${\sf p}$
by noting from expression (\ref{eq:defn.p}) that 
\begin{equation}
{\sf p}{\bf\cdot}\nabla{\bf u}=p_\perp\nabla{\bf\cdot}{\bf u}
+(p_\parallel-p_\perp) \bm{\tau}\bm{\tau}{\bf :}\nabla{\bf u} . 
\end{equation}
Then the co-moving equation (\ref{eq:2.16}) takes the form 
\begin{equation}
\deriv{\varepsilon}{t}+\nabla{\bf\cdot}\left(\varepsilon {\bf u}
+\sf{p}{\bf\cdot u}\right)={\bf u}{\bf\cdot}(\nabla{\bf\cdot}\sf{p}), 
\label{eq:2.18}
\end{equation}
which is analogous to the internal energy density equation in fluid dynamics and MHD. 

The total kinetic energy equation for the plasma is obtained by taking 
the scalar product of the momentum equation (\ref{eq:2.2}) with ${\bf u}$,
which yields:
\begin{equation}
\begin{aligned}
&\derv{t}\left(\frac{1}{2}\rho u^2+\rho\Phi\right)
+\nabla{\bf\cdot}\left({\bf u}\left(\frac{1}{2}\rho u^2+\rho\Phi\right)\right)\\
&=-{\bf u}{\bf\cdot}\nabla{\bf\cdot}\sf{p}
+{\bf J}{\bf\cdot}{\bf E}+{\bf u\cdot B}\frac{\nabla{\bf\cdot}{\bf B}}{\mu_0}, 
\end{aligned}
\label{eq:2.19}
\end{equation}
where $u = |{\bf u}|$, and 
\begin{equation}
{\bf E}=-{\bf u}\times{\bf B}\quad \hbox{and}\quad {\bf J}
=\frac{\nabla\times{\bf B}}{\mu_0}, 
\label{eq:2.20}
\end{equation}
are the electric field ${\bf E}$, and the electric current ${\bf J}$. 

Using Maxwell's equations (\ref{eq:2.5})--(\ref{eq:2.6}), we obtain Poynting's 
theorem (the electromagnetic energy equation) in the form:
\begin{equation}
\derv{t}\left(\frac{1}{2\mu_0}B^2\right)
+\nabla{\bf\cdot}\left(\frac{1}{\mu_0}{\bf E}\times {\bf B}\right)
= -{\bf J\cdot E} -\frac{1}{\mu_0} (\nabla{\bf\cdot}{\bf B})({\bf u\cdot B}). 
\label{eq:2.21}
\end{equation}

The equation for the total energy in conserved form is obtained by 
adding the internal energy equation (\ref{eq:2.18}), the kinetic energy 
equation (\ref{eq:2.19}) and the electromagnetic energy equation (\ref{eq:2.21}), 
which gives 
\begin{equation}
\derv{t}\left(\frac{1}{2}\rho u^2+\rho\Phi
+\varepsilon+\frac{1}{2\mu_0}B^2\right)
+\nabla{\bf\cdot}\left(\left(\frac{1}{2}\rho u^2+\rho \Phi+\varepsilon\right) {\bf u}
+{\sf p}{\bf\cdot}{\bf u} +\frac{1}{\mu_0} {\bf E}\times{\bf B}\right)=0. 
\label{eq:2.22}
\end{equation}
It is useful here to note that the Poynting electromagnetic energy flux is
\begin{equation}
{\bf F}_{\text{em}} = \frac{1}{\mu_0} {\bf E}\times{\bf B}
= \frac{B^2}{2\mu_0} {\bf u} + {\bf u}{\bf\cdot}{\sf M}_B. 
\end{equation}

The resulting energy balance equation 
on a volume $V(t)$ moving with the fluid is given by
\begin{equation}
\frac{d}{dt}\int_{V(t)} \left(\frac{1}{2}\rho u^2+\rho\Phi +\varepsilon+\frac{1}{2\mu_0}B^2\right)\,d^3x 
= -\oint_{\partial V(t)} {\bf u}{\bf\cdot} \left({\sf p}+ {\sf M}_B\right){\bf\cdot}\hat{\bf n}\, dA
\label{eq:2.23}
\end{equation}
in terms of the magnetic pressure tensor (\ref{eq:Bpressure}). 
The flux terms in (\ref{eq:2.23}) represent the rate at which the total pressure tensor 
${\sf p}+ {\sf M}_B$ does work on the fluid. 
Note that there is no contribution due to advection of the total energy density. 
Further discussion is provided in Section~\ref{sec:eulerianconslaw}
(see also \cite{Padhye98} and \cite{AncoDar09,AncoDar10}). 
In addition, note that if the total pressure tensor 
has no perpendicular component at the moving boundary, 
i.e. $({\sf p}+ {\sf M}_B){\bf\cdot}\hat{\bf n}=0$, 
then the moving energy integral will be conserved (i.e.\ a constant of the motion). 

%Abraham -Shrauner (1967a,b) in the case where $\Phi=0$ (no external gravity)
%has split the total energy equation (\ref{eq:2.28}) into separate energy 
%equations for the parallel and perpendicular energy components. 
%This split of the total energy equation into parallel and perpendicular 
%energy components was used in order to study CGL plasma shocks
%and Rankine Hugoniot conditions. The split up of the total energy 
%conservation equations was done in such a way as to conserve the total
%energy conservation equation (\ref{eq:2.28}) 
%for the CGL system. I have not been able to verify that this split 
%up of the total energy equation follows from the other CGL equations. 
%This may well be a good approximation for the analysis of CGL plasma 
%shocks analyzed  in detail in Abraham-Shrauner (1967b).  

\subsection{Cross Helicity}\label{sec:crosshelicity}
The cross helicity transport equation for a CGL plasma is obtained by taking 
the scalar product of ${\bf u}$ with Faraday's equation (\ref{eq:2.5}) 
plus the scalar product of ${\bf B}$ with the momentum equation (\ref{eq:2.2}), 
in the following form. 
By a standard cross-product identity, 
Faraday's equation can be written
\begin{equation}
\derv{t}{\bf B} - {\bf B}{\bf\cdot}\nabla{\bf u} + {\bf u}{\bf\cdot}\nabla{\bf B} + (\nabla{\bf\cdot}{\bf u}){\bf B} =0. 
\label{eq:2.5alt}
\end{equation}
The momentum equation minus ${\bf u}$ times the mass continuity equation (\ref{eq:2.1}) yields the velocity equation:
\begin{equation}
\frac{d}{dt}{\bf u}= -\frac{1}{\rho}\nabla{\bf\cdot}\left({\sf p} +{\sf M}_B\right)-\nabla\Phi , 
\label{eq:2.2.velocity}
\end{equation}
where $-(1/\rho)\nabla{\bf\cdot}{\sf p}$ is the acceleration  
due to the anisotropic pressure. 
Forming and adding the respective scalar products with ${\bf u}$ and ${\bf B}$
then gives the equation for cross-helicity density ${\bf u\cdot B}$ as:
\begin{equation}
\derv{t}({\bf u\cdot B}) +\nabla{\bf\cdot}\left[({\bf u\cdot} {\bf B}){\bf u}
+\left(\Phi -\frac{1}{2} u^2\right){\bf B}\right]
=-{\bf B}{\bf\cdot}\left(\frac{1}{\rho}\nabla{\bf\cdot}{\sf p}\right) ,
\label{eq:2.ch1}
\end{equation}
where $u = |{\bf u}|$. 

The thermodynamic form of this transport equation (\ref{eq:2.ch1}) 
is obtained by using the equation 
\begin{equation}
-\nabla{\bf\cdot}{\sf p}={\bf B}\times (\nabla\times{\bm\Omega})
-{\bm\Omega}(\nabla{\bf\cdot}{\bf B})+\rho(T\nabla S-\nabla h), 
\label{eq:2.ch2}
\end{equation}
which relies on the Pfaffian equation (first law of thermodynamics) (\ref{eq:2.16i})
for a CGL plasma with an internal energy $e(\rho,B,S)$,
where, from the thermodynamic relations (\ref{eq:2.16l}), 
$T=e_S$ is the gas temperature, 
and $h=\varepsilon_\rho=e+\rho e_\rho$ is the enthalpy of the fluid: 
\begin{equation}
h= \frac{3p_\parallel+2p_\perp}{2\rho} . 
%= e+\frac{p_\parallel}{\rho}
\label{eq:2.ch4}
\end{equation}
Here 
\begin{equation}
{\bm\Omega}=\frac{p_{\Delta}}{B} \bm{\tau} .
\label{eq:2.ch3}
\end{equation}
Note that the case ${\bm\Omega}=0$ corresponds to the MHD limit 
in which the anisotropy vanishes, $p_{\Delta}=p_\parallel-p_\perp=0$. 
A derivation of the pressure divergence equation (\ref{eq:2.ch2}) is given in Appendix~B.
An alternative derivation is provided within the Euler-Poincar\'e formulation 
of the CGL equations in Section 4.1 and Appendix E. 
The scalar product of this equation with $\frac{1}{\rho}{\bf B}$ 
reduces to:
\begin{equation}
-{\bf B}{\bf \cdot} \left(\frac{1}{\rho}\nabla{\bf\cdot}{\sf p}\right)
={\bf B}{\bf \cdot}\left(T\nabla S\right) -\nabla{\bf \cdot}\left(h{\bf B}\right)
\end{equation}
using Gauss' law (\ref{eq:2.6}). 
As a result, the cross helicity density transport equation in thermodynamic form is simply
\begin{equation}
\derv{t}({\bf u\cdot B}) +\nabla{\bf\cdot}\left[({\bf u\cdot} {\bf B}){\bf u}
+\left(\Phi+h-\frac{1}{2} u^2\right){\bf B}\right]
={\bf B}{\bf\cdot}(T \nabla S) . 
\label{eq:2.ch1b}
\end{equation}

The term on the righthand side of the cross helicity transport equation (\ref{eq:2.ch1b})
can be written in a conserved form through use of the equation 
\begin{equation}
\derv{t}\left(\phi\nabla S{\bf\cdot}{\bf B}\right)
+ \nabla{\bf\cdot}\left[\left(\phi\nabla S{\bf\cdot}{\bf B}\right){\bf u}\right]
= \left(\nabla S{\bf\cdot}{\bf B}\right)\frac{d}{dt}\phi
\label{eq:2.ch3a}
\end{equation}
which holds for any scalar variable $\phi$
and follows from Faraday's equation (\ref{eq:2.5}) 
and the entropy equation (\ref{eq:2.4}). 
This leads to the nonlocal cross helicity  density conservation law
\begin{equation}
\derv{t}({\bf w\cdot B}) +\nabla{\bf\cdot}\left[({\bf w\cdot B}){\bf u}
+\left(\Phi+h-\frac{1}{2} u^2\right){\bf B}\right]=0, 
\label{eq:2.ch5}
\end{equation}
in terms of 
\begin{equation}
{\bf w}={\bf u}+r \nabla S, 
\label{eq:2.ch6}
\end{equation}
where $r$ is a nonlocal variable obtained by integrating the temperature $T$ 
back along the path of the Lagrangian fluid element. 
Specifically, 
\begin{equation}
\frac{dr}{dt}=-T,
\quad 
r=-\int^t T\, dt, 
\label{eq:2.ch7}
\end{equation}
where $d/dt=\partial/\partial t+{\bf u}{\bf\cdot}\nabla$ is the Lagrangian time derivative. 

The nonlocal conservation law (\ref{eq:2.ch5}) yields a moving cross-helicity balance equation (cf Appendix~C):
\begin{equation}
\frac{d}{dt}\int_{V(t)} {\bf w\cdot B} \,d^3x
= -\oint_{\partial V(t)} \left(\Phi+h-\frac{1}{2} u^2\right){\bf B}{\bf\cdot}\hat{\bf n}\, dA
\end{equation}
on a volume $V(t)$ moving with the fluid. 
Note that if ${\bf B}$ is perpendicular to the boundary, i.e. ${\bf B\cdot}\hat{\bf n}=0$, 
then the cross-helicity integral $\int_{V(t)} {\bf w\cdot B} \,d^3{\bf x}$ is conserved in the flow. 

An alternative form of the nonlocal conservation law (\ref{eq:2.ch5}) 
is obtained by directly taking the divergence of the pressure tensor $\sf p$ 
using the gyrotropic expression (\ref{eq:defn.p}),
and combining it with the gradient of the enthalpy (\ref{eq:2.ch4}). 
As shown in Appendix~B, this gives
\begin{equation}
\begin{aligned}
{\bf B}{\bf\cdot}\left(\frac{1}{\rho} \nabla{\bf\cdot}{\sf p}-\nabla h\right)
& =-\left( \frac{p_\parallel}{2\rho} {\bf B}{\bf\cdot}\nabla\ln c_\parallel(S) 
+ \frac{p_\perp}{\rho} {\bf B}{\bf\cdot}\nabla\ln c_\perp(S) \right)
\\
& =-\left( \frac{p_\parallel}{2\rho} {\bf B}{\bf\cdot}\nabla{\bar S}_\parallel
+ \frac{p_\perp}{\rho} {\bf B}{\bf\cdot}\nabla{\bar S}_\perp \right) 
\end{aligned}
\label{eq:C14}
\end{equation}
in terms of the adiabatic integrals (\ref{eq:2.9a}) and (\ref{eq:S.eqns}). 
Expression (\ref{eq:C14}) is equivalent to $-{\bf B}{\bf\cdot}(T\nabla  S)$
through the temperature expression \eqref{eq:T}. 
Instead if we use the gas law temperatures (\ref{eq:2.16}) and the specific heats (\ref{eq:2.18})
associated with $p_\parallel$ and $p_\perp$, 
then we can write
\begin{equation}
{\bf B}{\bf\cdot}\left(\frac{1}{\rho} \nabla{\bf\cdot}{\sf p}-\nabla h\right)
=-\left( T_\parallel {\bf B}{\bf\cdot}\nabla S_\parallel
+ T_\perp {\bf B}{\bf\cdot}\nabla S_\perp \right) 
\label{eq:C15}
\end{equation}
in term of the gas law entropies (\ref{eq:S.eqns}). 
Substituting expression (\ref{eq:C15}) into the cross helicity {\bf density} 
transport equation (\ref{eq:2.ch1b}) yields
\begin{equation}
\derv{t}({\bf u\cdot B}) +\nabla{\bf\cdot}\left[({\bf u\cdot} {\bf B}){\bf u} 
+{\bf B}\left(\Phi+h-\frac{1}{2} u^2\right)\right]
={\bf B}{\bf\cdot}(T_\parallel \nabla S_\parallel +T_\perp\nabla S_\perp) . 
\label{eq:2.ch7a}
\end{equation}
This result (\ref{eq:2.ch7a}) can be also derived using the \cite{Newcomb62} 
action principle for CGL plasmas shown in Section 4.

The corresponding nonlocal cross helicity  density conservation law is given by 
\begin{equation}
\derv{t}(\tilde{\bf w}{\bf\cdot B}) 
+\nabla{\bf\cdot}\left[(\tilde{\bf w}{\bf\cdot B}){\bf u}
+{\bf B}\left(\Phi+h-\frac{1}{2} u^2\right)\right]=0, 
\label{eq:2.ch7b}
\end{equation}
where
\begin{equation}
\tilde{\bf w}={\bf u}+r_\parallel \nabla S_\parallel +r_\perp \nabla S_\perp, 
\label{eq:2.ch7c}
\end{equation}
with $r_\parallel$ and $r_\perp$ being nonlocal variables 
defined by the equations:
\begin{equation}
\frac{dr_\parallel}{dt}=-T_\parallel, \quad \frac{dr_\perp}{dt}=-T_\perp. 
\label{eq:2.ch7d}
\end{equation}
 
The nonlocal cross helicity {\bf density} conservation law for MHD analogous to equation (\ref{eq:2.ch5}) 
was developed in \cite{Webb14a,Webb14b}, \cite{Webb18}, \cite{Webb19}, \cite{Yahalom17a,Yahalom17b} and \cite{Yahalom21}. 
A topological interpretation of the generalized cross helicity conservation law 
has been found in terms of an MHD Aharonov-Bohm effect 
in \cite{Yahalom17a, Yahalom17b}. \cite{Yahalom13} discusses a topological 
interpetation of magnetic helicity as an Aharonov Bohm effect in MHD. 

If ${\bf B}{\bf\cdot}\nabla S=0$, then the nonlocal conservation law (\ref{eq:2.ch5})
reduces to a local cross helicity {\bf density} conservation law given by $r= 0$ and ${\bf w}={\bf u}$. 
This result is analogous to the local cross helicity conservation law in MHD (e.g. \cite{Webb18}; \cite{Webb19}).  

It is interesting to note that the velocity equation (\ref{eq:2.2.velocity})
can be written in the suggestive form:
\begin{equation}
\frac{d{\bf u}}{dt}
=T\nabla S-\nabla h 
+\frac{\tilde{\bf J}\times {\bf B}}{\rho}-\nabla\Phi , 
\label{eq:chaz10}
\end{equation}
where
\begin{equation}
\tilde{\bf J}\equiv {\bf J}-\nabla\times{\bm\Omega} 
=\frac{\nabla\times\tilde{\bf B}}{\mu_0}, 
\quad
\tilde{\bf B}\equiv {\bf B}\left(1-\frac{\mu_0p_{\Delta}}{B^2}\right). 
\label{eq:chaz12}
\end{equation}
This equation (\ref{eq:chaz10}) turns out to arise directly from the non-canonical Hamiltonian formulation presented in Section~\ref{sec:Hamiltonianeqns}.

\subsection{Magnetic Helicity}\label{sec:magnetichelicity}
In ideal MHD and in ideal CGL plasmas, Faraday's equation (\ref{eq:2.5}) 
written in terms of the electric field 
\begin{equation}
\deriv{\bf B}{t}+\nabla\times {\bf E}=0 
\label{eq:2.39}
\end{equation}
can be uncurled to give ${\bf E}=-{\bf u}\times{\bf B}$
in the form:
\begin{equation}
{\bf E}=-\nabla\phi_E-\deriv{\bf A}{t}=-({\bf u}\times{\bf B})\quad 
\hbox{where}\quad 
{\bf B}=\nabla\times {\bf A}. 
\label{eq:2.40}
\end{equation}
The uncurled form of Faraday's equation (\ref{eq:2.40}) implies
\begin{equation}
\deriv{\bf A}{t}+{\bf E}+\nabla\phi_E=0. 
\label{eq:2.41}
\end{equation}
Combining the scalar product of Faraday's equation (\ref{eq:2.39}) with ${\bf A}$ 
plus the scalar product of the uncurled equation (\ref{eq:2.41}) with ${\bf B}$ 
yields the magnetic helicity transport equation
\begin{equation}
\derv{t}({\bf A\cdot B})+\nabla{\bf\cdot}\left({\bf E}\times {\bf A}+\phi_E {\bf B}\right)=0 , 
\label{eq:2.42}
\end{equation}
which can also be written in the form:
\begin{equation}
\derv{t}({\bf A\cdot B})+\nabla{\bf\cdot}
\left[({\bf A\cdot B}){\bf u}+(\phi_E-{\bf u\cdot A}) {\bf B}\right]=0. 
\label{eq:2.43}
\end{equation}

The total magnetic helicity for a volume $V(t)$ moving with the fluid is given by 
\begin{equation}
H^{\text{M}}=\int_{V(t)} {\bf A}{\bf\cdot}{\bf B}\,d^3x . 
\label{eq:2.44}
\end{equation}
It satisfies the moving balance equation (cf Appendix~C)
\begin{equation}
\frac{dH^{\text{M}}}{dt}=-\oint_{\partial V(t)} (\phi_E -{\bf A\cdot u}){\bf B}{\bf \cdot}\hat{\bf n}\, dA , 
\label{eq:2.47}
\end{equation}
where $\hat{\bf n}$ is the outward unit normal of the moving boundary $\partial V(t)$. 
Thus, if ${\bf B}$ is perpendicular to the boundary, i.e. ${\bf B\cdot}\hat{\bf n}=0$, 
then $H^{\text{M}}$ is conserved in the flow. 
This discussion applies both to ideal MHD and also to ideal CGL plasmas.
(e.g. \cite{Woltjer58},\cite{Kruskal58}, \cite{Moffatt69, Moffatt92}, \cite{Berger84}; \cite{Arnold98}). 

It is beyond the scope of the present exposition to discuss the issues of how to define 
the relative magnetic helicity for volumes $V$ for which ${\bf B\cdot n}\neq 0$ on $\partial V$ 
(e.g. \cite{Berger84}; \cite{Finn85, Finn88}; \cite{Webb10}). 
Similar considerations apply to field line magnetic helicity (e.g. \cite{Prior14}) 
and absolute magnetic helicity (e.g. \cite{Low06, Low11}; \cite{Berger18}). 

It is of interest to note that, under gauge transformations
(i.e. ${\bf A}\to {\bf A}+\nabla\chi$ and $\phi_E\to \phi_E -\deriv{\chi}{t}$ for an arbitrary function $\chi(t,{\bf x})$), 
the moving balance equation is gauge invariant, but the magnetic helicity integral 
changes by addition of a boundary integral:
$H^{\text{M}}\to H^{\text{M}} + \oint_{\partial V(t)} \chi {\bf B}{\bf \cdot}\hat{\bf n}\, dA$. 
However, if the electric field potential $\phi_E$ satisfies the gauge
\begin{equation}
\phi_E={\bf A\cdot u}, 
\label{eq:2.49}
\end{equation}
then the moving flux of the magnetic helicity vanishes
and the resulting (gauge-dependent) magnetic helicity $H^{\text{M}}$ 
is conserved in the flow. 
This gauge turns out to hold precisely when ${\bf A}$ is advected by the flow, 
which will be shown in the next subsection.

\subsection{Lie dragged (advected) quantities}\label{sec:advected}
A quantity, $a$, is Lie dragged by the fluid flow if its advective 
Lie derivative vanishes:
\begin{equation}
\left(\derv{t}+{\cal L}_{\bf u}\right) a=0,
\label{eq:advect}
\end{equation}
where ${\cal L}_{\bf u}$ is the ordinary Lie derivative with respect to the vector field ${\bf u}$
(e.g. \cite{Tur93}; \cite{Webb14a}; \cite{Anco20}). 
For a scalar quantity, its advective Lie derivative reduces to its material derivative
\begin{equation}
\frac{d}{dt} = \derv{t}+{\bf u}{\bf \cdot}\nabla .
\label{eq:materialderiv}
\end{equation}
For quantities that are vector fields or differential forms, the advective Lie derivative 
also contains a rotation-shear term which involves $\nabla {\bf u}$. 

In CGL plasmas, $S$ is an advected scalar,  as are the 
double adiabatic integrals (\ref{eq:2.9a}). 
There are two basic advected non-scalar quantities: 
the differential form 
${\bm\alpha}=\nabla S{\bf\cdot}d{\bf x}\equiv \nabla_i S dx^i$ 
and the vector field ${\bf b}=(1/\rho){\bf B}\equiv b^i\ \nabla_i$ (note 
the vector field ${\bf b}$ is a directional derivative operator). 
The contraction of the vector field ${\bf b}$ with the co-vector
or differential form ${\bm\alpha}$ is a scalar (
note that $\nabla_i\hook dx^j=\delta^i_j$, and 
 ${\bf b}\hook {\bm\alpha}=b^i \alpha_i\equiv
b^i\nabla_iS$).  Notice that 
${\bf b}$ is Lie dragged with the flow, (this statement 
is equivalent to Faraday's equation, when one takes into account 
the mass continuity equation: i.e., 
$(\partial_t+{\cal L}_{\bf u}){\bf b}=0$ which implies 
$\partial_t {\bf b}+[{\bf u},{\bf b}]=0$). 
${\bm\alpha}$ is a one-form or co-vector, that is Lie dragged with the flow.
The inner product of the vector field ${\bf b}$ with the one-form ${\bm\alpha}$
is a scalar invariant which is advected with the 
flow (e.g. \cite{Tur93}). 
%Since the contraction of any advected vector field and advected 
%differential form yields an advected scalar quantity, 
Thus, the quantities 
\begin{equation}
\frac{{\bf B}{\bf\cdot}\nabla S}{\rho}, 
\quad 
\frac{{\bf B}{\bf\cdot}\nabla \bar S_\parallel}{\rho}, 
\quad
\frac{{\bf B}{\bf\cdot}\nabla \bar S_\perp}{\rho}
\label{eq:2.38}
\end{equation}
are advected scalars in CGL plasmas. 

%The corresponding conservation laws underlie the cross helicity transport equations (\ref{eq:2.ch1}) and (\ref{eq:2.ch7a}). 

The gauge (\ref{eq:2.49}) in which the magnetic helicity integral is conserved 
is equivalent to Lie dragging the one-form $\alpha={\bf A\cdot}d{\bf x}$ 
with the fluid flow: 
\begin{equation}
\left( \derv{t}+{\cal L}_{\bf u}\right)({\bf A\cdot}d{\bf x})
=\left(\deriv{\bf A}{t}-{\bf u\times B}+\nabla({\bf A\cdot u})\right){\bf\cdot} d{\bf x}=0 , 
\label{eq:2.51}
\end{equation}
which vanishes due to the uncurled form of Faraday's equation (\ref{eq:2.40}) 
combined with equation (\ref{eq:2.49}). 
\cite{Holm83a, Holm83b} used this advected-${\bf A}$ gauge in 
the formulation of non-canonical Poisson brackets for MHD and for multi-fluid plasmas 
(see also \cite{Gordin87, Gordin89}, \cite{Padhye96a,Padhye96b} and \cite{Padhye98}).

%section3
\section{The Lagrangian Map}\label{sec:Lagrangianmap}

In a Lagrangian formulation of MHD and CGL plasmas, 
fluid elements are given labels that are constant in the fluid flow. 
The simplest labeling consists of initial values ${\bf x}_0={\bf x}(0)$ 
for integrating the flow equations of a fluid element 
\begin{equation}
\frac{d{\bf x}(t)}{dt}={\bf u}({\bf x}(t), t), 
\label{eq:3.0a}
\end{equation}
in which the fluid velocity ${\bf u}$ is assumed to be a known function of ${\bf x}$ and $t$. 
In general, fluid labels are given by functions ${\bf a}={\bf a}({\bf x}_0)$. 

A Lagrangian map is an invertible mapping from the Lagrangian fluid labels ${\bf a}$ to the Eulerian position coordinates ${\bf x}(t)={\bf X}({\bf a},t)$ for the motion of a fluid element. 
For simplicity, we will take ${\bf x}$ and ${\bf a}={\bf x}_0$ to be expressed 
in terms of Cartesian coordinates ${\bf x}=(x,y,z)$ and ${\bf x}_0=(x_0,y_0,z_0)$. 
Then, the Lagrangian map takes the form 
\begin{equation}
{\bf x}={\bf X}({\bf x}_0) 
\label{eq:3.0b}
\end{equation}
with the $t$ dependence being suppressed in the notation. 
Invertibility implies that the Jacobian of this map is non-degenerate:
\begin{equation}
J=\det(X_{ij})\neq 0
\quad\hbox{where}\quad 
X_{ij}\equiv\deriv{X^i({\bf x}_0)}{x_0^j} .
\label{eq:3.3}
\end{equation}
We will write $Y_{ij}$ to denote the inverse of the matrix $X_{ij}$, whereby 
\begin{equation}
Y_{ik} X_{kj}=X_{ik} Y_{kj}=\delta_{ij} \label{eq:3.3a}
\end{equation}
with $\delta_{ij}$ denoting the components of the identity matrix. 
Recall the standard formulae:
\begin{equation}
Y_{ij}=J^{-1} A_{ji}, 
\quad
X_{ji}A_{ki} =A_{ik}X_{ij}=J\delta_{jk}, 
\quad
\deriv{J}{X_{ij}} = A_{ij} ,
\label{eq:3.5}
\end{equation}
where $A_{ij}={\rm cofac}(X_{ij})$ is the cofactor matrix of $X_{ij}$. 
In addition, note that 
\begin{equation}
\dot J = \deriv{J}{X_{ij}} \dot X_{ij} = A_{ij}\deriv{\dot X^i}{x_0^j} . 
\label{eq:3.4}
\end{equation}

\subsection{Map Formulas}\label{sec:mapformula}
We  now formulate the maps between the CGL plasma variables 
$\rho$, $S$, ${\bf B}$, $p_\parallel$, $p_\perp$, 
and their Lagrangian counterparts 
$\rho_0({\bf x}_0)$, $S_0({\bf x}_0)$, ${\bf B}_0({\bf x}_0)$, $p_\parallel({\bf x}_0)$, $p_\perp({\bf x}_0)$. 

The density $\rho_0$ of a fluid element with label ${\bf x}_0$ 
is related to the Eulerian density by the mass conservation equation 
\begin{equation}
\rho d^3x=\rho_0d^3x_0, 
\label{eq:3.1}
\end{equation}
where $d^3x = J d^3x_0$. 
This implies 
\begin{equation}
\rho=\frac{\rho_0}{J} . 
\label{eq:3.2}
\end{equation}

From \cite{Newcomb62} the Cartesian components of $B^i$ in the 
Eulerian frame are related to the Lagrangian magnetic field component 
$B_0^k$ by the equation:
\begin{equation}
B^i=\frac{X_{ik} B_0^k}{J}. \label{eq:3.9a}
\end{equation}
The derivation of (\ref{eq:3.9a}) follows by noting:
\begin{align}
B^id\sigma_i=&B^k_0d\sigma_{0k}, \label{eq:3.10a}\\
d^3x=&dx^id\sigma_i=J d^3x_0=Jdx^k_0 d\sigma_{ok}
\equiv JY_{ki} dx^id\sigma_{0k}. \label{eq:3.10b}
\end{align}
Equation (\ref{eq:3.10a}) describes the conservation of magnetic flux, 
where $d\sigma_i$ is the flux tube area normal to the $x^i$ coordinate surface,
and $d\sigma_{0k}$ is the flux tube area normal to the $x^k_0$ surface
 in the Lagrangian frame. 
Equation (\ref{eq:3.10b}) relates the volume elements 
$d^3x$ to $d^3x_0$. From (\ref{eq:3.10b}) one obtains
\begin{equation}
d\sigma_i=A_{ik}d\sigma_0^k, \label{eq:3.10c}
\end{equation}
for the transformation between the area elements $d\sigma_i$ and $d\sigma_0^k$.
The magnetic flux conservation equation (\ref{eq:3.10a}) now gives the 
transformation $B_0^k=A_{ik} B^i$ which in turn implies the transformation
(\ref{eq:3.9a}).  
%The magnetic field ${\bf B}_0$ attached to a fluid element with label ${\bf x}_0$ 
%is related to the Eulerian magnetic field by using the frozen-in property of 
%magnetic flux (e.g. \cite{Parker79}) expressed as 
%\begin{equation}
%{\bf B}{\bf\cdot}\hat{\bf n}dA
%= {\bf B}_0{\bf\cdot}\hat{\bf n}_0dA_0
%\label{eq:advect.Bflux}
%\end{equation}
%where $dA$ is the area element of any surface moving with the fluid flow,
%and $\hat{\bf n}$ is the outward unit normal. 
%Since $d^3x=Jd^3x_0$ implies $\hat{\bf n}{\bf\cdot}d{\bf x}dA=J\hat{\bf n}_0{\bf\cdot}d{\bf x}_0dA_0$, 
%where $dA=dA_0$ due to the surface being frozen-in, 
%we see that $\hat{n}_0^i = J^{-1} X_{ij} \hat{n}{}^j$. 
%Substituting this relation into (\ref{eq:advect.Bflux}) gives 
%\begin{equation}
%B^i=J^{-1} X_{ij}B_0^j . 
%\label{eq:3.6}
%\end{equation}

An alternative form of the relation (\ref{eq:3.9a}) arises from the Lie dragged or frozen-in vector field
\begin{equation}
{\bf b}=\frac{B^i}{\rho} \derv{x^i}=\frac{B_0^k}{\rho_0} \derv{x_0^k}\equiv 
\frac{B_0^k}{\rho_0} X_{ik}\derv{x^i}, \label{eq:3.7}
\end{equation}
which implies the transformation (\ref{eq:3.9a}) between $B^i$ and $B_0^k$.
In (\ref{eq:3.7}) we use the modern differential geometry 
notion of a vector field as a directional derivative 
operator (e.g. \cite{Misner73}).  
%\frac{\bf B}{\rho} = \frac{{\bf B}_0}{\rho_0} . 
%\label{eq:3.7} 
%\end{equation}
%By considering any curve that moves with the fluid flow, 
%it follows that 
%$({\bf B}/\rho){\bf\cdot}d{\bf s} = ({\bf B}_0/\rho_0) {\bf\cdot}d{\bf s}_0$
%is an advected scalar, 
%where $d{\bf s}=d{\bf s}_0$ is the line element of the curve. 
%(e.g. \cite{Tur93}; \cite{Webb14a};  \cite{Webb18}). 
%The line element can be expressed as $d{\bf s} = \hat{\bf s}\,ds$, 
%with $ds = \hat {\bf s}{\bf\cdot}d{\bf x}$, 
%in terms of a unit tangent vector $\hat{\bf s}$ to the curve. 
%Then, $ds=ds_0$ implies $\hat{s}^i X_{ij} = \hat{s}_0^j$, 
%and thus 
%$B^i \hat{s}^i/\rho = B_0^i\hat{s}_0^i/\rho_0 = B_0^i X_{ji} \hat{s}^j J^{-1}/\rho$
%through (\ref{eq:3.2}), which yields (\ref{eq:3.6}). 

From the relation (\ref{eq:3.9a}), we have 
\begin{equation}
B^2=\frac{\zeta^2 B_0^2}{J^2} 
\quad\hbox{where}\quad
\zeta^2=X_{ij}\tau_0^j X_{ik}\tau_0^k 
\label{eq:3.9}
\end{equation}
in terms of the unit vector $\tau_0=B_0^{-1}{\bf B}_0$ along the magnetic field ${\bf B}_0$ (cf (\ref{eq:2.3})),
with $B_0=|{\bf B}_0|$ being the magnetic field strength at ${\bf x}_0$. 
Hence, we obtain:
\begin{equation}
B=\frac{\zeta\,B_0}{J},
\quad 
\tau^i =\frac{B^i}{B} =\frac{X_{ij} \tau_0^j}{\zeta} . 
\label{eq:3.11}
\end{equation}
The latter relation can be inverted to obtain $\bm{\tau}_0$ 
in terms of $\bm{\tau}$ through the formulae (\ref{eq:3.5}):
\begin{equation}
\tau_0^i = \frac{\zeta}{J} A_{ji} \tau^j . 
\label{eq:3.11b}
\end{equation}

Next, from the double adiabatic conservation laws (\ref{eq:2.9}), we have 
the frozen-in quantities 
\begin{equation}
\frac{p_\perp}{\rho B}=\frac{p_{\perp 0}}{\rho_0B_0}, 
\quad 
\frac{p_\parallel B^2}{\rho^3}=\frac{p_{\parallel 0} B_0^2}{\rho_0^3} . 
\label{eq:3.12}
\end{equation}
By combining these equations with the previous ones (\ref{eq:3.11}) and (\ref{eq:3.2}), 
we obtain 
\begin{equation}
p_\parallel=\frac{p_{\parallel 0}}{J\zeta^2}, 
\quad p_\perp=\frac{p_{\perp 0}\zeta}{J^2}, 
\label{eq:3.13}
\end{equation}
which gives the respective relations between the pressures $p_{\parallel 0}$ and $p_{\perp 0}$
at a fluid element with label ${\bf x}_0$ 
and the Eulerian pressures $p_\parallel$ and $p_\perp$. 

Finally, we note that the Eulerian entropy $S$ is equal to the entropy $S_0$ 
at a fluid element with label ${\bf x}_0$, 
since $S$ is frozen-in by the transport equation (\ref{eq:2.4}).

\subsection{Lagrangian Variational Principle and 
Euler-Lagrange Equations}\label{sec:Lagrangian}
The Lagrangian variational principle for MHD and for CGL plasmas was obtained in \cite{Newcomb62}): 
\begin{equation}
{\mathcal A}=\int_{t_0}^{t_1} \int_{V} L\, d^3x\, dt = \int_{t_0}^{t_1}\int_{V_0} L_0\, d^3x_0\, dt , 
\label{eq:3.15}
\end{equation}
where $L$ is the Lagrangian in Eulerian variables and $L_0$ is its counterpart
arising through the Lagrangian map. 
Here $V$ is a fixed spatial domain and $[t_0,t_1]$ is a fixed time interval. 
In general, $L$ will be given by the kinetic energy density minus the potential energy density for a fluid element. 
The kinetic energy density for both MHD and CGL plasmas is simply 
$\frac{1}{2}\rho u^2$, with $u=|{\bf u}|$. 
Subtracting this expression from the total energy density for a CGL plasma
(cf the moving energy balance equation (\ref{eq:2.23})) 
\begin{equation}
H \equiv \frac{1}{2}\rho u^2+\rho\Phi +\varepsilon+\frac{1}{2\mu_0} B^2
\label{eq:3.28}
\end{equation}
yields the potential energy density 
$\rho\Phi +\varepsilon+\frac{1}{2\mu_0}B^2$, 
where $\varepsilon$ is the internal energy density (\ref{defn.energydens}).
Thus, a CGL plasma has the Lagrangian
\begin{equation}
L=\frac{1}{2}\rho u^2 -\rho\Phi -\varepsilon -\frac{1}{2\mu_0}B^2 , 
\label{eq:3.16}
\end{equation}
where, for the sake of completeness, we have included the energy contributed by 
an external gravitational potential $\Phi({\bf x})$
(e.g. which would arise from the gravitational field of the Sun in the case of solar and interplanetary physics). 
The corresponding Lagrangian $L_0$ is obtained from the relation 
$Ld^3x=L_0d^3x_0 = LJ d^3x_0$, which gives
\begin{equation}
L_0=J L . 
\label{eq:3.18}
\end{equation}
Substituting the Eulerian expressions (\ref{eq:3.16}), (\ref{eq:3.13}), (\ref{eq:3.6}), (\ref{eq:3.2}),  and (\ref{eq:3.0b}) into $L_0$, 
and then using expression (\ref{defn.energydens}) for the internal energy density
in terms of the pressures $p_\parallel$ and $p_\perp$,  
we obtain 
\begin{equation}
L_0 =\rho_0({\bf x}_0)\left(\frac{1}{2}  |\dot{\bf X}|^2-\Phi({\bf X})\right)
-\left(\frac{p_{\parallel 0}({\bf x}_0)}{2\zeta^2}
+\frac{p_{\perp 0}({\bf x}_0)\zeta}{J}\right)
-\frac{\zeta^2 B_0({\bf x}_0)^2}{2\mu_0 J} . 
%-\frac{X_{ij} B_0^j({\bf x}_0) X_{ik} B_0^k}{2\mu_0 J}
\label{eq:3.19}
\end{equation}

The Euler-Lagrange equations of the resulting action principle (\ref{eq:3.15})
for a CGL plasma are obtained by variation of the Cartesian components $X^i$ of the Lagrangian map (\ref{eq:3.0b}), 
modulo boundary terms:
$\delta{\mathcal A}/\delta X^i =0$. 
In essence, we work in a reference frame moving with the fluid flow, 
where the dynamics is described in terms of how the Eulerian position ${\bf x}$
of the frame varies with the fluid labels ${\bf x}_0$ and the time $t$
through the Lagrangian map (\ref{eq:3.0b}). 
Equivalently, the Euler-Lagrange equations are given by applying to $L_0$ 
the Euler operator (variational derivative) $E_{X^i}$ in the calculus of variations,
where $X^i(x_0^j,t)$ is a function of $x^j_0$ and $t$. 
Thus, 
\begin{equation}
\frac{\delta{\mathcal A}}{\delta X^i} \equiv E_{X^i}(L_0) 
=\deriv{L_0}{X^i} -\derv{t}\left(\deriv{L_0}{\dot X^i}\right)
-\derv{x_0^j}\left(\deriv{L_0}{X_{ij}}\right)=0 ,
\label{eq:3.20}
\end{equation}
where $\partial/\partial_{t}$ and $\partial/\partial_{x_0^j}$ act as total derivatives. 
(See section~\ref{sec:ConsLaws} for more details of variational calculus.)
In Appendix~C, 
we show that when this equation (\ref{eq:3.20}) is simplified and expressed in terms of Eulerian variables, 
it reduces to the Eulerian momentum equation (\ref{eq:2.2}).

\subsection{Hamilton's Equations}\label{sec:Hamiltonian}
We now derive the corresponding Hamilton formulation of the Euler-Lagrange equation (\ref{eq:3.20}) for a CGL plasma. 
Here we will write $X^i=x^i$ for simplicity. 

First, the generalized momentum is defined as:
\begin{equation}
\pi_i=\deriv{L_0}{\dot{x}^i}=\rho_0 \dot x^i. 
\label{eq:3.24}
\end{equation}
Next, the Hamiltonian density is defined by the Legendre transformation:
\begin{equation}
H_0=\pi_k\dot x^k -L_0. 
\label{eq:3.25}
\end{equation}
After substitution of the CGL Lagrangian (\ref{eq:3.19}) along with $\dot x^i$ in terms of the momentum (\ref{eq:3.24}), 
we then obtain the expression 
\begin{equation}
H_0=\frac{\pi_i \pi_i}{2\rho_0}+\rho_0\Phi({\bf x}) 
+ \frac{p_{\parallel 0}}{2\zeta^2}+\frac{p_{\perp 0}\zeta}{J}
+\frac{X_{ij}X_{ik}B_0^jB_0^k}{2\mu_0 J} . 
\label{eq:3.26}
\end{equation}
In Eulerian variables, this expression is given by 
\begin{equation}
H_0 = J H
\label{eq:3.27}
\end{equation}
where $H$ is the Eulerian total energy density (\ref{eq:3.28}). 
The resulting Hamiltonian equations of motion consist of 
\begin{equation}
\dot{x}^i=\frac{\delta \Ham}{\delta\pi^i}, 
\quad 
\dot{\pi}^i =-\frac{\delta \Ham}{\delta x^i} ,
\label{eq:3.33}
\end{equation}
which yield the momentum relation (\ref{eq:3.24})
and the Euler-Lagrange equation (\ref{eq:3.20}), 
where the Hamiltonian function is defined as 
\begin{equation}
\Ham=\int_{V} H_0\, d^3x_0 =\int_{V} H\, d^3x . 
\label{eq:3.29}
\end{equation}

As in ordinary Hamiltonian mechanics, 
with the use of the canonical variables $q^i= x^i$ and $p^i=\pi^i$, 
Hamilton's equations (\ref{eq:3.33}) can be written in a canonical Poisson bracket formulation: 
\begin{equation}
\dot{\F}=\left\{\F,\Ham\right\}
=\int\left( 
%\F_{\bf q} {\bf\cdot}\Ham_{\bf p} -\F_{\bf q}{\bf\cdot}\Ham_{\bf p}
\F_{q^i} \Ham_{p^i} -\F_{q^i}\Ham_{p^i}
\right)\, d^3x_0 . 
\label{eq:3.34}
\end{equation}
Here $\F$ is a general functional of $q^i$ and $p^i$ 
(see \cite{Morrison82}). 
There is an equivalent, non-canonical Poisson bracket that employs only Eulerian variables, 
which we will present in Section~\ref{sec:Poissonbracket}.

%section 4
\section{Euler-Poincar\'e Action Principle}\label{sec:EulerPoincare}

In this section, 
we formulate the Euler-Poincar\'e action principle for the CGL plasma equations.  
We follow the developments in \cite{Holm98}, \cite{Webb14b}, 
which provide the Euler-Poincar\'e formulation for MHD. 
The main difference for a CGL plasma compared to MHD is that 
the internal energy density $\varepsilon=\rho e$ (cf (\ref{eq:2.16j}))
depends on $\rho$, $S$, and $B$, 
whereas for MHD it depends only on $\rho$ and $S$. 

The Lagrangian and Hamiltonian action principles in Section~\ref{sec:Lagrangianmap}
are based on a reference frame that is attached to fluid elements, whereby 
the Eulerian position ${\bf x}$ of the frame is a function of 
fluid element labels ${\bf x}_0$ and time $t$,
as given by the Lagrangian map (\ref{eq:3.0b}). 
In contrast, the Euler-Poincar\'e action principle is based on the use of a fixed Eulerian reference frame, 
in which the motion of a fluid element is given by inverting the Lagrangian map 
so that the fluid element label ${\bf x}_0$ becomes a function of ${\bf x}$ and $t$. 
To set up this formulation, 
it is convenient to introduce the notation 
\begin{equation}
g{\bf x}_0 = {\bf x} 
\label{eq:4.0}
\end{equation}
with $g$ at any fixed time $t$ representing an element in the group of diffeomorphisms on Euclidean space in Cartesian coordinates ${\bf x}$ and ${\bf x}_0$. 
Note that $g$ has an inverse $g^{-1}$ defined by $g^{-1}{\bf x} = {\bf x}_0$, 
while $g|_{t=0}$ is just the identity map. 
In (\ref{eq:4.0}) and the sequel, 
the $t$ dependence of $g$ and ${\bf x}_0$ is suppressed for simplicity of notation; 
an overdot stands for $\dot{}=\partial/\partial t$. 
The diffeomorphism group will be denoted $G\equiv {\rm Diff}({\mathbb R}^3)$. 

The coordinate components of a fluid element label, $x_0^i$, 
represent advected quantities: 
\beqn
\left(\derv{t}+{\bf u}{\bf\cdot}\nabla\right)x_0^i=0 . 
%=\left(\derv{t}+{\cal L}_{\bf u}\right)x_0^i
\label{eq:4.3}
\eeqn
%where ${\cal L}_{\bf u}$ is the Lie derivative with respect to the Eulerian vector field ${\bf u}$. 
From equations (\ref{eq:4.0}) and (\ref{eq:4.3}), 
we see that 
\begin{equation}
\dot{x}_0^i=(\dot{(g^{-1})}{\bf x})^i=-(g^{-1}\dot{g}g^{-1} {\bf x})^i
=-(g^{-1}\dot{g}{\bf x}_0)^i=-{\bf u}{\bf\cdot}\nabla x_0^i . 
\end{equation}
Hence, we obtain 
\begin{equation}
g^{-1}\dot{g} = {\bf u}{\bf\cdot}\partial_{\bf x} ,
\label{eq:4.4}
\end{equation}
which can be viewed as both a left-invariant vector field in the tangent space of $G$ 
and the directional derivative along the fluid flow in Euclidean space. 
The property of left-invariance means that, for any fixed element $h$ in $G$, 
$g\to hg$ implies $\eta\to (hg)^{-1}\dot{(hg)}=g^{-1}h^{-1} h\dot{g}=g^{-1}\dot{g} =\eta$, 
with $\dot{h}\equiv 0$. 

Compared to the Lagrangian formulation,
whose dynamical variable is ${\bf x}({\bf x}_0,t)$, 
the Euler-Poincar\'e formulation uses the dynamical variables $\rho$, $S$, $B$, and ${\bf u}$, 
which are functions of ${\bf x}$ and $t$. 
Variations of the Euler-Poincar\'e variables are defined through 
variation of the inverse Lagrangian map, $g$, in the following way. 

We consider a general variation $\delta g$ and define 
\beqn
\eta \equiv g^{-1}\delta g 
\label{eq:4.5}
\eeqn
which represents a left invariant vector field on the group $G$. 
Similarly to the identification (\ref{eq:4.4}), 
we can write
\begin{equation}
\eta = \bm{\eta}{\bf\cdot}\partial_{\bf x}
\label{eq:4.6}
\end{equation}
viewed as the directional derivative associated with $\delta g$,
where $\bm{\eta}$ is an Eulerian vector field. 
(In general, 
${\bf\cdot}\partial_{\bf x}$ identifies Eulerian vector fields in Euclidean space 
with left-invariant vector fields on the diffeomorphism group $G$.)

Now, the variation $\delta g$ is intended to leave the fixed Eulerian reference frame unchanged, 
whereby $\delta {\bf x}\equiv 0$. 
Taking the corresponding variation of the inverse Lagrangian map (\ref{eq:4.0}) 
yields 
$\delta x_0^i = \delta (g^{-1} x)^i = -(g^{-1}(\delta g)g^{-1} x)^i = -\eta x_0^i$. 
Thus, since $x_0^i$ is a function of $x^j$ and $t$, we obtain 
\begin{equation}
\delta x_0^i = -{\bm \eta}{\bf\cdot}\nabla x_0^i =-\lieder{\bm \eta} x_0^i ,
\end{equation}
where $\lieder{\bm \eta}$ is the Lie derivative with respect to the Eulerian vector field $\bm{\eta}$. 

More generally, the same Lie derivative operation is used to define 
the variation of any advected quantity (a scalar, or a vector, or a differential form), $a$:
\begin{equation}
\delta a = -\lieder{\bm \eta} a,
\label{eq:4.7}
\end{equation}
where $a$ satisfies the advection equation (\ref{eq:advect}). 
Expressions for the induced variations of $\rho$, $S$, ${\bf B}$ 
can then be deduced by considering advected quantities in terms of those variables
(cf Section~\ref{sec:advected} and Section~\ref{sec:mapformula}). 
The basic advected quantities in a CGL plasma are:
\begin{equation}
S,
\quad
(1/\rho){\bf B},
\quad
\rho\, d^3x,
\quad
{\bf B}{\bf\cdot}\hat{\bf n}dA,
\label{eq:4.8}
\end{equation}
where $dA$ is the area element on a surface moving with the fluid. 
(These quantities are also referred to as a Cauchy invariant.)
Since $S$ is advected, its variation is given directly by equation (\ref{eq:4.7}). 
Advection of $\rho\, d^3x$ combined with $\delta x^i=0$ implies that 
$(\delta\rho)\, d^3x = \delta(\rho\, d^3x)  = -\lieder{\bm \eta} (\rho\, d^3x)  
= -(\lieder{\bm \eta} \rho + \rho\nabla{\bf\cdot}{\bm \eta})\, d^3x$
due to the well known expansion/contraction formula 
$\lieder{\bf v}(d^3x) = \nabla{\bf\cdot}{\bf v} d^3x$ 
holding for any vector field ${\bf v}$. 
Thus, we have
\begin{align}
\delta S & = -{\bm \eta}{\bf\cdot}\nabla S, 
\label{eq:4.9a}
\\
\delta \rho & = -\nabla{\bf\cdot}(\rho{\bm \eta}) . 
\label{eq:4.9b}
\end{align}
Next, since $(1/\rho){\bf B}$ is advected, 
this gives
\begin{equation}
\begin{aligned}
\delta\left(\frac{{\bf B}}{\rho}\right)
&  = -\lieder{\bm \eta}\left(\frac{{\bf B}}{\rho}\right)
= \frac{(\lieder{\bm \eta}\rho){\bf B}}{\rho^2}
- \frac{\lieder{\bm \eta}{\bf B}}{\rho} \\
& = -\frac{(\delta\rho){\bf B}}{\rho^2}  +\frac{\delta{\bf B}}{\rho} . 
\end{aligned}
\end{equation}
Thus, after substituting $\delta\rho$ from (\ref{eq:4.9b}), we obtain 
\begin{equation}
\delta{\bf B} = -(\nabla{\bf\cdot}{\bm \eta}){\bf B} -\lieder{\bm \eta} {\bf B} 
= {\bf B}{\bf\cdot}\nabla \bm{\eta} - \nabla {\bf\cdot}(\bm{\eta}{\bf B})
=\nabla\times({\bm \eta}\times{\bf B}) ,
\label{eq:4.9c}
\end{equation}
where $\lieder{\bm \eta}{\bf B} = [{\bm\eta},{\bf B}]$ is the commutator of vector fields. 

Finally, the induced variation of ${\bf u}$ is given by combining the equations
\begin{equation}
(\delta{\bf u}){\bf\cdot}\partial_{\bf x}
= \delta(g^{-1}\dot{g}) = -g^{-1}(\delta g)g^{-1}\dot{g} +g^{-1}(\dot{\delta g})
= -\bm{\eta}{\bf\cdot}\partial_{\bf x} \otimes {\bf u}{\bf\cdot}\partial_{\bf x} +g^{-1}(\dot{\delta g})
\end{equation}
and
\begin{equation}
(\dot{\delta g}) = \dot{(g\eta)} = \dot{g}\eta + g\dot{\eta} 
= g({\bf u}{\bf\cdot}\partial_{\bf x}\otimes \bm{\eta}{\bf\cdot}\partial_{\bf x} 
+\dot{\bm{\eta}}{\bf\cdot}\partial_{\bf x}) . 
\end{equation}
This yields
\begin{equation}
\delta{\bf u}{\bf\cdot}\partial_{\bf x}  
=  \dot{\bm{\eta}}{\bf\cdot}\partial_{\bf x} 
- [\bm{\eta}{\bf\cdot}\partial_{\bf x},{\bf u}{\bf\cdot}\partial_{\bf x}]
=\left( \dot{\bm{\eta}} - \lieder{\bm \eta} {\bf u}\right){\bf\cdot}\partial_{\bf x}
\end{equation}
where $[\cdot,\cdot]$ is the commutator of vector fields 
which coincides with the Lie bracket.
Thus, we obtain the variation: 
\begin{equation}
\delta{\bf u} = \dot{\bm{\eta}} - \lieder{\bm \eta} {\bf u}
= (\partial/\partial t + \lieder{{\bf u}})\bm{\eta}
= \frac{d}{dt} \bm{\eta}
\equiv \deriv{\bm\eta}{t}+\left[{\bf u},{\bm\eta}\right].
\label{eq:4.10}
\end{equation}
%in terms of the material derivative (\ref{eq:materialderiv}).

\subsection{The Euler-Poincar\'e equation}
For CGL plasmas, the Euler-Poincar\'e action principle is given by 
\beqn
{\cal J}=\int_{t_0}^{t_1} \int_{V} L\, d^3x\, dt ,
\label{eq:EPaction}
\eeqn
where $L$ is the Lagrangian density (\ref{eq:3.16}). 
Here $V$ is a fixed spatial domain and $[t_0,t_1]$ is a fixed time interval. 
The stationary points of this action principle, $\delta {\cal J}=0$, 
under the variations (\ref{eq:4.10}), (\ref{eq:4.9a}), (\ref{eq:4.9b}), (\ref{eq:4.9c})
of the respective variables ${\bf u}$, $S$, $\rho$, ${\bf B}$, 
turn out to yield the Eulerian momentum equation (\ref{eq:2.2}),
as we will now show. 

The derivation of the stationary points is non-trivial because of the form of the variations
in terms of $\eta$ given by equations (\ref{eq:4.5}) and (\ref{eq:4.6}). 
Specifically, the variation of ${\cal J}$ needs to be put into the form
\beqn
\delta{\cal  J}=\int_{t_0}^{t_1} \left\langle \eta, F\right\rangle\, dt
=\int_{t_0}^{t_1} \int_{V} {\bm \eta}{\bf\cdot} {\bf F}\,d^3x\, dt
\label{eq:4.12}
\eeqn
modulo boundary terms,
where $F={\bf F}{\bf\cdot} \partial_{\bf x}$ is both 
a left-invariant vector field in the tangent space of $G$ 
and the directional derivative along a vector field ${\bf F}$ in Euclidean space. 
Then, since ${\bm \eta}$ is an arbitrary vector field 
(corresponding to an arbitrary variation $\delta g$), 
the equation yielding the stationary points of ${\cal J}$ is ${\bf F}=0$. 

We now proceed to find $\delta{\cal J}$ and ${\bf F}$. 
A general expression is available for ${\bf F}$ in \cite{Holm98}, 
which involves the diamond operator, $\diamond$, defined by property:    
\beqn
\left\langle\frac{\delta L}{\delta a}\diamond a,\eta\right\rangle
=-\left\langle\frac{\delta L}{\delta {a}},{\cal L}_{\eta}(a)\right\rangle , 
\label{eq:4.13}
\eeqn
where, as earlier, $a$ is an advected quantity. 
It will be instructive to instead show how to obtain ${\bf F}$ from the variation 
(\ref{eq:4.10}), (\ref{eq:4.9a}), (\ref{eq:4.9b}), (\ref{eq:4.9c})
directly using the Eulerian variables. 

From the Lagrangian density (\ref{eq:3.16}), we obtain:
\begin{equation}
\begin{aligned}
&
\frac{\delta L}{\delta{\bf u}} 
=\rho{\bf u}, 
\quad
\frac{\delta L}{\delta S}
=-\varepsilon_S
=-\rho T, 
\quad
\frac{\delta L}{\delta\rho}
=\frac{1}{2}u^2-\varepsilon_\rho-\Phi
=\frac{1}{2}u^2-h-\Phi,
\\
&
\frac{\delta L}{\delta{\bf B}}
=-\varepsilon_B \bm{\tau} -\frac{\bf B}{\mu_0} 
%= -\frac{\bf B}{\mu_0}+\frac{p_{\Delta}}{B} \bm{\tau} 
=\left(\frac{p_{\Delta}}{B^2}  -\frac{1}{\mu_0}\right) {\bf B} 
\equiv - \frac{1}{\mu_0}\tilde {\bf B} , 
\end{aligned}
\label{eq:4.15}
\end{equation}
using the thermodynamic relations (\ref{eq:2.16l}),  
where $T$ is the temperature and $h$ is the enthalpy (\ref{eq:2.ch4}),
and $p_{\Delta}=p_\parallel-p_\perp$ is the pressure anisotropy. 
Here we have also used the identity $\delta B = \bm{\tau}{\bf\cdot}\delta{\bf B}$,
with ${\bm \tau}={\bf B}/B$ being the unit vector along ${\bf B}$.
Note that $\tilde{\bf B}$ is related ${\bf B}$ by equation (\ref{eq:chaz12}). 
The main difference in these variational derivative expressions (\ref{eq:4.15})
compared to the MHD case is the addition of the anisotropy term in $\delta L/\delta{\bf B}$, i.e.\ ${\bf B}$ is replaced by $\tilde{\bf B}$. 

The total variation of $L$ is given by combining the products of 
the variational derivatives (\ref{eq:4.15}) and the corresponding variations 
(\ref{eq:4.10}), (\ref{eq:4.9a}), (\ref{eq:4.9b}), (\ref{eq:4.9c}). 
This yields
\begin{equation}
\begin{aligned}
\delta L 
& = \frac{\delta L}{\delta{\bf u}}{\bf\cdot}\delta{\bf u}
+ \frac{\delta L}{\delta S}{\delta S}
+ \frac{\delta L}{\delta\rho}{\delta\rho}
+ \frac{\delta L}{\delta{\bf B}}{\bf\cdot}\delta{\bf B} \\
& = \rho{\bf u}{\bf\cdot} \left( \dot{\bm{\eta}} - \lieder{\bm \eta} {\bf u}\right)
-\rho T \left( -{\bm \eta}{\bf\cdot}\nabla S \right)
+\left( \frac{1}{2}u^2-h-\Phi\right)\left( -\nabla{\bf\cdot}(\rho{\bm \eta}) \right)
\\&\qquad
-\frac{1}{\mu_0}\tilde{\bf B}{\bf\cdot} \left( \nabla\times({\bm \eta}\times{\bf B}) \right) .
\end{aligned}
\label{eq:4.16}
\end{equation}
The next step is to bring each term into the form (\ref{eq:4.12}), 
modulo a total time derivative and a total divergence, 
using integration by parts. 
The first term in expression (\ref{eq:4.16}) expands out to give:
\begin{equation}
\begin{aligned}
\rho{\bf u}{\bf\cdot} \dot{\bm{\eta}}
+\rho{\bf u}{\bf u}{\bf :}\nabla\bm{\eta} 
-\rho  (\bm{\eta} {\bf\cdot}\nabla{\bf u}){\bf\cdot}{\bf u}
& = \bm{\eta} {\bf\cdot}\left( 
-\deriv{(\rho{\bf u})}{t} 
-\frac{1}{2} \rho \nabla(u^2) -\nabla{\bf\cdot}(\rho{\bf u}{\bf u}) 
\right)
\\&\qquad
+\deriv{(\rho{\bf u}{\bf\cdot}\bm{\eta})}{t}
+\nabla{\bf\cdot}\left( \rho ({\bf u}{\bf\cdot}\bm{\eta}) {\bf u} \right) .
\end{aligned}
\label{eq:4.17a}
\end{equation}
Similarly, the third term in (\ref{eq:4.16}) yields
\begin{equation}
{\bm \eta}{\bf\cdot} \left( \rho\nabla \left(\frac{1}{2}u^2-h-\Phi\right) \right)
+\nabla{\bf\cdot}\left( 
-\left(\frac{1}{2}u^2-h-\Phi\right)\rho{\bm \eta} 
\right) .
\label{eq:4.17b}
\end{equation}
The second term in (\ref{eq:4.16}) simply gives
\begin{equation}
{\bm \eta}{\bf\cdot}\left( \rho T \nabla S \right) .
\label{eq:4.17c}
\end{equation}
The last term in (\ref{eq:4.16}) can be rearranged by cross-product identities:
\begin{equation}
({\bf B}\times {\bm \eta}){\bf\cdot}\left( \frac{1}{\mu_0}\nabla\times \tilde{\bf B} \right) 
%\left(\frac{p_{\Delta}}{B^2} -\frac{1}{\mu_0}\right){\bf B} 
+ \nabla{\bf\cdot}\left( 
\frac{1}{\mu_0}\tilde{\bf B} \times({\bm \eta}\times{\bf B})
\right) ,
\end{equation}
and
\begin{equation}
({\bf B}\times {\bm \eta}){\bf\cdot}\left( \frac{1}{\mu_0}\nabla\times \tilde{\bf B} \right)
= 
{\bm \eta}{\bf\cdot}\left( -\frac{1}{\mu_0} {\bf B}\times\left( \nabla\times \tilde{\bf B} \right) \right) . 
\label{eq:4.17d}
\end{equation}
Now, combining the four terms (\ref{eq:4.17d}), (\ref{eq:4.17c}), (\ref{eq:4.17b}), (\ref{eq:4.17a}), 
and discarding the total derivatives, 
we obtain 
\begin{equation}
\begin{aligned}
\delta L = {\bm \eta}{\bf\cdot}\bigg( 
& -\deriv{(\rho{\bf u})}{t} 
-\nabla{\bf\cdot}(\rho{\bf u}{\bf u}) 
+\rho( T \nabla S -\nabla h-\nabla \Phi ) 
-\frac{1}{\mu_0}{\bf B}\times\left( \nabla\times\tilde{\bf B} \right)
\bigg)
\end{aligned}
\label{eq:4.18}
\end{equation}
modulo total time derivatives and total divergences. 
This yields the desired relation 
\begin{equation}
\int_{V} \delta L \,d^3x
=\int_{V} \bm{\eta}{\bf \cdot} {\bf F} \,d^3x 
= \left\langle \eta, F\right\rangle,
\end{equation}
with
\begin{equation}
{\bf F} = 
-\deriv{(\rho{\bf u})}{t} 
-\nabla{\bf\cdot}(\rho{\bf u}{\bf u}) 
+\rho( T \nabla S -\nabla h-\nabla \Phi ) 
-\frac{1}{\mu_0}{\bf B}\times\left( \nabla\times\tilde{\bf B} \right) . 
\label{eq:4.19}
\end{equation}

The equation ${\bf F}=0$ resulting from the variational principle (\ref{eq:4.12}) 
is called the Euler-Poincar\'e equation. 
It is equivalent to the Eulerian momentum equation (\ref{eq:2.2}),
after the pressure divergence identity (\ref{eq:2.ch2}) is used. 
A direct derivation of this identity is provided in Appendix~B. 
Alternatively, equation (\ref{eq:4.19}) can be viewed as showing how
the pressure divergence identity arises from the variational 
principle (\ref{eq:4.12}). 
Appendix~E derives the identity (\ref{eq:2.ch2}) and the 
Euler-Poincar\'e equation ${\bf F}=0$
using the approach of \cite{Holm98}. 

\cite{Holm86} studied a corresponding Hamiltonian form of the CGL plasma equations 
for both relativistic and non-relativistic flows; 
however, details of the Euler-Poincar\'e formulation were not covered.

%section5
\section{Non-Canonical Poisson Bracket}\label{sec:Poissonbracket}

\cite{MorrisonGreene80, MorrisonGreene82} and \cite{Holm83a, Holm83b}
obtained the non-canonical Poisson bracket for ideal MHD,
which involves the basic variables
$\rho$, $\sigma\equiv\rho S$, ${\bf M}\equiv\rho{\bf u}$, ${\bf B}$:
\begin{equation}
\begin{aligned}
\left\{\F,\G\right\}^{\scriptscriptstyle\text{MHD}}
=-\int_{V} \big\{ 
& \rho\left(\F_{\bf M}{\bf\cdot}\nabla \G_\rho -\G_{\bf M}{\bf\cdot}\nabla \F_\rho\right)
+\sigma\left(\F_{\bf M}{\bf\cdot}\nabla \G_\sigma -\G_{\bf M}{\bf\cdot}\nabla \F_\sigma\right)
\\&
+{\bf M}{\bf\cdot}\left(\F_{\bf M}{\bf\cdot}\nabla \G_{\bf M}-\G_{\bf M}{\bf\cdot}\nabla \F_{\bf M}\right)
+{\bf B}{\bf\cdot}\left( \F_{\bf M}{\bf\cdot}\nabla \G_{\bf B} -\G_{\bf M}{\bf\cdot}\nabla \F_{\bf B}\right)
\\&
+{\bf B}{\bf\cdot}\left( (\nabla \F_{\bf M}){\bf\cdot}\G_{\bf B} -(\nabla \G_{\bf M}){\bf\cdot}\F_{\bf B}\right)
\big\}\, d^3x ,
\end{aligned}\label{eq:5.3}
\end{equation}
where $\F$ and $\G$ are arbitrary functionals,
and subscripts denote a variational derivative. 
This bracket is bilinear, antisymmetric, and obeys the Jacobi identity. 

 It turns out the CGL plasma Poisson bracket obtained by \cite{Holm86},  
$\{\F,\G\}^{CGL}$ has the 
the same form as the MHD plasma Poisson bracket  
 except that the thermodynamics and internal energy $e(\rho,S,B)$
are completely different in the two cases. 
%We indicate this difference
%by using the subscript $*$ to indicate this difference in the evaluation 
%of the bracket. 
\cite{Holm86} 
obtained the bracket:
\begin{equation}
\begin{aligned}
\left\{\F,\G\right\}^{\scriptscriptstyle\text{CGL}}
=-\int_{V} \big\{
& \rho\left(\F_{\bf M}{\bf\cdot}\nabla \G_\rho -\G_{\bf M}{\bf\cdot}\nabla \F_\rho\right)
+\sigma\left(\F_{\bf M}{\bf\cdot}\nabla \G_\sigma -\G_{\bf M}{\bf\cdot}\nabla \F_\sigma\right)
\\&
+{\bf M}{\bf\cdot}\left(\F_{\bf M}{\bf\cdot}\nabla \G_{\bf M}-\G_{\bf M}{\bf\cdot}\nabla \F_{\bf M}\right)
+{\bf B}{\bf\cdot}\left( \F_{\bf M}{\bf\cdot}\nabla \G_{\bf B} -\G_{\bf M}{\bf\cdot}\nabla \F_{\bf B}\right)
\\&
+{\bf B}{\bf\cdot}\left( (\nabla \F_{\bf M}){\bf\cdot}\G_{\bf B} -(\nabla \G_{\bf M}){\bf\cdot}\F_{\bf B}\right)
\big\}\, d^3x .
\end{aligned}\label{eq:5.3a}
\end{equation}

%An alternative formulation of the non-canonical CGL Poisson bracket
%can be constructed in which 
%has essentially the same form as the MHD bracket (\ref{eq:5.3}). 
%The main difference is that instead of one entropy $S$, 
%there two additional entropy variables 
%$\bar S_\parallel$ and $\bar S_\perp$ besides $S$  
%that are advected by the flow. This leads to two extra integrals for 
%$\bar S_\parallel$ and $\bar S_\perp$ come from the double 
%adiabatic equations (\ref{eq:2.9})
%and are given by (\ref{eq:2.9a}) in terms of 
%$p_\parallel$, $p_\perp$, $\rho$, and $B$. We will not use this form of
%of the Poisson bracket in the present paper.
%For a general equation of state $e=e(\rho,S,B)$, 
%$\bar S_\parallel$ and $\bar S_\perp$ have a functional relation (\ref{eq:2.9b}) to $S$. 
%Nevertheless, it will be useful to develop the CGL Poisson bracket by introducing 
%\begin{equation}
%\sigma\equiv \rho S,
%\quad
%\sigma_\parallel\equiv\rho \bar S_\parallel,
%\quad
%\sigma_\perp\equiv\rho \bar S_\perp
%\label{eq:5.0}
%\end{equation}
%as independent variables.
%This will lead to the bracket being degenerate, 
%but it does have the advantage of allowing for different functional dependencies
%to be specified (cf (\ref{eq:2.9b})) among these variables. 
%In general, degenerate Poisson brackets arising from singular Lagrangians 
%can be regularized by the introduction of Dirac brackets. 

An overview of Hamiltonian systems is given in \cite{Morrison98}.   
\cite{Banerjee16} provides a Dirac bracket approach to the MHD Poisson 
bracket. 
The property that the MHD bracket is linear in the dynamical variables $\rho$, $\sigma$, ${\bf M}$, ${\bf B}$ has an important mathematical relationship to semi-direct product Lie algebras, 
which is explained in \cite{Holm83c} and \cite{Holm98}
for general fluid systems.

%Section 4.1
%\subsection{CGL Poisson Bracket}
The variables appearing in the CGL non-canonical Poisson bracket consist of:
\begin{equation}
\rho,
\quad
\sigma, 
\quad
\quad
{\bf M}, 
\quad 
{\bf B} . 
\label{eq:5.1}
\end{equation}

The analysis in Appendix D starts from the canonical bracket 
\begin{equation}
\left\{\F,\G\right\}=\int_{V} \left(
\F_{\bf q}{\bf\cdot}\G_{\bf p} -\F_{\bf p}{\bf\cdot}\G_{\bf q}
%\F_{q^i} \G_{p^i} -\F_{p^i}\G_{q^i}
\right)\, d^3 x_0
\label{eq:5.5}
\end{equation}
involving canonical Hamiltonian variables 
$({\bf q},{\bf p})\equiv ({\bf x}({\bf x}_0,t), \bm{\pi}({\bf x}_0,t))$ 
where $\bm{\pi}=\rho_0 \dot{\bf x}$ is the canonical momentum 
(cf Section~\ref{sec:Hamiltonian}). 
The main steps consist of using the Lagrangian map (\ref{eq:3.0b}) to obtain 
a transformation to the non-canonical variables (\ref{eq:5.1}), 
followed by applying a variational version of the chain rule to the variational derivatives with respect to $({\bf q},{\bf p})$ 
(see e.g. \cite{Zakharov97}). 
These steps are carried out by working in a fixed Lagrangian frame, 
while the corresponding Eulerian frame given by the Lagrangian map 
undergoes a variation, 
which includes varying Cartesian basis vectors associated to the components of ${\bf x}$. 
This general approach is described  
in a short communication by \cite{Holm83c}.

Below, we give an alternative more succinct derivation of the CGL Poisson 
bracket (\ref{eq:5.3a}). 
  It leads to the same 
Poisson bracket as that obtained in Appendix D.   
Note that the canonical bracket (\ref{eq:5.5}) can be written in the form:
\begin{equation}
\left\{\F,\G\right\}=\int\left({\F}_{\bf x}{\bf\cdot}\G_{\bf p}
-{\F}_{\bf p}{\bf\cdot}\G_{\bf x}\right)\ \frac{d^3{\bf x}}{J}. \label{eq:5.5a}
\end{equation}
Here ${\bf q}\equiv {\bf x}$.  The transformation of variational derivatives between the canonical variables 
and the new variables (\ref{eq:5.1}) is effected by noting that:
\begin{align}
\delta \F=&\int \frac{1}{J} \left(\F_{\bf x}{\bf\cdot}\Delta {\bf x}
+\F_{\bf p}{\bf\cdot}\Delta{\bf p}\right)\ d^3{\bf x}\nonumber\\
=&\int \biggl(\hat{\F}_\rho\delta\rho+\hat{\F}_{\bf M}{\bf\cdot}\delta{\bf M}
+\hat{\F}_\sigma \delta \sigma
%+\hat{\F}_{\sigma_\parallel} \delta \sigma_\parallel 
%+\hat{\F}_{\sigma_\perp} \delta \sigma_\perp
+\hat{\F}_{\bf B}{\bf\cdot}\delta{\bf B}\biggr)\ d^3{\bf x}, \label{eq:5.5b}
\end{align}
where $\hat{\F}(\rho,{\bf M},\sigma,%\sigma_\parallel,\sigma_\perp,
{\bf B})\equiv 
\F({\bf q},{\bf p})$ is the functional $\F({\bf q},{\bf p})$ 
expressed in terms of the new variables 
$(\rho,{\bf M}^T,\sigma,
%\sigma_\parallel,\sigma_\perp,
{\bf B}^T)^T$. Here 
$\delta\psi$ denotes the Eulerian variation of $\psi$ 
and $\Delta \psi$ denotes the Lagrangian variation of $\psi$. 

Using the transformations:
\begin{align}
\Delta J=&J\nabla{\bf\cdot}\Delta{\bf x},\quad 
\Delta{\bf M}=\frac{\Delta{\bf p}}{J}
-{\bf M}(\nabla{\bf\cdot}\Delta{\bf x}), \nonumber\\
\delta {\bf M}=&\frac{\Delta{\bf p}}{J} 
-\nabla_j\left(\Delta x^j{\bf M}\right), 
\quad \delta\rho=-\nabla{\bf\cdot}(\rho\Delta{\bf x}),\quad  
\delta \sigma=-\nabla{\bf\cdot}(\sigma\Delta {\bf x}),\nonumber\\
%\quad  \delta\sigma_\parallel=-\nabla{\bf\cdot}(\sigma_\parallel\Delta{\bf x}), 
%\quad \delta\sigma_\perp=-\nabla{\bf\cdot}(\sigma_\perp\Delta{\bf x})
%\nonumber\\
\delta{\bf B}=& \nabla\times(\Delta{\bf x}\times{\bf B})
-\Delta{\bf x} (\nabla{\bf\cdot}{\bf B}), \label{eq:5.5c}
\end{align}
in (\ref{eq:5.5b}) gives the formulae:
\begin{align}
\F_{\bf x}=&J\biggl[(\nabla\hat{\F}_{\bf M}){\bf\cdot}{\bf M}
+\sigma\nabla \hat{\F}_\sigma
%+\sigma_\parallel\nabla {\hat{\F}}_{\sigma_\parallel}
%+\sigma_\perp\nabla {\hat{\F}}_{\sigma_\perp}+
+\rho\nabla\hat{\F}_\rho\nonumber\\
&+(\nabla \hat{\F}_{\bf B}){\bf\cdot}{\bf B}-({\bf B}{\bf\cdot}\nabla) 
\hat{\F}_{\bf B}
-(\nabla{\bf\cdot}{\bf B}) \hat{\F}_{\bf B}\biggr], \nonumber\\
\F_{\bf p}=&\hat{\F}_{\bf M}, \label{eq:5.5d}
\end{align}
for the transformation of variational derivatives from the old variables 
$({\bf x},{\bf p})$ to the new variables ${\bf M}$,$\sigma$, 
 $\rho$ and ${\bf B}$. Using the expressions (\ref{eq:5.5d}) for 
$\F_{\bf x}$ and $\F_{\bf p}$ in (\ref{eq:5.5a}) and dropping the hat accents, 
the Poisson bracket (\ref{eq:5.5a}) reduces to
\begin{equation}
\{\F,\G\}=\left\{\F,\G\right\}^{\scriptscriptstyle\text{CGL}}
-\int_V\nabla{\bf\cdot}
\left[{\bf B}
\left(\G_{\bf M}{\bf\cdot}\F_{\bf B}
-\G_{\bf B}{\bf\cdot}\F_{\bf M}\right)\right]\ d^3{\bf x}, 
\label{eq:5.5e}
\end{equation}
Dropping the last pure divergence term in (\ref{eq:5.5e}) (which converts
to a surface integral over the boundary by Gauss's theorem), 
(\ref{eq:5.5e}) reduces to the CGL Poisson bracket (\ref{eq:5.3a}), 
which applies in the general case where $\nabla{\bf\cdot}{\bf B}\neq 0$
(i.e. the Jacobi identity applies for the bracket (\ref{eq:5.3a})), 
which is the 
analog of the \cite{MorrisonGreene82} bracket in MHD.

The CGL bracket (\ref{eq:5.3a}) shares the same main features as the MHD bracket:
it is bilinear, antisymmetric, and obeys the Jacobi identity. 
A verification of the Jacobi identity can be given using functional multi-vectors
(\cite{Olver93}) 
in the same way as for the MHD bracket (e.g. \cite{Webb18} Chapter 8). 
An alternative verification of the Jacobi identity for the MHD bracket is given in \cite{Morrison82}; 
see also \cite{Chandre13a, Chandre13b}, 
as well as \cite{Holm83a, Holm83b} which uses the magnetic vector potential 
${\bf A}$ in the advected gauge (\ref{eq:2.51}). 

In general, any Poisson bracket can be expressed in a co-symplectic form 
which defines a corresponding Hamiltonian (co-symplectic) operator. 
Using non-canonical Eulerian variables ${\sf Z}$, the co-symplectic form is given by 
\begin{equation}
\left\{\F,\G\right\}=\int_{V} \F_{{\sf Z}^{\sf T}} {\bm {\mathcal D}} \G_{{\sf Z}}\, d^3 x, 
\label{eq:5.6}
\end{equation}
in which $\bm{\mathcal D}$ is the Hamiltonian (matrix) operator,
where $\sf T$ denotes the transpose.  
Note that, in the case ${\sf Z} =({\bf q},{\bf p})$, 
this operator reduces to the skew matrix 
$\bm{\mathcal D} = \begin{pmatrix} 0 & {\sf I} \\ -{\sf I} & 0 \end{pmatrix}$. 
Antisymmetry of the bracket (\ref{eq:5.6}), 
$\{\F,\G\}+\{\G,\F\}=0$, 
corresponds to $\bm{\mathcal D}$ being skew-adjoint; 
the Jacobi identity,  
$\{\{\F,\G\},{\mathcal H}\}+\{\{{\mathcal H},\F\},\G\}+\{\{\G,{\mathcal H}\},\F\}=0$, 
corresponds to $\bm{\mathcal D}$ having a vanishing Schouten bracket with itself \cite{Olver93}.  

Taking 
\begin{equation}
{\sf Z}=\big(\rho,\sigma,M^i,B^i\big),
\label{eq:5.7}
\end{equation}
with $M^i$ and $B^i$ respectively denoting the components of ${\bf M}$ and ${\bf B}$ in Cartesian coordinates $x^i$, 
we see that the cosymplectic form (\ref{eq:5.6}) of the 
CGL bracket (\ref{eq:5.3a}) 
after integration by parts is given by the Hamiltonian operator:
\begin{equation}
\begin{aligned}
\bm{\mathcal D} = -\begin{pmatrix}
0 & 0    &\nabla^j \circ \rho & 0 \\
0 & 0   & \nabla^j \circ \sigma & 0 \\
%0 & 0  & 0 & 0  & \nabla^j \circ \sigma_\parallel & 0 \\
%0 & 0  & 0 & 0  & \nabla^j \circ \sigma_\perp & 0 \\
\rho \nabla^i & \sigma \nabla^i  & 
M^j\nabla^i + \nabla^j \circ M^i & B^j\nabla^i - \delta^{ij}B^k \nabla_k \\
0 & 0   & \nabla^j \circ B^i - \delta^{ij}\nabla_k\circ B^k & 0 
\end{pmatrix} 
\label{eq:5.8}
\end{aligned}
%\label{eq:5.8}
\end{equation}
where $\circ$ denotes operator composition
(i.e.\ $\nabla \circ a = (\nabla a) + a\nabla$ ). 
Here $\nabla^k$ acts as the total derivative with respect to $x^k$. 
Note that the minus sign in (\ref{eq:5.8}) is due to the overall minus sign 
in the CGL Poisson bracket (\ref{eq:5.3a}) 
which follows the sign convention used in the MHD bracket (\ref{eq:5.3}) 
in \cite{MorrisonGreene80, MorrisonGreene82}. 

We can convert (\ref{eq:5.8}) into vector notation by identifying
$\nabla^i F= \Grad F = \nabla F$, 
$\nabla^j F^j= \Div {\bf F} =\nabla{\bf \cdot}{\bf F}$, 
and $\nabla^j F^k - \nabla^k F^j = \Curl {\bf F}= \nabla\times{\bf F}$.

\subsection{Non-canonical Hamiltonian equations}\label{sec:Hamiltonianeqns}
A non-canonical Poisson bracket (\ref{eq:5.6}) provides a Hamiltonian formulation once a Hamiltonian functional $\Ham$ is chosen. 
The formulation involves expressing the dynamical variables ${\sf Z}({\bf x},t)$ 
formally as functionals
\begin{equation}
{\mathcal Z}({\bf x}',t) \equiv \int_{V} {\sf Z}({\bf x}',t) \delta({\bf x}-{\bf x}')\, d^3x
\end{equation} 
and writing Hamilton's equations in the form 
\begin{equation}
\deriv{\mathcal Z}{t} = \{ {\mathcal Z}, \Ham \}.
\label{eq:5.9}
\end{equation}
Using the explicit expression (\ref{eq:5.6}) for the bracket yields
\begin{equation}
\deriv{{\sf Z}}{t} = \bm{\mathcal D} \frac{\delta \Ham}{\delta {\sf Z}}
\label{eq:5.9a}
\end{equation}
in terms of the Hamiltonian operator $\bm{\mathcal D}$. 

The appropriate Hamiltonian for describing CGL plasmas is given by the conserved total energy (cf Section~\ref{sec:totalenergy}):
\begin{equation}
\Ham=\int_{V} H\, d^3x ,
\label{eq:5.9b}
\end{equation}
where $H$ is the Eulerian total energy density (\ref{eq:3.28}). 
Substituting this Hamiltonian into the non-canonical Hamilton's equations (\ref{eq:5.9a})
can be shown to yield the CGL plasma equations (\ref{eq:2.1})--(\ref{eq:2.5}) and (\ref{eq:2.7})--(\ref{eq:2.8}).

\subsection{Casimirs}\label{sec:Casimirs}
In a non-canonical Hamiltonian system, 
a functional $\C = \int_{V} C\, d^3x$ satisfying the equation 
\begin{equation}
\{\C,\F\}\equiv 0
\quad\text{ for all functionals $\F$ }
\label{eq:5.10a}
\end{equation}
is called a Casimir. 
Existence of a non-trivial Casimir $\C\not\equiv 0$ indicates that the Poisson bracket
is degenerate. 
Correspondingly, the Hamiltonian operator $\bm{\mathcal D}$ 
in the co-symplectic form of the bracket (\ref{eq:5.6}) 
will have a non-trivial kernel: 
\begin{equation}
\bm{\mathcal D} \C_{\sf Z} \equiv 0
\label{eq:5.10b}
\end{equation}
where $\C_{\sf Z} \equiv \frac{\delta\C}{\delta {\sf Z}}$. 
Note that a Casimir is a conserved integral, since the time evolution of any functional is given by the Hamiltonian equations (\ref{eq:5.9}):
\begin{equation}
\frac{d}{dt}\C = \deriv{\C}{t}=\{\C,\Ham\}=0
\label{eq:5.11}
\end{equation}
 Casimirs are useful in stability analysis of steady flows and plasma equilibria (e.g. \cite{Holm85};  \cite{Hameiri04}). 
The conservation (\ref{eq:5.11}) holds modulo boundary integrals, 
and an investigation of boundary conditions is needed 
for a Casimir to be a strictly conserved integral (i.e.\ a constant of motion). 

All Casimirs can be determined by solving the equation (\ref{eq:5.10a}),
or alternatively the equation (\ref{eq:5.10b}) (see e.g.
 \cite{Hameiri04}, 
\cite{Padhye96a, Padhye96b} 
 for the MHD case).  
For CGL plasmas, we obtain the Casimir determining equations:
%equation (\ref{eq:5.10b}) with $\bm{\mathcal D}$ given by the Hamilton (matrix) operator (\ref{eq:5.8}) yields the following system of Casimir determining equations:
\begin{equation}
\begin{aligned}
& \nabla^j(\rho \C_{M^j}) = \nabla{\bf\cdot}(\rho \C_{\bf M})=0, 
\quad 
\nabla^j(\sigma \C_{M^j})= \nabla{\bf\cdot}(\sigma \C_{\bf M})=0, 
\\
%& \nabla^j(\sigma_{\parallel} \C_{M^j})= \nabla{\bf\cdot}(\sigma_\parallel \C_{\bf M})=0, 
%\quad 
%\nabla^j(\sigma_\perp \C_{M^j}) = \nabla{\bf\cdot}(\sigma_\perp \C_{\bf M})=0, 
%\\
&\begin{aligned} 
& \rho \nabla^i \C_{\rho} 
+ \sigma \nabla^i \C_{\sigma} 
%+\sigma_\parallel \nabla^i \C_{\sigma_\parallel}
%+\sigma_\perp \nabla^i \C_{\sigma_\perp}
%\\& 
+ M^j\nabla^i \C_{M^j} + \nabla^j ( M^i \C_{M^j}) 
+ B^j\nabla^i \C_{B^j} - B^j \nabla_j \C_{B^i} 
\\&
\equiv\rho \nabla \C_{\rho} + \sigma \nabla \C_{\sigma} 
%+\sigma_\parallel \nabla \C_{\sigma_\parallel}
%+\sigma_\perp \nabla \C_{\sigma_\perp}
%\\&\quad
+ (\nabla \C_{\bf M}){\bf\cdot}{\bf M} + \nabla{\bf\cdot}(\C_{\bf M}{\bf M}) 
+ (\nabla \C_{\bf B}){\bf\cdot}{\bf B} - {\bf B}{\bf\cdot} \nabla \C_{\bf B} 
=0 , 
\end{aligned}
\\
& \nabla^j(B^i \C_{M^j}) - \nabla_j(B^j \C_{M^i})
= \nabla\times({\bf B}\times\C_{\bf M})
=0 ,
\end{aligned}
\label{eq:5.12}
\end{equation}
where $\nabla$ ($\nabla^k$) acts as the total derivative with respect to ${\bf x}$ ($x^k$). 
This is an overdetermined system of linear partial differential equations for 
\begin{equation}
\C = \int_{V} C(t,{\bf x},{\sf Z},\nabla{\sf Z},\ldots,\nabla^l{\sf Z})\,d^3x,
\end{equation}
with ${\sf Z}$ denoting the dynamical variables (\ref{eq:5.7}). 
The system can, in principle, be integrated to find $\C$ explicitly, 
once a differential order $l$ for the dependence of $\C$ 
on derivatives of the variables is chosen. 
Note that solutions of the divergence form $C=\nabla{\bf\cdot}{\bf F}$ lead to $\C$ 
being a boundary integral which can be assumed to be trivial 
if suitable boundary conditions are imposed. 
This classification problem is beyond the scope of the present work.

\subsection{Mass, cross helicity, magnetic helicity Casimirs}
The well-known Casimirs for ideal barotropic MHD are the mass integral, 
the cross helicity integral, and the magnetic helicity integral. 
CGL plasmas with an isentropic equation of state $e=e(\rho,B)$ 
possess these same Casimirs:
\begin{equation}
\C_1=\int_V \rho\, d^3x,
\quad
\C_2=\int_V {\bf u}{\bf\cdot}{\bf B}\, d^3x,
\quad
\C_3=\int_V {\bf A}{\bf\cdot}{\bf B}\, d^3x, 
\label{eq:5.21}
\end{equation}
In the more physically realistic case with an equation of state $e=e(\rho,S,B)$, 
$\C_1$ and $\C_3$ still hold as Casimirs, 
but $\C_2$ turns out to be a Casimir only in the case ${\bf B}{\bf\cdot}\nabla S=0$,
as we will show shortly. 

Physically, 
the mass integral $\C_1$ is the total mass of the plasma; 
the cross helicity integral $\C_2$ describes the linking of the fluid vorticity and magnetic field flux tubes; 
and the magnetic helicity integral $\C_3$ 
describes the knotting, linking, and twist and writhe of the magnetic flux tubes
(see e.g.\, \cite{Berger84}; \cite{Moffatt92};  \cite{Hameiri04}). 
(Also see e.g. \cite{Yoshida16}, for interesting applications).  
In a fixed volume $V$ with a boundary $\partial V$, 
$\C_1$ is conserved,  $\deriv{\C_1}{t}=0$, 
if ${\bf u}$ has no normal component at the boundary 
(cf Section~\ref{sec:totalenergy}). 
Conservation of $\C_3$, $\deriv{\C_3}{t}=0$, holds 
if in addition ${\bf B}$ has no normal component at the boundary,
while the cross helicity integral $\C_2$ is conserved, $\deriv{\C_2}{t}=0$,
only with the further condition ${\bf B}{\bf\cdot}\nabla S=0$
(cf Sections~ \ref{sec:crosshelicity}, \ref{sec:magnetichelicity}).

\subsection{Advected Casimirs}
In general, Casimirs can be sought be looking among advected scalars, $\theta$, 
since the corresponding scalar integral $\int_{V(t)} \rho \theta\,d^3x$  
will be conserved on volumes $V(t)$ moving with the flow. 
This implies that, on a fixed volume $V$, 
the integral $\int_{V} \rho \theta\,d^3x$ will be conserved up to a flux integral 
$-\oint_{\partial V} \rho \theta{\bf u}{\bf\cdot}\hat{\bf n}\,dA$ 
that vanishes if ${\bf u}$ has no normal component at the boundary $\partial V$,
and therefore $\int_{V} \rho \theta\,d^3x$ will satisfy the Casimir property $\deriv{\C}{t}=0$. 
Note that this property is necessary but not sufficient for $\int_{V} \rho \theta\,d^3x$ to be a Casimir, 
since there are conserved integrals such as the total energy and angular momentum 
that do not belong to the kernel of the Poisson bracket. 

For CGL plasmas, the basic advected scalars are $S$, $\bar S_\parallel$, $\bar S_\perp$. 
Additional advected scalars are provided by Ertel's theorem:
if $\theta$ is an advected scalar, then so is 
$\frac{{\bf B}{\bf\cdot}\nabla\theta}{\rho}$. 
This yields the advected quantities (\ref{eq:2.38}) shown in Section~\ref{sec:advected}.
 
One can  show that  
\begin{equation}
\C_4 =\int_V \rho f(S,\theta)\, d^3{\bf x}
\quad\text{ where }\quad
\theta=\frac{{\bf B}{\bf\cdot}\nabla S}{\rho} ,
\label{eq:5.24a}
\end {equation}
is an advected Casimir, 
for any function $f(S,\theta)$. We omit the proof. 
\section{Noether's Theorem and Conservation Laws}\label{sec:ConsLaws}

Conservation laws of the CGL plasma equations (\ref{eq:2.1}) to (\ref{eq:defn.p})
can be derived from Noether's theorem applied to the Lie point symmetries of 
the Lagrangian variational principle (\ref{eq:3.15}). 
A brief discussion of the MHD case was outlined in \cite{Webb19}. 
The CGL plasma conservation laws differ in comparison with the MHD conservation laws
mainly in the form of pressure tensor:
in particular, the MHD isotropic gas pressure tensor ${\sf p}=p\, {\sf I}$ is replaced by 
the non-isotropic CGL plasma pressure tensor 
${\sf p}=p_\perp {\sf I}+(p_\parallel-p_\perp)\bm{\tau}\bm{\tau}$.

Appendix H gives a description of Noether's first theorem, 
for a differential equation system described by an action principle 
as developed by \cite{Bluman89}.  The analysis uses 
canonical 
Lie symmetry operators $\X$ in which both the dependent and independent 
variables change in the Lie transformation, and also the evolutionary form 
of the symmetry operator denoted by $\hat{\X}$ in which the independent 
variables do not change, but the dependent variables, and the derivatives 
of the dependent variables change. The prolongation 
operators $\pr \X$ and $\pr\hat{\X}$ are described, and used to 
derive Noether's first theorem similar to \cite{Bluman89}.
More recent derivations of Noether's theorem using $\pr\hat{\X}$, 
are given by \cite{Bluman02} and \cite{Olver93}. 

In Section 6.1 
we use a recent form of Noether's theorem to derive conservation laws
using the evolutionary form of the prolonged symmetry operator 
$\pr\hat{\X}$. In Section 6.2 we obtain the same results
 from the classical form of Noether's theorem given in 
Appendix H. Although the recent form of Noether's theorem, 
is conceptually more appealing, it is not  
any simpler than the classical form of Noether's theorem. 
The classical form of Noether's theorem is perhaps easier 
to understand, as it relates directly back to the invariance 
of the action integral under Lie and divergence transformations.
 
%Application  of the classical  
% Noether's theorem to the CGL equations is given in Appendix H.

\subsection{Noether's Theorem and Evolutionary Symmetries}
To use Noether's theorem, 
we need to obtain the Lie point symmetries of the Lagrangian variational principle (\ref{eq:3.15}). 
Since this variational principle employs the variables $x^i=X^i(x_0^j,t)$ 
given by the Cartesian components of the Lagrangian map (\ref{eq:3.0b}), 
a Lie point symmetry acting on the coordinate space $(t,x_0^j,x^i)$ 
has the form 
\beqn
t\to t + \epsilon \xi^t + O(\epsilon^2),
\quad
x^i\to x^i +\epsilon \xi^i + O(\epsilon^2),
\quad
x_0^i\to x_0^i +\epsilon \xi_0^i + O(\epsilon^2)
\label{eq:6.0}
\eeqn
with $\epsilon$ denoting the parameter in the point symmetry transformation,
where $\xi^i$, $\xi_0^i$, $\xi^t$ are functions of $t,x_0^j,x^i$. 
The infinitesimal transformation corresponds to the generator 
\beqn
\X =\xi^i\derv{x^i} +\xi_0^i\derv{x_0^i}+\xi^t\derv{t}
\label{eq:6.0a}
\eeqn
while the finite transformation (\ref{eq:6.0}) is given by exponentiation of the generator, 
$(t,x^i,x_0^i)\to \exp(\epsilon\X)(t,x^i,x_0^i)$. 

To be a symmetry, a generator (\ref{eq:6.0a}) must leave 
the variational principle invariant modulo boundary terms.
This is equivalent to the condition that the 
change in the Lagrangian must satisfy
\cite{Ovsjannikov78,Ibragimov85}):
\beqn
\pr\X L_0 = \xi^t D_t L_0 + \xi_0^i D_{x^i_0} L_0 
+ D_t \Lambda_0^t + D_{x_0^i} \Lambda_0^i 
\label{eq:6.1a}
\eeqn
where the operators 
$D_t=\partial/\partial_{t} + \dot x^i\partial/\partial_{x^i} +\cdots$
and $D_{x_0^i} = \partial/\partial_{x_0^i} + x^{ji}\partial/\partial_{x^j} +\cdots$
denote total derivative operators with respect to the independent variables 
$t$ and $x_0^i$ and $D_{x^i}$ denotes the partial derivative with respect
to $x^i$ keeping ${\bf x}_0$ and $t$ constant.  Here $\Lambda^t_0$ and $\Lambda^i_0$ 
are arbitrary potentials, that arise in Noether's theorem, because 
the variational derivative of a perfect derivative term has 
zero variational derivative. 
The first two terms on the righthand~side of (\ref{eq:6.1a}) 
represent the Lie derivative ${\cal L}_{\bf X} L_0$
using the chain rule for differentiation.   
The terms involving total derivatives of $L_0$ on the 
righthand side of (\ref{eq:6.1a}) 
can be understood to arise from the change in the spatial domain $V_0$ and the time interval $[t_0,t_1]$ in the variational principle (\ref{eq:3.15}) 
under the action of an infinitesimal point transformation (\ref{eq:6.0}). 
On the lefthand side of (\ref{eq:6.1a}), 
$\pr$ denotes prolongation to the extended coordinate space $(t,x_0^j,x^i,\dot x^i,x^{ij})$ (i.e. jet space) 
in which $\dot x^i$ and $x^{ij}$ are coordinates that correspond to 
$\dot X^i(x_0^j,t)$ and $\partial X^i(x_0^j,t)/\partial x_0^j$ 
on solutions $x^i=X^i(x_0^j,t)$ of the Euler-Lagrange equations (\ref{eq:3.20}) of the variational principle. 
An explicit formula for the components of $\pr\X$ will not be needed 
if we express the invariance condition (\ref{eq:6.1a}) 
by using the characteristic form of the generator 
in which only the dependent variables $x^i$ undergo a transformation:
\beqn
\hat\X =\hat\xi^i \derv{x^i},
\quad
\hat\xi^i = \xi^i -\xi^t \dot x^i - \xi_0^j x^{ij}
\label{eq:6.0b}
\eeqn
which arises from how $\X$ acts on solutions $x^i=X^i(x_0^j,t)$ 
The relationship between the two forms (\ref{eq:6.0b}) and (\ref{eq:6.0a}) is that 
$\pr\X= \pr\hat\X + \xi^t D_t + \xi_0^i D_{x_0^i}$. 
Hence, (\ref{eq:6.1a}) becomes (\cite{Olver93,Bluman02})
\beqn
\pr\hat\X (\hat L_0) = D_t \Lambda_0^t + D_{x_0^i} \Lambda_0^i
\label{eq:6.1b}
\eeqn
where 
\beqn
\hat L_0 = L_0\big|_{{\bf X}={\bf x}}
\eeqn
is a function of $t,x_0^j,x^i,\dot x^i,x^{ij}$. 
The prolongation $\pr\hat\X$ is given by simply extending $\hat\X$ to act on $\dot x^i$ and $x^{ij}$ through the total derivative relations
$\pr\hat\X(\dot x^i) = D_t(\hat\X x^i) = D_t \xi^t$ 
and
$\pr\hat\X(x^{ij}) = D_{x_0^j}(\hat\X x^i) = D_{x_0^j}\xi^i$. 

In turn, this form (\ref{eq:6.1b}) of the invariance condition can be expressed succinctly as 
\beqn
E_{x^i}(\pr\hat\X(\hat L_0)) =0
\label{eq:6.1c}
\eeqn
using the Euler-Lagrange operator $E_{x^i}$ which has the property that 
it annihilates a function iff the function is given by a total divergence with respect to $t$ and $x_0^j$. 
The formulation (\ref{eq:6.1c}) can be used as a determining equation to find 
all Lie point symmetries of the variational principle (\ref{eq:3.15}). 
(See \cite{Olver93} and \cite{Bluman02} for a general discussion.)

Each Lie point symmetry (\ref{eq:6.0}) gives rise to a conservation law
through combining the condition (\ref{eq:6.1b}) and Noether's identity 
\begin{equation}
\pr\hat\X(\hat L_0) 
=\hat \xi^i E_{x^i} (L_0) +D_t W^t + D_{x_0^i} W^i 
\label{eq:6.2a}
\end{equation}
where 
\begin{equation}
W^t = \hat\xi^j \deriv{\hat L_0}{\dot x^j}, 
\quad
W^i= \hat\xi^j \deriv{\hat L_0}{x^{ji}} . 
\label{eq:6.2b}
\end{equation} 
Thus, we obtain the following statement of Noether's theorem. 

\begin{proposition}\label{prop:noether}
If the Lagrangian variational principle (\ref{eq:3.15}) is invariant up to boundary terms
under an infinitesimal point transformation (\ref{eq:6.0}), 
then the Euler-Lagrange equations (\ref{eq:3.20}) of the variational principle 
possess a conservation law 
\begin{equation}
D_t \mathcal{I}^t_0 +D_{x_0^i} {\mathcal{I}}^i_0  =0
\label{eq:6.3}
\end{equation} 
in which the conserved density and the spatial flux are given by 
\begin{equation}
\mathcal{I}^t_0 = \left(\Lambda_0^t  -W^t\right)\big|_{x={\bf X}} ,
\quad
\mathcal{I}^i_0  = \left(\Lambda_0^i  -W^i\right)\big|_{x={\bf X}} ,
\label{eq:6.4}
\end{equation} 
using (\ref{eq:6.1b}) and (\ref{eq:6.2b}).
\end{proposition}

Explicit expressions for 
$\partial \hat L_0/\partial \dot x^j = \left(\partial L_0/\partial \dot X^j\right)\big|_{{\bf X}={\bf x}}$
and 
$\partial \hat L_0/\partial x^{ji} = \left(\partial L_0/\partial X^{ji}\right)\big|_{{\bf X}={\bf x}}$
are provided by (\ref{eq:D1}) in appendix~C. 
Note that the conservation law is a local continuity equation 
which holds when $x^i=X^i(x_0^j,t)$ satisfies the Euler-Lagrange equations (\ref{eq:3.20}). 
On solutions, $D_t|_{x={\bf X}} = \partial/\partial t$ and $D_{x_0^i}|_{x={\bf X}} = \partial/\partial x_0^i$ 
(acting as total derivatives). 

We remark that (\ref{eq:6.4}) can be derived alternatively using the canonical form of a symmetry generator (\ref{eq:6.0a}), which requires computing the prolongation. 
See \cite{Webb07,Webb19} for the MHD case. The canonical symmetry approach 
to conservation laws for the CGL system is also described in Appendix H. 

%Subsection 6.1.2
\subsubsection{Eulerian form of a Lagrangian conservation law}\label{sec:eulerianconslaw}
A Lagrangian conservation law (\ref{eq:6.4}) can be expressed equivalently 
as an Eulerian conservation law (\cite{Padhye98})
\beqn
\partial_t \Psi^t +\nabla_i \Psi^i=0 
\label{eq:6.5a}
\eeqn
whose conserved density and spatial flux are given by 
\beqn
\Psi^t =\frac{1}{J} \mathcal{I}_0^t\big|_{x={\bf X}} ,
\quad
\Psi^i =\frac{1}{J}\left( u^i \mathcal{I}_0^t +X_{ik} \mathcal{I}_0^k \right)\big|_{x={\bf X}}  
\label{eq:6.5b}
\eeqn
on solutions $x^i=X^i(x_0^j,t)$ of the Euler-Lagrange equations (\ref{eq:3.20}). 
This form (\ref{eq:6.5a})--(\ref{eq:6.5b}) of a conservation law is appropriate for considering 
conserved integrals 
\beqn
\frac{d}{dt}\int_{V} \Psi^t \,d^3x = - \oint_{\partial V} {\bm \Psi}\cdot \hat{\bf n}\,dA
\label{eq:6.5c}
\eeqn
on a fixed spatial domain $V$ in the CGL plasma. 

In general, 
an Eulerian conservation law is a continuity equation of the form (\ref{eq:6.5a}) 
holding on the solution space of the CGL plasma equations (\ref{eq:2.1}) to (\ref{eq:defn.p}). 
The corresponding conserved integral (\ref{eq:6.5c}) 
has the physical content  that the rate of change of the integral quantity 
$\int_{V} \Phi^t\, d^3x$ in $V$ is balanced by the net flux leaving the boundary of $V$. 
It is often physically useful to consider instead a spatial domain $V(t)$ that moves with the plasma. 
The form of the conservation law for moving domains is given by 
\beqn
\frac{d}{dt}\int_{V(t)} \Psi^t \,d^3x = - \oint_{\partial V(t)} {\bm \Gamma}\cdot \hat{\bf n}\,dA
\label{eq:movingfluxconslaw}
\eeqn
in terms of the moving flux 
\beqn
{\bm \Gamma} = {\bm \Psi} - {\bf u}\Psi^t = \frac{1}{J} {\bf X}\cdot{\bm{\mathcal I}}_0 .
\label{eq:movingflux}
\eeqn
Note that the moving integral quantity $\int_{V(t)} \Psi^t\,d^3x$ will be an invariant (i.e. a constant of motion) when the net moving flux vanishes on the boundary $\partial V (t)$. 
(See \cite{Anco20} for a discussion of moving domain conservation laws and invariants 
in fluid mechanics.)
The equivalence between the conservation laws (\ref{eq:6.5c}) and (\ref{eq:movingfluxconslaw})
can be derived by writing the continuity equation (\ref{eq:6.5a}) 
in terms of the material (co-moving) derivative (\ref{eq:materialderiv}):
$d\Phi^t/dt = -(\nabla{\bf\cdot}{\bf u})\Phi^t - \nabla{\bf\cdot}\left(\bm{\Phi}-\Phi^t{\bf u}\right)$,
where $(\nabla{\bf\cdot}{\bf u})$ represents the expansion or contraction of 
an infinitesimal volume $d^3x$ moving with the fluid 
(see e.g. \cite{AncoDar09}; \cite{AncoDar10}). 

Now, rather than seeking to find all Lie point symmetries, 
we will consider two main classes: kinematic and fluid relabelling. 
Kinematic symmetries are characterized by the generator (\ref{eq:6.0}) having $\xi_0^i=0$,
with $\xi^t$ and $\xi^i$ being functions only of $t,x^i$. 
Fluid relabelling symmetries have $\xi^i=\xi^t=0$ in the generator (\ref{eq:6.0}),
with $\xi_0^i$ being a function only of $t,x_0^i$. 

A useful general remark is that any Lie point symmetry of a variational principle
corresponds to a Lie point symmetry of the Euler-Lagrange equations, 
because invariance of a variational principle means that its extremals are preserved. 
Thus, one way to find all Lie point symmetries of a variational principle is by 
firstly obtaining the Lie point symmetries of the Euler-Lagrange equations, 
and secondly checking which of those symmetries leaves invariant the variational principle.

Before deriving conservation laws using the above 
analysis, it is useful to note that there are three basic steps 
in the analysis. 
\begin{description}
\item{(\romannumeral1)} First it is necessary to determine 
for a given Lie symmetry, whether the Lie invariance condition (\ref{eq:6.1b})
 for the action can be satisfied by choosing the potentials 
$\Lambda^t_0$ and $\Lambda^i_0$. Here the left handside of (\ref{eq:6.1b}) 
for the action ${\rm pr}\hat{X}(L_0)$ is evaluated for the symmetry 
operator $X$.

\item{(\romannumeral2)}\ Determine the surface vector components
$W^t$ and $W^i$ that occur in the Noether identity (\ref{eq:6.2a}), 
where $W^t$ and $W^i$
are given by (\ref{eq:6.2b}) 
(recall $\dot{x}^j=\partial x^j({\bf x}_0,t)/\partial t=u^j$
and $x^{ji}=\partial x^j/\partial x_0^i$). Then using  
Proposition \ref{prop:noether}  one can obtain the Lagrangian conservation 
law (\ref{eq:6.3}) 
with conserved density $\mathcal{I}_0^t$ and flux $\mathcal{I}_0^i$ 
given in (\ref{eq:6.4}). 

\item{(\romannumeral3)}\ Determine the Eulerian form of the conservation law 
using the results of Padhye (1998) described by (\ref{eq:6.5a}) 
and (\ref{eq:6.5b}). 
\end{description}

%subsection6.1.7
\subsubsection{Lie Invariance Condition}

The Lie invariance condition for the action in (\ref{eq:6.1b}) 
%and (\ref{eq:6.23}) 
may be written in the form:
\begin{align}
&J\biggl\{ \nabla{\bf\cdot}(\rho \hat{\bm\xi})
\left[\Phi+h-\frac{1}{2}u^2\right]
+\rho {\bf u}{\bf\cdot} \left(\frac{d\hat{\bm\xi}}{dt} 
-\hat{\bm\xi}{\bf\cdot}\nabla {\bf u}\right)
 +\rho T\hat{\bm\xi}{\bf\cdot}\nabla S\nonumber\\
&-\frac{\tilde{\bf B}}{\mu_0}{\bf\cdot} 
\left[\nabla\times(\hat{\bm\xi}\times{\bf B})
-\hat{\bm\xi} \nabla{\bf\cdot}{\bf B}\right]\nonumber\\
&+\nabla{\bf\cdot}\left\{\rho \hat{\bm\xi}
\left[\frac{1}{2} u^2 -(h+\Phi)\right]
+\hat{\bm\xi}{\bf\cdot}\left(\sf{p}+\sf{M}_B\right)
+\frac{1}{\mu_0}
\left(\hat{\bm\xi}\times{\bf B}\right)
\times\tilde{\bf B}\right\}\biggr\}\nonumber\\
&=D_t\Lambda^t_0+ D_{x_0^j} \left(\Lambda^j_0\right), \label{eq:6.23a}
\end{align}
Here we use the notation $\hat{\bm\xi}\equiv\hat{\bm\xi}_0$ in the 
fluid re-labelling symmetry case. 
The Lie invariance condition (\ref{eq:6.23a}) also applies 
for the general Lie symmetry case, including the Lie point Galilean 
symmetry cases, the fluid relabelling symmetry cases, and other more 
general cases. 
The derivation of (\ref{eq:6.23a}) from (\ref{eq:6.1b}) 
is outlined in Appendix G. 
In the general case 
${\hat\xi}^i=\xi^i-\left(\xi^tD_t+\xi^s_0 D_{x_0^s}\right)x^i$. 
The Lie invariance condition (\ref{eq:6.23a}) is similar to 
the Eulerian Lie invariance condition for the action used by
\cite{Webb19} for the case of MHD. Here  
 the evolutionary form of the symmetry operator $X$ is used
 rather than the canonical symmetry operator used by \cite{Webb19}. 

%subsection6.1.3
\subsubsection{Galilean symmetries}\label{sec:Galileansymms}
The Eulerian form of the CGL plasma equations (\ref{eq:2.1}) to (\ref{eq:defn.p})
clearly indicates that they possess the Galilean group of Lie point symmetries when $\Phi=0$,
which are generated by (\cite{Fuchs91},\cite{RogersAmes89}):
\beqn
P_0=\derv{t},
\quad 
P_i=\derv{x^i},
\quad
K_i=t\derv{x^i}+\derv{u^i},
\quad
J_i=\epsilon_{ijk}\left(x^j\derv{x^k}+u^j\derv{u^k}+B^j \derv{B^k} \right).
\label{eq:6.6a}
\eeqn
These generators respectively describe 
a time translation ($P_0$), 
space translations ($P_i$, $i=1,2,3$), 
Galilean boosts ($K_i$, $i=1,2,3$), 
and rotations ($J_i$, $i=1,2,3$) about the $x$, $y$, $z$ axes. 
As shown in appendix~F, 
the only additional Lie point symmetries admitted by equations (\ref{eq:2.1}) to (\ref{eq:defn.p})
consist of scalings. 
Hence, Galilean symmetries and scaling symmetries comprise all kinematic Lie point symmetries of the CGL plasma equations. 

The Galilean symmetries (\ref{eq:6.6a}) have a corresponding Lagrangian form
\beqn
\X_{P_0} = \derv{t},
\quad
\X_{P_i} = \derv{x^i},
\quad
\X_{K_i}=t\derv{x^i}, 
\quad
\X_{J_i}=\epsilon_{ijk} x^j\derv{x^k} . 
\label{eq:6.6b}
\eeqn
Note that the prolongation of these Lagrangian generators to $\dot x^i=\dot X^i = u^i$
through the Lagrangian relation (\ref{eq:3.0a}) for the fluid velocity 
yields the corresponding Eulerian generators (\ref{eq:6.6a}) acting on the variables $(t,x^i,u^i)$. 
Since $\xi_0^i=0$ for all of the generators (\ref{eq:6.6b}), 
they are of kinematic type. 

In the case when the gravitational potential $\Phi$ is non-zero, 
then the preceding generators must satisfy the condition 
$\X \Phi(x^i) =0$ to be admitted as symmetries. 
For example, 
if $\Phi$ is invariant under $z$-translation, 
then $\X_{P_3}$ and $\X_{K_3}$ are symmetries; 
if $\Phi$ is invariant under $z$-rotation, 
then $\X_{J_3}$ is a symmetry;
and if $\Phi$ is spherically symmetric, 
then $\X_{J_i}$, $i=1,2,3$, are symmetries. 

%subsection6.1.4
\subsubsection{Galilean conservation laws}\label{sec:Galileanconslaws}
It is straightforward to show that 
each Galilean generator (\ref{eq:6.6b}) satisfies the condition (\ref{eq:6.1c}) 
for invariance of the Lagrangian variational principle (\ref{eq:3.15}) when $\Phi=0$ 
and thereby yields a conservation law (\ref{eq:6.3}). 

Specifically, 
time translation ($X_{P_0}$) has $\hat\xi^i=-\dot x^i$, 
and thus 
$\pr\hat\X_{P_0} L_0 = - \ddot x^i \partial L_0/\partial_{\dot x^i} - \dot x^{ij}\partial L_0/\partial_{x^{ij}} = -D_t L_0$ 
due to $\partial L_0/\partial_t =0$. 
This gives $\Lambda_0^t = -L_0$, $\Lambda_0^i=0$,
which yields conservation of energy 
\beqn\label{eq:6.7a}
\mathcal{I}^t_0 = \dot X^j \deriv{L_0}{\dot X^j} -L_0 =H_0 , 
\quad
\mathcal{I}^i_0  = \dot X^j \deriv{L_0}{X^{ji}} ,
\eeqn
with $H_0$ being the Hamiltonian (\ref{eq:3.26}).

Space translations ($X_{P_j}$) have $\hat\xi^i=\delta^i{}_j$, 
and thus 
$\pr\hat\X_{P_j} L_0 = 0$ due to $\partial L_0/\partial_{x^i}=0$ 
(where we neglect gravity)
combined with $D_t\hat\xi^i=0$ and $D_{x_0^j}\hat\xi^i=0$. 
Hence, $\Lambda_0^t =0$,  $\Lambda_0^i=0$, 
which yields conservation of momentum
\beqn\label{eq:6.7b}
\mathcal{I}^t_0 = -\deriv{L_0}{\dot X^j}, 
\quad
\mathcal{I}^i_0  = -\deriv{L_0}{X^{ji}} . 
\eeqn

Galilean boosts ($X_{K_j}$) have $\hat\xi^i=t\delta^i{}_j$, 
and thus 
$\pr\hat\X_{K_j} L_0 = \rho_0 \dot x^j = D_t (\rho_0 x^j)$. 
This gives $\Lambda_0^t =\rho_0 x^j$,  $\Lambda_0^i=0$, 
yielding conservation of Galilean momentum (center of mass) 
\beqn\label{eq:6.7c}
\mathcal{I}^t_0 = \int \deriv{L_0}{\dot X^j}dt -t\deriv{L_0}{\dot X^j}, 
\quad
\mathcal{I}^i_0  = -t\deriv{L_0}{X^{ji}} .
\eeqn

Rotations ($X_{J_j}$) have $\hat\xi^i=\epsilon^i{}_{kj}x^k$. 
This leads to $\pr\hat\X_{J_j} L_0 = 0$, since $L_0$ depends on $x^i$ 
only through the scalars $|\dot x^i|$ and $|X_{ij}\tau_0^j|$ which are rotationally invariant. 
Thus $\Lambda_0^t =0$,  $\Lambda_0^i=0$, 
which yields conservation of axial angular momentum
\beqn\label{eq:6.7d}
\mathcal{I}^t_0 = \epsilon_{jkl}x^k \deriv{L_0}{\dot X^l}, 
\quad
\mathcal{I}^i_0  = \epsilon_{jkl}x^k\deriv{L_0}{X^{li}} .
\eeqn

The set of Galilean conservation laws (\ref{eq:6.7a}) to (\ref{eq:6.7d}) 
can be written in a unified form through use of the observation that 
\beqn
\Lambda_0^t =\dot\xi^j \int \deriv{L_0}{\dot X^j}dt -\xi^t L_0,
\quad
\Lambda_0^i= 0 ,
\eeqn
with
\beqn
\xi^t = a_0,
\quad
\xi^i = a_1^i + a_2^i t + a_3^k \epsilon_{ijk}x^j ,
\eeqn
where $a_0$, $a_1^i$, $a_2^i$, $a_3^i$ ($i=1,2,3$) are arbitrary constants
(parameterizing the respective Galilean symmetry generators (\ref{eq:6.6b})). 
Hence, we have 
\beqn\label{eq:6.8}
\mathcal{I}_0^t = \dot\xi^j \int \deriv{L_0}{\dot X^j}dt -\xi^t L_0 -(\xi^j -\xi^t\dot X^j) \deriv{L_0}{\dot X^j}, 
\quad
\mathcal{I}_0^i = -(\xi^j -\xi^t\dot X^j) \deriv{L_0}{X^{ji}} . 
\eeqn
The resulting set of Eulerian conservation laws (\ref{eq:6.5a})--(\ref{eq:6.5b})
 takes the form 
\begin{align}
\Psi^t & = 
H \xi^t  -\rho u^j \xi^j +\rho x^j \dot\xi^j,
\label{eq:6.9a}
\\
\Psi^i  & = 
u^i \Psi^t -(\xi^j -\xi^t u^j) \left(p^{ij} +M_B^{ij}\right), 
\label{eq:6.9b}
\end{align}
where we have used 
\beqn
\dfrac{\partial L_0}{\partial\dot X^j} =\rho_0 u^j,
\quad
\int \dfrac{\partial L_0}{\partial \dot X^j}dt =\rho_0 x^j,
\quad
\dfrac{\partial L_0}{\partial X^{jk}} = \left(p^{jl}+M_B^{jl}\right) A_{lk} , 
\label{eq:L0rels}
\eeqn
which follow from expressions (\ref{eq:D1}) and (\ref{eq:D1b}) in appendix~C, 
along with the formulae (\ref{eq:3.5}) and (\ref{eq:3.2}). 
Here $p^{ij}$ and $M_B^{ij}$ respectively represents the components of 
the CGL non-isotropic pressure tensor (\ref{eq:2.3}) 
and the magnetic pressure tensor (\ref{eq:Bpressure}). 
Note that $H =\rho |{\bf u}|^2 -L$ is the Hamiltonian (\ref{eq:3.28}). 

In the MHD case, $p^{ij}$ is replaced by the components of the isotropic MHD gas pressure tensor $p\,\delta^{ij}$. 

To summarize the preceding derivations:

(i) 
Time translation symmetry ($a_0$) yields energy conservation, for which:
\beqn
\begin{aligned}
& \Psi^t = H 
= \rho \left( \frac{1}{2}|{\bf u}|^2+\Phi({\bf x}) \right)
+\varepsilon +\frac{B^2}{2\mu_0} ,
\quad
{\bm \Psi} 
= \left(\frac{1}{2}\rho |{\bf u}|^2 + \varepsilon+\rho\Phi\right){\bf u} 
+ {\sf p}{\bf\cdot u} +\frac{1}{\mu_0}{\bf E}\times{\bf B} ,
\\
& {\bm \Gamma} = {\bm \Psi} -{\bf u}H 
=\left({\sf p}+{\sf M}_B\right)\cdot {\bf u} ,
\end{aligned}
\label{eq:6.11}
\eeqn
where ${\bf E}=-{\bf u}\times{\bf B}$ is the electric field strength, 
and ${\bf E}\times{\bf B}/\mu_0$ is the Poynting flux. 

(ii) 
Space translation symmetry ($a_1^i$, $i=1,2,3$) yields momentum conservation
 with:
\beqn
\begin{aligned}
& \Psi^t = -{\bf M}
= -\rho {\bf u},
\quad
{\bm\Psi} = -{\sf T} , 
\\
& {\bm \Gamma} = {\bm\Psi} -{\bf u}{\bm \Psi}^t
= -\left({\sf p} +{\sf M}_B\right) ,  
\end{aligned}
\label{eq:6.12}
\eeqn
where
\beqn
{\sf T} = \rho {\bf u}{\bf u} + {\sf p}+ {\sf M}_B
\eeqn
is the CGL plasma stress tensor. 

(iii) 
Galilean boost symmetry yields center of mass conservation, with:
\beqn
\begin{aligned}
& \Psi^t = \rho {\bf x} -t{\bf M} 
= \rho ({\bf x} -t{\bf u}) ,
\quad
{\bm\Psi} = \rho {\bf u} {\bf x}  -t{\sf T}, 
\\
& 
{\bm \Gamma} = {\bm\Psi}-{\bf u}(\rho {\bf x} -t{\bf M})
=-t \left({\sf p} +{\sf M}_B\right) . 
\end{aligned}
\label{eq:6.12a}
\eeqn

(iv)
Rotational symmetry yields angular momentum conservation, with:
\beqn
\begin{aligned}
& \Psi^t = {\bf x}\times {\bf M}
= \rho {\bf x}\times {\bf u},
\quad
{\bm\Psi} = {\bf x}\times {\sf T} ,
\\
& {\bm\Gamma} = {\bm\Psi} -{\bf u}({\bf x}\times {\bf M})
= {\bf x}\times \left({\sf p} + {\sf M}_B\right) , 
\end{aligned}
\label{eq:6.13}
\eeqn
where ${\bf M} =\rho{\bf u}$ is the momentum density vector. 

%subsection6.1.4
\subsubsection{Fluid Relabelling Symmetries}\label{sec:relabellingsymms}
Fluid relabelling symmetries correspond to transformations (\ref{eq:6.0})
that change the Lagrangian fluid labels ${\bf x}_0$, 
but leave the Eulerian variables and ${\bf x}$ and $t$ invariant:
$x_0^i \to x_0^i +\epsilon \xi_0^i +O(\epsilon^2)$, 
where $\xi_0^i=\xi_0^i(t,x_0^j)$. 
The symmetry generator has the form 
\beqn
\X =\xi_0^i\derv{x_0^i} .
\label{eq:6.14}
\eeqn
For evaluating the condition for invariance of 
the Lagrangian variational principle (\ref{eq:3.15}), 
we need the characteristic form of the generator:
\beqn
\hat\X =\hat\xi_0^i\partial_{x^i} , 
\quad
\hat\xi_0^i = - \xi_0^j x^{ij} . 
\label{eq:6.15}
\eeqn
To proceed we will use the formulation (\ref{eq:6.1b}) instead of (\ref{eq:6.1c}), 
which allows for consideration of functions 
$\Lambda_0^t$, $\Lambda_0^i$ that can involve nonlocal potentials. 
This generality is necessary to derive the cross helicity conservation law (\ref{eq:2.ch5})--(\ref{eq:2.ch6}) 
which involves the temperature potential (\ref{eq:2.ch7}).

%subsection6.1.6
\subsubsection{Noether's theorem for fluid-relabelling conservation laws}\label{sec:relabellingconslaw}

The Lie invariance condition (\ref{eq:6.23a}) in the fluid relabelling 
symmetry cases,  may be written 
in the form:
\begin{align}
&J\biggl\{ \nabla{\bf\cdot}(\rho \hat{\bm\xi})
\left[\Phi+h-\frac{1}{2}u^2\right]
+\rho {\bf u}{\bf\cdot} \left(\frac{d\hat{\bm\xi}}{dt}
-\hat{\bm\xi}{\bf\cdot}\nabla {\bf u}\right)\nonumber\\
%+\rho T\hat{\bm\xi}{\bf\cdot}\nabla S\nonumber\\
&-\frac{\tilde{\bf B}}{\mu_0}{\bf\cdot}
\left[\nabla\times(\hat{\bm\xi}\times{\bf B})
-\hat{\bm\xi} \nabla{\bf\cdot}{\bf B}\right]\biggr\}\nonumber\\
&=D_t\left(\Lambda^t_0-\rho_0r {\bm\xi}_0{\bf\cdot}\nabla_0S\right)
+D_{x_0^i}\left(\Lambda^i_0-A_{ki} G^k\right), \label{eq:6.23d}
\end{align}
where
\begin{equation}
{\bf G}=\rho \hat{\bm\xi}
\left[\frac{1}{2} u^2 -(h+\Phi)\right]
+\hat{\bm\xi}{\bf\cdot}\left(\sf{p}+\sf{M}_B\right)
+\frac{1}{\mu_0}
\left(\hat{\bm\xi}\times{\bf B}\right)
\times\tilde{\bf B}. \label{eq:6.23e}
\end{equation}
For cross helicity conservation, 
$\hat{\bm\xi}=-{\bf B}/\rho$ and ${\bm\xi}_0={\bf B}_0/\rho_0$. 
Setting $\nabla{\bf\cdot}{\bf B}=0$ (Gauss's law), the left handside of
(\ref{eq:6.23d}) vanishes. The condition that the right handside of 
(\ref{eq:6.23d}) vanish is satisfied by the choices:
\begin{equation}
\Lambda^t_0=\rho_0r{\bf\xi}_0{\bf\cdot}\nabla S, \quad 
\Lambda^i_0=A_{ki} G^k, \label{eq:6.23f}
\end{equation}
where ${\bf G}$ is given by (\ref{eq:6.23e}). Evaluating formulae 
(\ref{eq:6.23e}) and (\ref{eq:6.23f})  gives the formulae:
\begin{align}
{\bf G}=&{\bf B}\left[e+\Phi-\frac{1}{2}u^2+\frac{B^2}{2\mu_0\rho}\right], 
\label{eq:6.23g}\\
\Lambda^t_0=&r {\bf B}_0{\bf\cdot}\nabla_0 S
\equiv r (J {\bf B}{\bf\cdot} \nabla S), 
\label{eq:6.23h}\\
\Lambda^i_0=&B_0^i\left[h+\Phi-\frac{1}{2}u^2+\frac{B^2}{2\mu_0\rho}-
\frac{p_\parallel}{\rho}\right], \label{eq:6.23i}
\end{align}
where $h=e+p_\parallel/\rho$ is the enthalpy of the CGL plasma. 
Here $\hat{\bm\xi}=-{\bf B}/\rho$, and ${\bf\xi}_0={\bf B}_0/\rho_0$,  
which gives rise to the 
cross helicity conservation laws described 
by (\ref{eq:6.30}) and (\ref{eq:6.31}). 

%Equation (\ref{eq:6.23}) provides the condition for 
%an infinitesimal fluid relabelling transformation (\ref{eq:6.14}) 
%to be a symmetry of the Lagrangian variational principle (\ref{eq:3.15}). 
%By examining the form of the terms on the lefthand side of equation (\ref{eq:6.23}),
%we find that those terms vanish for 
%\beqn
%\xi_0^i = B_0^i/\rho_0 ,
%\label{eq:6.25a}
%\eeqn
%which is equivalent to 
%\beqn
%\hat\xi_0^i\big|_{{\bf x}={\bf X}} = -B^i/\rho . 
%\label{eq:6.25b}
%\eeqn
%To verify this result, 
%we note that the Faraday term 
%$\frac{d}{dt} (\rho \hat{\bm\xi}_0)  -(\rho \hat{\bm\xi}_0{\bf\cdot}\nabla) {\bf u}
%+ (\nabla{\bf\cdot}{\bf u})\rho\hat{\bm\xi}_0$ 
%reduces to Faraday's equation (\ref{eq:2.5alt}), 
%and that the curl term 
%${\bm\tau}{\bf\cdot}\left(\nabla\times({\bf B} \times \hat{\bm\xi}_0)\right)$
%vanishes since ${\bf B} \times \hat{\bm\xi}_0=0$, 
%while remaining terms on the lefthand side vanish due to 
%Gauss' law $D_{x_0^i} B_0^i=J\nabla{\bf\cdot}{\bf B}=0$ holding, 
%and $\xi_0^i D_{x_0^i} S = (B_0^i/\rho_0) D_{x_0^i} S = ({\bf B}/\rho){\bf\cdot}\nabla S$
%being an advected scalar. 
%
%The righthand side of equation (\ref{eq:6.23}) then yields 
%\beqn
%\Lambda_0^t  =  %r\rho_0\xi_0^i D_{x_0^i} S 
%r B_0^i D_{x_0^i} S, 
%\quad
%\Lambda_0^i  = 
%\left( h+\Phi-\frac{1}{2} u^2 
%-\frac{1}{\rho} \left(p_\parallel-\frac{B^2}{2\mu_0}\right)  \right) B_0^i . 
%\label{eq:6.27}
%\eeqn
%
Noether's theorem given by Proposition~\ref{prop:noether} now gives the following main result. 

\begin{proposition}
The Lagrangian variational principle (\ref{eq:3.15}) is invariant up to boundary terms
under the infinitesimal fluid relabelling transformation 
\beqn
x_0^i \to x_0^i + \epsilon B_0^i/\rho_0 +O(\epsilon^2) . 
\label{eq:6.28}
\eeqn
The resulting conservation law (\ref{eq:6.3})--(\ref{eq:6.4}) of the 
Euler-Lagrange equations \ref{eq:3.20})  
is obtained by using (\ref{eq:6.23h})-(\ref{eq:6.23i}) for 
$\Lambda^t_0$ and $\Lambda^i_0$ and using the results: 
\beqn
W_0^t = \hat\xi_0^j \rho_0 \dot x^j
= -J {\bf B}{\bf\cdot}{\bf u} ,
\quad
W_0^i  = \hat\xi_0^j A_{ki}\left(p^{jk} +M_B^{jk}\right) 
= -\frac{1}{\rho} \left(p_\parallel-\frac{B^2}{2\mu_0}\right)  B_0^i .
\label{eq:6.29}
\eeqn
Here  $W^t_0$ and $W^i_0$  are given by (\ref{eq:6.2b}),
in which the derivatives of $L_0$ are given by (\ref{eq:I2}). 
This yields the conserved density and the flux 
\begin{equation}
\mathcal{I}^t_0 = 
J {\bf B}{\bf\cdot}({\bf u} + r\nabla S) , 
\quad
\mathcal{I}^i_0  = B_0^i \left( h+\Phi -\tfrac{1}{2}u^2 \right) . 
\label{eq:6.30}
\end{equation} 
\end{proposition}

The corresponding Eulerian conservation law (\ref{eq:6.5a})--(\ref{eq:6.5b}) is given by 
\begin{equation}
\begin{aligned}
& \Psi^t = {\bf B}{\bf\cdot}({\bf u} + r\nabla S) , 
\quad
{\bm\Psi} = \left( h+\Phi -\tfrac{1}{2}u^2 \right) {\bf B} + \left( {\bf B}{\bf\cdot}({\bf u} + r\nabla S) \right){\bf u} , 
\\& 
{\bm\Gamma} = \left( h+\Phi -\tfrac{1}{2}u^2 \right) {\bf B} . 
\end{aligned}
\label{eq:6.31}
\end{equation} 
This is the cross helicity  density conservation law (\ref{eq:2.ch5})--(\ref{eq:2.ch7}).

%subsection6.2
%\appendix
\subsection{Classical Noether Approach}

In this subsection  we use a classical version of Noether's theorem 
(see e.g. \cite{Bluman89} and Appendix H) 
to derive conservation laws for the CGL plasma action (\ref{eq:3.15})
that uses the canonical Lie symmetry operator ${\tilde X}={\rm pr}X$
rather than the evolutionary form ${\rm pr}{\hat X}$. 
The canonical symmetry operator form of Noether's theorem was 
used by \cite{Webb07} and \cite{Webb19} for the MHD fluid case. 
In this approach one searches for Lie transformations and divergence 
transformations that leave the
action (\ref{eq:6.0}) invariant, where 
\beqn
L_0'=L_0+\epsilon D_\alpha{\bar\Lambda}^\alpha_0+O(\epsilon^2), \label{eq:H1}
\eeqn
is the divergence transformation.
Here $D_0=\partial/\partial t$ and $D_i=\partial/\partial x_0^i$ are total 
partial derivatives with respect to $t$ and $x_0^i$. 
Note we use ${\bar\Lambda}^\alpha_0$ and ${\bar\Lambda}^\alpha$ 
 to denote the potentials, 
in order to distinguish them from the 
potentials used for the evolutionary potentials in Section 6.1.

  The condition for the action to remain invariant under  
 (\ref{eq:6.0}) and (\ref{eq:H1}) may be written in the form
(cf. \cite{Bluman89}):
\begin{equation}
{\rm pr}X L_0+L_0\left[D_t\xi^t+D_{x_0^j}\left(\xi^j_0\right)\right]
+D_t{\bar\Lambda}^0_0+D_{x_0^j}{\bar\Lambda}^j_0=0, \label{eq:H2}
\end{equation}
where
\begin{equation}
{\rm pr}X L_0=\xi^t\derv{t}+\xi^s_0\derv{x_0^s}+\xi^k\derv{x^k}+\xi^{x^k_t}\derv{x^k_t} +\xi^{x_{kj}}\derv{x^{kj}}+\ldots, \label{eq:H3}
\end{equation}
is the prolonged, canonical Lie symmetry operator.   

${\rm pr}X$ is related to ${\rm pr}{\hat X}$ by the 
equations (\cite{Ovsjannikov78}, \cite{Ibragimov85} and \cite{Bluman89}): 
\begin{align}
{\rm pr}X=&{\rm pr}{\hat X}+\xi^\alpha_0 D_\alpha,\nonumber\\
{\rm pr}{\hat X}=&{\hat\xi}^k\derv{x^k}+D_\alpha\left({\hat\xi}^k\right)
\derv{x^k_\alpha}+D_{\alpha} D_{\beta}\left({\hat\xi}^k\right)
\derv{x^k_{\alpha\beta}}+\ldots, \label{eq:H4}
\end{align}
where the evolutionary symmetry generator ${\hat\xi}^i$ is 
given by (\ref{eq:6.0b}) and  $\xi^t\equiv \xi^0_0$.

Noether's theorem follows from Noether's identity:
\begin{equation}
{\rm pr}X L_0+L_0 D_\alpha\xi^\alpha_0+D_\alpha{\bar\Lambda}^\alpha_0
=\hat{\xi}^i E_{x^i}\left(L_0\right) 
+D_\alpha\left(W^\alpha+L_0\xi^\alpha_0\right) 
+D_\alpha\left({\bar\Lambda}^\alpha_0\right), \label{eq:H5}
\end{equation}
where $E_{x^i}(L_0)\equiv \delta{\mathcal J}/\delta x^i$ is 
the variational derivative  
of the action ${\mathcal J}$ with respect to $x^i$. 
For the case of CGL plasmas, the surface terms  $W^\alpha$ are given by:
\beqn
W^t\equiv W^0=\hat{\xi}^j \deriv{L_0}{x^j_t},
\quad W^i=\hat{\xi}^j\deriv{L_0}{x^{ji}}, \label{eq:H6}
\eeqn
(see \cite{Bluman89}, \cite{Ibragimov85} for more general cases).

If the Lie invariance condition (\ref{eq:H2}) is satisfied, then 
the left handside of (\ref{eq:H5}) vanishes, 
and consequently the right handside 
of (\ref{eq:H5}) must vanish. If in addition, the Euler Lagrange equations 
$E_{x^i}(L_0)=0$ are satisfied, then (\ref{eq:H5}) implies 
\beqn
D_\alpha\left(W^\alpha+L_0\xi^\alpha_0+{\bar\Lambda}^\alpha_0\right)=0, 
\label{eq:H7}
\eeqn
which is the conservation law of Noether's first theorem, which applies 
if the Euler~-Lagrange equations $E_{x^i}(L_0)=0$ are independent, 
which is the case for a finite Lie algebra of Lie point symmetries.

In the more general case where the symmetries depend on  continuous 
functions  $\left\{ \phi^k ({\bf x}_0,t): 1\leq k\leq N\right\}$,
then the Lie-pseudo algebra of symmetries is infinite dimensional. 
In this case,  
Noether's second theorem implies that the Euler-Lagrange equations 
are not all independent, and that there exists differential relations 
between the Euler-Lagrange equations (see e.g. \cite{Olver93}
and \cite{Hydon11} for details). Noether's second theorem, in some cases 
results in mathematically trivial conservation laws. \cite{Charron18} 
discuss Ertel's theorem and Noether's second theorem.  

For the CGL plasma model, the $W^\alpha$ from (\ref{eq:H6}) are given by:
\beqn
W^t={\hat{\xi}}^j\rho_0 u^j, 
\quad W^i=\hat{\xi}^j \left(p^{js}+M_B^{js}\right) A_{si}. \label{eq:H8}
\eeqn
Substitution of (\ref{eq:H8}) for the $W^\alpha$ into Noether's theorem 
(\ref{eq:H7}) gives the Lagrangian conservation law:
\beqn
\deriv{{\cal I}^0}{t}+\deriv{{\cal I}^i}{x_0^i}=0, \label{eq:H9}
\eeqn
where
\begin{align}
{\cal I}^0=&\rho_0u^j \hat{\xi}^j +\xi^tL_0+{\bar\Lambda}^t_0, \nonumber\\
{\cal I}^i=&\hat{\xi}^j\left(p^{js}
+M_B^{js}\right) A_{si}+\xi^i_0 L_0+{\bar\Lambda}^i_0, \label{eq:H10}
\end{align}
are the conserved density and flux. 

The Lagrangian conservation law 
(\ref{eq:H9}) corresponds to an Eulerian conservation law 
of the form (\ref{eq:6.5a}) (e.g. \cite{Padhye98}) 
with conserved density $\Psi^t$, and flux $\Psi^j$ 
given by:
\begin{align}
\Psi^t=&\rho u^k\hat{\xi}^k+\xi^t L+{\bar\Lambda}^0, \nonumber\\
\Psi^j=& \hat{\xi}^k\left(T^{jk}-L\delta^{jk}\right)+\xi^j L+{\bar\Lambda}^j, 
\label{eq:H11}
\end{align}
where
\begin{align}
T^{jk}=&\rho u^j u^k+p^{jk}+M_B^{jk}, \nonumber\\
{\bar\Lambda}^0=&{\bar\Lambda}^0_0/J,\quad 
{\bar\Lambda}^j=\left(u^j {\bar\Lambda}^0_0+x^{js} 
{\bar\Lambda}^s_0\right)/J. \label{eq:H12}
\end{align}
The Lagrangian and Eulerian conservation laws for the Galilean group 
(Sections 6.1.2-6.1.3) and the cross helicity conservation law associated 
with the fluid relabelling symmetry $\hat{\bm\xi}=-{\bf B}/\rho$ 
(Sections 6.1.4~-6.1.5) 
now follow from (\ref{eq:H10})-(\ref{eq:H12}) for appropriate 
choices of the potentials ${\bar\Lambda}^t_0$ and ${\bar\Lambda}^j_0$ 
and of the symmetry generators $\xi^i$, $\xi^i_0$ and $\xi^t$. 
Note that
the evolutionary symmetry potentials used in Section 6.1 
 are 
different than those used in this subsection (Section 6.2). 

\leftline{\bf Example 1.}
%\begin{proposition}
The time translation symmetry of the Lagrangian action (\ref{eq:3.15}), 
satisfies the Lie invariance condition (\ref{eq:H2}) by choosing:
\beqn
\xi^t=1,\quad \xi^i=0, \quad \xi^s_0=0,\quad \hat{\xi}^{i}=- u^i, 
\quad {\bar\Lambda}_0^\alpha=0, 
\label{eq:5.21a}
\eeqn
where $i,s=1,2,3,$ and $\alpha=0,1,2,3$ is a variational symmetry of the 
action (\ref{eq:3.15}). The corresponding conservation law 
using Noether's theorem results (\ref{eq:H11}) is the energy conservation 
law:
\begin{equation}
\derv{t}\left[\frac{1}{2}\rho |{\bf u}|^2+\varepsilon 
+\frac{B^2}{2\mu_0} +\rho\Phi({\bf x})\right] 
+\nabla{\bf\cdot}\left[\rho {\bf u}\left(\frac{1}{2}|{\bf u}|^2
+\Phi({\bf x})\right)
+\frac{{\bf E}\times{\bf B}}{\mu_0}
+\varepsilon {\bf u}+{\sf p}{\bf\cdot u}\right]=0, 
\label{eq:5.22a}
\end{equation}
where ${\sf p}$ is the CGL pressure tensor (\ref{eq:2.3}), 
 ${\bf E}=-{\bf u}\times{\bf B}$ is the electric field strength 
and ${\bf E}\times{\bf B}/\mu_0$ is the Poynting flux. 

\leftline{\bf Example 2.}
If the gravitational potential $\Phi({\bf x})$ is independent of $x^{j_1}$
say, then the Lie invariance condition (\ref{eq:H2}) 
for a divergence symmetry of the action is satisfied, by the choice
\begin{equation}
\xi^i=\delta^{ij_1},\quad \xi^s_0=\xi^t=0,\quad  
\quad {\hat \xi}^{i}=\delta^{ij_1}, 
\quad {\bar\Lambda}_0^0={\bar\Lambda}_0^i=0, 
\label{eq:5.23a}
\end{equation}
and condition (\ref{eq:H2}) reduces to the equation:
\begin{equation}
-\rho_0\deriv{\Phi}{x^{j_1}}=0. \label{eq:5.24a}
\end{equation}
Then using (\ref{eq:H1}) we obtain the momentum conservation
equation in the $x^{j_1}$ direction in the form:
\begin{equation}
\left\{\derv{t}\left(\rho {\bf u}\right)
+\nabla{\bf\cdot}\left(\rho {\bf u}\otimes {\bf u}+{\sf p}
+\frac{B^2}{2\mu_0} {\sf I}
-\frac{{\bf B}\otimes {\bf B}}{\mu_0}\right)\right\}^{j_1}=0, 
\label{eq:5.25a}
\end{equation}
 In the case 
where gravity can be neglected (i.e. $\Phi=0$)) the superscript 
$j_1$ can be dropped in (\ref{eq:5.25a}). Technically, the conservation 
of momentum law appears more complicated in non-Cartesian coordinates
(e.g. in spherical geometry), where the metric tensor, the 
covariant derivative and the affine connection play an important role.

\leftline{\bf Example 3.}
The Galilean boost  symmetry, with infinitesimal 
generators:
\beqn
\xi^i=\Omega^i t,
\quad \xi^s_0=0,\quad \xi^t=0, 
\quad {\bar\Lambda}^0_0=-\rho_0({\bf x}_0)\boldsymbol{\Omega}{\bf\cdot}{\bf x},
\quad {\bar\Lambda}_0^i=0, \label{eq:5.26a}
\eeqn
 ($i,s=1,2,3$) and (\ref{eq:H11})  
gives rise to the center of mass conservation law:
\begin{equation}
\derv{t}[\boldsymbol{\Omega}{\bf\cdot}\rho ({\bf u}t-{\bf x})] 
+\nabla{\bf\cdot}
\left[\boldsymbol{\Omega}{\bf\cdot}
\left\{ 
\rho ({\bf u}t-{\bf x})\otimes{\bf u}
+t\left[{\sf p}+\frac{B^2}{2\mu_0} {\sf I} 
-\frac{{\bf B}\otimes{\bf B}}{\mu_0}\right]
\right\}
\right]=0, \label{eq:5.27a}
\eeqn
provided
\begin{equation}
{\bm\Omega}t{\bf\cdot}\nabla\Phi\equiv {\bm\xi}{\bf\cdot}
\nabla\Phi=0.
\label{eq:5.28a}
\end{equation}
Thus one obtains the center of mass conservation law for Galilean boosts
perpendicular to the external gravitational field. Thus, for a spherically 
symmetric gravitational potential (e.g. for the Sun), 
a Galilean boost conservation law exists for a 
boost ${\bm\xi}=\bm{\Omega}t$ perpendicular 
to the radial direction. 

\leftline{\bf Example 4}

The Lie transformation generators:
\begin{equation}
\xi^{i}=\epsilon_{ijk} \Omega^j x^k,\quad \xi^s_0=\xi^t=0,\quad  
\hat{\xi}^{i}=\xi^{x^i},\quad {\bar\Lambda}^{\alpha}_0=0, \label{eq:5.29a}
\end{equation}
give rise to the angular momentum conservation equation:
\begin{equation}
\derv{t}\left[{\bm\Omega}{\bf\cdot}({\bf x}\times {\bf M})\right]
+\nabla \left[ {\bm\Omega}{\bf\cdot}({\bf x}\times {\sf T}\right]=0, 
\label{eq:5.30a}
\end{equation}
where
\begin{equation}
{\bf M}=\rho {\bf u}\quad \hbox{and}\quad  \left({\bf x}\times 
{\sf T}\right)^{pj}=\epsilon_{pqk} x^q T^{kj}, \label{eq:5.31a}
\end{equation}
define the mass flux or momentum density ${\bf M}$ 
and ${\bf x}\times {\sf T}$ respectively. 
The invariance condition (\ref{eq:H2})
for a divergence symmetry of the action, requires:
\begin{align}
{\tilde X} L_0=&-\rho_0({\bm \Omega}\times{\bf x}){\bf\cdot}\nabla\Phi
+\epsilon_{kps}\Omega^p \left[p^{sk}+\frac{B^2}{2\mu_0}\delta^{sk}
- \frac{B^s B^k}{\mu_0}\right]\nonumber\\
\equiv & 
-\rho_0({\bm \Omega}\times{\bf x}){\bf\cdot}\nabla\Phi=0, \label{eq:5.32a}
\end{align}
Note that the second term in the first line of (\ref{eq:5.32a}) vanishes 
because the large square brackets term is symmetric in $s$ and $k$.

For the case where ${\bm\Omega}=\Omega {\bf e}_z$ the condition 
(\ref{eq:5.32a})
reduces to the equation:
\begin{equation}
{\tilde X} L_0=-\rho_0 \Omega r\sin\theta \deriv{\Phi}{\phi}=0, 
\label{eq:5.33a}
\end{equation}
where $(r,\theta,\phi)$ are spherical polar coordinates. 
Condition (\ref{eq:5.33a}) is satisfied for $\Phi=\Phi(r,\theta)$. 

\leftline{\bf Example 5}

The fluid relabelling symmetry transformations:
\begin{equation}
\xi^i=0,\quad\xi^t=0,\quad \xi^s_0=-\frac{B_0^s}{\rho_0},\quad \hat{\xi}^i
=\frac{B^i}{\rho}\equiv b^i, \label{eq:5.34a}
\end{equation}
and the choices:
\begin{equation}
{\bar\Lambda}^0=r \left({\bf B}{\bf\cdot}\nabla S\right), 
\quad {\bar\Lambda}^i= u^i {\bar\Lambda}^0, \quad \frac{dr}{dt}=-T, \label{eq:5.35a}
\end{equation}
for the gauge potentials ${\bar\Lambda}^\alpha$ ($\alpha=0,1,2,3$),  leave 
the action invariant (i.e. (\ref{eq:H2}) is satisfied). 
Using (\ref{eq:H11}) to calculate $\Psi^t$
and $\Psi^j$ gives the results (\ref{eq:6.31}) 
for the cross helicity conservation law 
(\ref{eq:2.ch5})-(\ref{eq:2.ch7}).  

The above examples, illustrate the use of the classical 
version of Noether's first theorem in obtaining 
conservation laws of the CGL equations. 

%section 7
\section{Summary and Concluding Remarks}\label{sec:conclude}

In this paper an investigation was carried out of the ideal CGL plasma
equations, based in part on the Lagrangian variational formulation of 
\cite{Newcomb62}, in which there is an anisotropic pressure tensor, with 
pressure components $p_\parallel$ and $p_\perp$ parallel and perpendicular
to the magnetic field ${\bf B}$, which satisfy the so-called double adiabatic
equations which in turn can be described by using entropy components: 
$S_\parallel$ and $S_\perp$ parallel and perpendicular to ${\bf B}$
(e.g. \cite{Du20}).

The total energy conservation law, and the cross helicity 
and magnetic helicity conservation laws were obtained 
(Section 2). The total energy equation was decomposed into the sum of 
three energy equations, namely the internal energy equation or co-moving
energy equation, the total kinetic energy equation and Poynting's theorem 
(the electro-magnetic energy equation). 
%The coupling terms describing the 
%energy transfer between the three component systems was delineated. 
The cross helicity transport equation involves the effective enthalpy 
$h=(p_\parallel+\varepsilon)/\rho$ of the gas associated with 
pressure work terms parallel to ${\bf B}$. If 
the internal energy density of the plasma per unit mass has 
 the form $e=e(\rho,S,B)\equiv\varepsilon/\rho$
(\cite{Hazeltine13}; \cite{Holm86}) one obtains a cross 
helicity transport equation of the form:
\begin{equation}
\derv{t}({\bf u\cdot B}) +\left[({\bf u\cdot B}){\bf u}+{\bf B}\left(\Phi+h-
\frac{1}{2} u^2\right)\right]=T {\bf B}{\bf\cdot}\nabla S, \label{eq:7.1}
\end{equation}
where $T=e_S$ is the temperature of the gas.  
If ${\bf B}{\bf\cdot}\nabla S=0$, then (\ref{eq:7.1}) is a local conservation 
law. More generally, if the source term $Q=T{\bf B}{\bf\cdot}\nabla S\neq 0$, 
(\ref{eq:7.1}) can be reduced to a nonlocal conservation law of the form:
\begin{equation}
\derv{t}({\bf w\cdot B}) +\left[({\bf w\cdot B}){\bf u}+{\bf B}\left(\Phi+h-
\frac{1}{2} u^2\right)\right]=0, \label{eq:7.2}
\end{equation}
where
\begin{equation}
{\bf w}={\bf u}+r\nabla S \quad\hbox{and}\quad 
\frac{dr}{dt}=-T, \label{eq:7.3}
\end{equation}
Here $d/dt=\partial/\partial t+{\bf u}{\bf\cdot}\nabla$ is the advective 
time derivative following the flow.
The CGL plasma conservation law (\ref{eq:7.2}) 
 generalizes the non-local, MHD cross helicity 
conservation law of \cite{Webb14a, Webb14b} and \cite{Yahalom17a,Yahalom17b}).
Here $\Phi({\bf x})$ is the potential for an external 
gravitational field.  
Similar transport equations for cross helicity to (\ref{eq:7.1}) 
and (\ref{eq:7.3}) were also developed in terms of the parallel 
and perpendicular temperatures $T_\parallel$ and $T_\perp$ 
defined as $T_\parallel=p_\parallel/(\rho R)$ and $T_\perp=p_\perp/(\rho R)$. 

There are two different, but equivalent forms of the momentum (or force) equation 
for the CGL system. 
The form of the momentum equation obtained
by \cite{Newcomb62} Lagrangian action principle reduces to the form:
\begin{equation}
\frac{d{\bf u}}{dt}=-\frac{1}{\rho}\nabla{\bf\cdot}{\sf p}+\frac{{\bf J}\times{\bf B}}{\rho} -\nabla\Phi, \label{eq:7.4}
\end{equation}
where ${\bf J}=\nabla\times{\bf B}/\mu_0$ is the current. 
This equation is essentially the same as the MHD momentum equation, 
except that the isotropic gas pressure tensor $p{\sf I}$ is replaced by 
the anisotropic CGL pressure tensor 
${\sf p}=p_\parallel {\bm\tau}{\bm\tau} +p_\perp ({\sf I}-{\bm\tau}{\bm\tau})$. 
However:
\begin{equation}
\nabla{\bf\cdot}{\sf p}=-\left[{\bf B}\times (\nabla\times {\bm\Omega})
+\rho(T\nabla S-\nabla h)\right]. \label{eq:7.5}
\end{equation}
(e.g. \ref{eq:2.ch2}, \ref{eq:C10}, \ref{eq:G13})
where
\begin{equation}
{\bm\Omega}=\frac{p_{\Delta}}{B^2}{\bf B},\quad 
p_{\Delta}=p_\parallel-p_\perp, 
\quad h=\frac{\varepsilon+p_\parallel}{\rho}. \label{eq:7.6}
\end{equation}
Using $\nabla{\bf\cdot}{\sf p}$ from (\ref{eq:7.5}) and (\ref{eq:7.6})
in (\ref{eq:7.4}) results in the CGL momentum equation in the form:
\begin{equation}
\frac{d{\bf u}}{dt}=T\nabla S-\nabla h
+\frac{\tilde{{\bf J}}\times{\bf B}}{\rho}
-\nabla\Phi, \label{eq:7.7}
\end{equation}
where
\begin{align}
\tilde{\bf J}=&{\bf J}-\nabla\times{\bm\Omega}
=\frac{\nabla\times\tilde{\bf B}}{\mu_0}, \label{eq:7.8}\\
\tilde{\bf B}=& {\bf B}\left[1-\frac{\mu_0 p_{\Delta}}{B^2}\right]. 
\label{eq:7.9}
\end{align}
In the form (\ref{eq:7.7}) the anisotropic pressure force term 
$-\nabla{\bf\cdot}{\sf p}/\rho$ has been partly transformed into the 
modified force ${\tilde J}\times {\bf B}/\rho$ where 
$\tilde{\bf J}=(\nabla\times\tilde{\bf B})/\mu_0$ in which $\tilde{\bf B}$ is
the modified magnetic induction $\tilde{\bf B}$. The form of $\tilde{\bf J}$
suggests that $-\nabla\times{\bm\Omega}$ could be interpreted 
as a magnetization current. 

It is interesting to note that $\tilde{B}=B(1-\mu_0 p_\Delta/B^2)<0$ if 
$p_\Delta=(p_\parallel-p_\perp)>B^2/\mu_0$, which  
corresponds to the firehose instability threshold (e.g. \cite{Stix92}, 
\cite{Hunana16, Hunana17}).
The mirror instability threshold for the CGL plasma model does not 
correspond to plasma kinetic theory.  
For the CGL plasma model with zero electron pressure (i.e. cold electrons)
 the mirror instability occurs if: 
$p_\perp-p_\parallel>(5/6)p_\perp+p_\parallel p_B/ p_\perp$ whereas 
kinetic theory gives the threshold for the instability as $p_\perp-p_\parallel>
p_\parallel p_B/ p_\perp$ where $p_B=B^2/(2\mu)$ is the magnetic pressure.
 Note the firehose instability occurs
if the parallel pressure dominates the perpendicular 
pressure. Similarly if $p_\perp$ dominates $p_\parallel$ one obtains the 
mirror instability. 
%\cite{Hunana17} have revisited the CGL plasma description and give further 
%analysis of the firehose and mirror instabilities
%study dispersion equations for CGL plasma instabilities, including Hall 
%current and finite Larmor radius corrections. 

In the \cite{Hazeltine13} and \cite{Holm86} 
 approaches, the internal energy density per unit mass, $e$ 
satisfies the first law of thermodynamics, in the form:
\begin{equation}
TdS=de+p_\parallel d\tau+\left(\frac{p_{\Delta}}{\rho B}\right) dB. 
\label{eq:7.10}
\end{equation}
%There are good physical reasons for the form of the first law 
%of thermodynamics in the analysis of \cite{Hazeltine13}. 
The variational formulation of \cite{Newcomb62} 
does not explicitly require (\ref{eq:7.10}) to apply. 
The \cite{Newcomb62} Lagrangian action principle  leads to 
the correct momentum equation for the CGL plasma, 
and to a Hamiltonian formulation of the equations
in Lagrangian variables.

The Lagrangian variational principle for ideal CGL plasmas 
obtained by \cite{Newcomb62} was used in Section 3 to obtain 
a canonical Hamiltonian formulation
of the equations based on the canonical coordinates 
${\bf q}={\bf x}({\bf x}_0,t)$  
and the canonical momentum ${\bm\pi}=\rho_0\dot{\bf x}({\bf x}_0,t)$,
and also to establish that stationary variations of the action 
give the CGL momentum equation in both its Lagrangian and Eulerian 
forms. 

Section 4 provides an Euler-Poincar\'e action principle derivation of the 
CGL momentum equation (see also Appendix E, which uses the approach of 
\cite{Holm98}). 

By transforming the canonical Poisson bracket for the CGL system
of Section 3, to non-canonical physical variables leads to 
the non-canonical Poisson bracket for the CGL system of Section 5. 
The detailed transformation
formulae from the canonical coordinates to the physical variables 
$\psi=(\rho,\sigma, {\bf M}, {\bf B})^T$
is described in Appendix~D. 
By writing the non-canonical Poisson bracket in co-symplectic form, leads to a system of equations for 
the Casimirs $C$, as solutions of the Poisson bracket
equation: $\{C,K\}=0$ where $K$ is an arbitrary functional of the 
physical variables. The Casimirs satisfy $\C_ t=\{C,K\}=0$. 
Hamiltonian dynamics of the system takes place on the symplectic
leaves $C=const.$ of the system.  

The classical Casimirs for ideal fluids and plasmas are 
the mass conservation integral, 
the cross helicity integral for barotropic flows, and the magnetic helicity.
For the CGL plasma case, there is effectively one entropy function 
that is Lie dragged with the flow, because $S_\parallel$ and $S_\perp$ 
are assumed to be functions only of $S$.  In principle, a more complicated 
bracket would arise, if one used the possibility that the adiabatic 
integrals for $p_\perp$ and $p_\parallel$ depended on other scalar invariants
that are advected with the flow. For example the integrals (\ref{eq:2.9a}) 
could also depend on the scalar invariant ${\bf B}{\bf\cdot}\nabla S/\rho$.  
This possibility was not explored in the present paper. 
%for $p_\perp$are two entropies $S_\perp$ and $S_\parallel$
%that are Lie dragged with the flow, as opposed to MHD where there 
%is only the fluid entropy $S$ to take into account.  

 The links between Noether's theorem and conservation laws 
for CGL plasmas are  developed (Section 6).
 The evolutionary symmetry 
form of the Lie invariance condition for the action is used. 
This approach differs
from the  
 canonical symmetry operator
form of the invariance condition (e.g. \cite{Webb19}, see also Appendix H). 
In the evolutionary form of the symmetry operator, 
the independent variables are frozen 
and all the Lie transformation changes are restricted to 
changes in the dependent variables and their derivatives 
(see e.g. \cite{Olver93}, \cite{Ibragimov85}, \cite{Bluman02}). 
In the canonical symmetry approach, 
both the dependent and independent variables and their derivatives 
change. 

The CGL plasma equations admit the 
Galilean Lie point symmetry group, and three extra scaling symmetries 
(Appendix F).   
The Galilean group leads to (\romannumeral1)\ 
the energy conservation law due to time translation invariance of the action;
(\romannumeral2)\ the momentum conservation equations due 
to space translation invariances; (\romannumeral3)\ 
the conservation of angular momentum due to rotational invariance 
 about some given rotation axis; and (\romannumeral4)\ 
the uniform center of mass conservation law which is due to 
invariance under Galilean boosts. These conservation laws
are derived using the evolutionary form of Noether's theorem. 

The nonlocal cross helicity conservation law 
(\ref{eq:6.30})-(\ref{eq:6.31}) (see also (\ref{eq:2.ch5}) 
arises from a fluid relabelling symmetry with generators 
$\hat{\bm\xi}=-{\bf B}/\rho$ and ${\bm\xi}_0={\bf B}_0/\rho_0$ 
and with non-trivial potentials $\Lambda^t_0$ and $\Lambda^i_0$. 
It is a nonlocal conservation law  
that depends on the Lagrangian time integral of the temperature back along the 
fluid flow trajectory. The cross helicity conservation 
law for MHD, is a local conservation law for the case of a barotropic 
gas (i.e. $p=p(\rho)$), but is a nonlocal conservation law for 
the non-barotropic case where $p=p(\rho,S)$ (see also \cite{Yahalom17a, 
Yahalom17b, Yahalom21} for a topological interpretation). 
The CGL entropies $S_\parallel$ and $S_\perp$ are not constants,
but are nontrivial  scalars that are advected with the background 
flow). 
%The fluid relabelling symmetry responsible for the cross helicity
%conservation law (described in Propositions \ref{prop6.6} and \ref{prop6.7})
%relies on the identity:
%\begin{equation}
%\frac{dB}{d\epsilon}={\cal L}_{\tilde{X}}\left(B\right)
%={\bf e}_{B}{\bf\cdot}\left[ \nabla\times(\hat{V}^{\bf x}\times{\bf B}) 
%-\hat{V}^{\bf x} (\nabla{\bf\cdot}{\bf B})\right].  \label{eq:7.11}
%\end{equation}
%for the rate of change of $B$ along the Lie trajectory 
%with canonical Lie generator 
%${\hat X}={\hat V}^{\bf x}{\bf\cdot}\nabla$. In this case the
%symmetry of interest ${\hat V}^{\bf x}={\bf B}/\rho$, 
%and $dB/d\epsilon=0$ (note $\nabla{\bf\cdot}{\bf B}=0$) which allows the use of Noether's theorem
%to obtain the conservation law. 

\cite{Lingam20} study extended variational principles of MHD (and CGL) 
type including gyro-viscous effects (i.e. 
higher order finite Larmor radius effects in the collisionless limit).
A single gyro-viscous term is added to the usual action.  
 The gyro-viscous term alters the total 
momentum density, but it does not alter the divergence of the mass flux.
The total momentum density has the form: 
${\bf M}^c={\bf M}+{\bf M}^*$, where ${\bf M}=\rho {\bf u}$ is the mass flux 
and ${\bf M}^*$ has the form ${\bf M}^*=\nabla\times {\bf L}^*$,
where ${\bf L}^*$ is the internal angular momentum of the particle
(${\bf L}^*=(2m/e){\bm\mu}$ and ${\bm\mu}=\mu{\bm\tau}$ 
is the vector form of the particle adiabatic moment.  
These ideas are related to papers by \cite{Newcomb72, Newcomb73, 
 Newcomb83}, \cite{Morrison14} and others. 
Our analysis can be extended in principle
 to include gyro-viscosity in the collisionless limit. 
However, to what extent the action principle approach,
reproduces the kinetic plasma and fluid approaches 
to gyro-viscosity
 requires further investigation.     

Analysis of the CGL plasma equations using 
Clebsch potentials  
 (e.g.  \cite{Zakharov97};
\cite{Webb18}, Ch. 8; 
\cite{Yahalom17a,Yahalom17b}; \cite{Yahalom21}) may yield further insights.
Similarly, further study of the Lie  symmetries the CGL plasma 
equations would be useful.     
Investigation of conservation laws for the CGL equations 
by using  Lie dragging remains open for further investigation. 

\medskip
\centerline{\bf Acknowledgements}
The work of GMW was supported in part by NASA grant 80NSSC19K0075.
GMW acknowledges stimulating discussions with Darryl Holm on the CGL 
plasma equations. SCA is supported by an NSERC Discovery research grant.
We acknowledge  discussions of the CGL plasma equations, 
and the anisotropic plasma moment equations and their 
closures with Peter Hunana.   
%The work of PH was supported in part by the European Research Council 
%in the frame of consolidating grant ERC-2017-CoG771310-P12A
%`Partial Ionisation: Two-Fluid Approach', led by Elena Khomenko.
We acknowledge the partial support of an NSF EPSCoR RII-Track-1 
cooperative agreement OIA-1655280 and a NASA IMAP-subaward under NASA
contract 80GSFC19C0027.

%GPZ is supported in part by an NSF EPSCOR project OIA-1655280:
%Connecting the Plasma Universe to Plasma Technology in AL: The Science and Technology of Low-Temperature Plasma. 

\appendix
\section*{Appendix A}
\setcounter{section}{1}
\setcounter{equation}{0}

In this appendix, 
we briefly discuss the derivation of the CGL plasma equations. 
%based mainly on the work of \cite{Hunana19a, Hunana19b}.
These equations were originally derived by \cite{Chew56} and later by many authors 
(e.g. \cite{Kulsrud83}; \cite{Ramos05a, Ramos05b}). 
Some of the derivations  use the adiabatic drift approximation, 
but others simply involve taking moments of the collisionless Vlasov equation 
or Boltzmann equation over the particle momenta. 
We will use the latter approach. 

Following \cite{Hunana19a, Hunana19b}, 
we introduce the velocity phase space distribution function 
$f({\bf x}, {\bf v},t)$ of the particles 
where $dN=f({\bf x},{\bf v},t) d^3v\ d^3{\bf x}$ is the number of particles 
in a volume of phase space at the point $({\bf x},{\bf v})$ at time $t$,
with velocity volume element $d^3v$ and position volume element $d^3{\bf x}$. 
The lower order moments of the velocity distribution function 
averaged over all velocities ${\bf v}$ are defined as:
\begin{equation}
\begin{aligned}
n=&\int f\ d^3{\bf v}, \\
n{\bf u}=&\int f{\bf v}\ d^3{\bf v}, \\
p_{ij}=&\int f (v^i-u^i)(v^j-u^j)\ d^3{\bf v}, \\
q_{ijk}=&\int f (v^i-u^i)(v^j-u^j)(v^k-u^k)\ d^3{\bf v}, 
\end{aligned}\label{eq:A1}
\end{equation}
where $n$ is the particle number density, 
${\bf u}$ is the fluid velocity, 
and in Cartesian coordinates, 
$p_{ij}$ represents the components of the pressure tensor $\sf p$, 
and $q_{ijk}$ represents the components of the heat flux tensor $\sf q$. 
Note that both of these tensors are symmetric. 

From adiabatic motion of guiding center theory, 
 the pressure tensor $\sf{p}$ at lowest order can be written in the form:
\begin{equation}
{\sf p}=p_\parallel \bm{\tau}\bm{\tau}+p_\perp\left(\sf{I}-\bm{\tau}\bm{\tau}\right)+ {\sf\Pi}, 
\label{eq:A2}
\end{equation}
where $\bm{\tau}=\bm{B}/B$ is the unit vector along the magnetic field, 
$p_\parallel$ and $p_\perp$ are the gyrotropic components of $\sf{p}$ 
parallel and perpendicular to the magnetic field ${\bf B}$, 
and $\sf{\Pi}$ represents the non-gyrotropic components of $\sf{p}$. 
In the limit of a strong background magnetic field ${\bf B}$ 
in which the particle gyro-radius is $r_g\ll L$ where $L$ is the scale length for variation of ${\bf B}$, 
and for times $T\gg T_\Omega$ where $T_\Omega$ is the gyro-period,  
the particle distribution is approximately gyrotropic
while the non-gyrotropic pressure $\sf{\Pi}$ can be neglected to first 
order in $r_g/L$ and $T_\Omega/T$. 

Hereafter, $\sf A^S=A+A^T$ denotes the symmetrized form of ${\sf A}$, 
where $\sf T$ denotes the transpose. 
%Details of the calculations can be found in (\cite{Hunana19a}).

Taking the first moment of the Vlasov equation with respect to ${\bf v}$
gives the pressure tensor equation
\begin{equation}
\begin{aligned}
&\deriv{{\sf p}}{t}+\nabla{\bf\cdot}\left({\bf u}\otimes {\sf p}+{\sf q}\right)
+\left({\sf p}{\bf\cdot}\nabla{\bf u}\right)^{\sf S}
+\frac{q}{mc} \left({\bf B}\times {\sf p} \right)^{\sf S}=0, 
\end{aligned}\label{eq:A3}
\end{equation}
where
\begin{equation}
\left({\bf B}\times {\sf p}\right)_{ik}=\epsilon_{ijl}B_jp_{lk}.
 \label{eq:A4}
\end{equation}
The equations for the parallel pressure $p_\parallel$ and perpendicular pressure $p_\perp$
obtained from (\ref{eq:A3}) respectively reduce to:
\begin{equation}
\begin{aligned}
&\deriv{p_\parallel}{t}+\nabla{\bf\cdot}\left(p_\parallel{\bf u}\right)
+2p_\parallel \bm{\tau}\bm{\tau}{\bf:}\nabla{\bf u} 
+\bm{\tau}\bm{\tau}{\bf :}\left(\nabla{\bf\cdot}{\sf q}\right)
\\&\quad
-{\sf \Pi}{\bf :}\frac{d}{dt}({\bm{\tau}\bm{\tau})}
+\left({\sf\Pi}{\bf\cdot}\nabla{\bf u}\right)^{\sf S}{\bf :}\bm{\tau}\bm{\tau}=0 ,
\end{aligned}\label{eq:A5}
\end{equation}
and 
\begin{equation}
\begin{aligned}
&\deriv{p_\perp}{t}+\nabla{\bf\cdot}\left(p_\perp {\bf u}\right) 
+p_\perp \nabla{\bf\cdot}{\bf u} 
-p_\perp \bm{\tau}\bm{\tau} {\bf :}\nabla {\bf u}%\\&\quad
+\frac{1}{2}\left({\sf Tr}\left(\nabla{\bf\cdot}{\sf q}\right)
-\bm{\tau}\bm{\tau}{\bf:}\left(\nabla{\bf\cdot}{\sf q}\right)\right)\\&\quad
+\frac{1}{2}\left({\sf Tr}\left(\sf{\Pi}{\bf\cdot}\nabla{\bf u}\right)^{\sf S}
+{\sf\Pi}{\bf :} \frac{d}{dt}(\bm{\tau}\bm{\tau})
-\left(\sf{\Pi}{\bf\cdot}\nabla{\bf u}\right)^S{\bf :}\bm{\tau}\bm{\tau}\right)=0. 
\end{aligned}\label{A6}
\end{equation}

Neglecting the non-gyrotropic component $\sf{\Pi}$ of the pressure tensor, 
and neglecting the heat flux tensor components ${\sf q}$ gives the simplified 
CGL plasma equations for $p_\parallel$ and $p_\perp$ as
\begin{align}
\deriv{p_\parallel}{t}+\nabla{\bf\cdot}\left(p_\parallel{\bf u}\right)
+2p_\parallel \bm{\tau}\bm{\tau}{\bf :}\nabla{\bf u}=&0,
\label{eq:A7}\\
\deriv{p_\perp}{t}+\nabla{\bf\cdot}\left(p_\perp {\bf u}\right) 
+p_\perp \nabla{\bf\cdot}{\bf u} -p_\perp\bm{\tau}\bm{\tau}{\bf :}\nabla {\bf u}=&0.  
\label{eq:A8}
\end{align}
The double adiabatic equations (\ref{eq:2.7}) for $p_\parallel$ and $p_\perp$
follow by combining these transport equations (\ref{eq:A7}) and (\ref{eq:A8}) 
with the mass continuity equation (\ref{eq:2.1}) in the form 
\begin{equation}
\frac{d}{dt} \rho =-\rho\nabla{\bf\cdot}{\bf u} , 
\label{eq:A9}
\end{equation}
and the magnetic field strength equation 
\begin{equation}
\frac{d}{dt}B = -(\nabla{\bf u}){\bf:}\left({\sf I} -\bm{\tau}\bm{\tau}\right)B
\label{eq:A10}
\end{equation}
which comes from Faraday's equation (\ref{eq:2.5}).

\appendix
\section*{Appendix B}
\setcounter{section}{2}
\setcounter{equation}{0}

In this appendix, 
we derive the two equivalent forms of the pressure divergence equation (\ref{eq:2.ch2}) and (\ref{eq:C14}) 
used in the derivation of the corresponding forms of the cross helicity conservation law
(\ref{eq:2.ch5}) and (\ref{eq:2.ch7b}). 
Throughout we take $\nabla{\bf\cdot}{\bf B}=0$. 

We start from the non-zero terms on the righthand side of equation (\ref{eq:2.ch2}):
\begin{equation}
{\bf B}\times(\nabla\times{\bm\Omega}) + \rho(T\nabla S -\nabla h) .
\label{eq:2.ch2.rhs}
\end{equation}
By applying a standard cross-product identity on the first term in (\ref{eq:2.ch2}), 
we expand 
\begin{equation}
\begin{aligned}
{\bf B}\times(\nabla\times{\bm\Omega}) 
& = \left[\nabla\left(\frac{p_\Delta}{B}\bm{\tau}\right)\right]{\bf \cdot}\bm{\tau}B
- B\bm{\tau}{\bf \cdot}\nabla\left(\frac{p_\Delta}{B}\bm{\tau}\right) \\
%& =B\nabla\left(\frac{p_\Delta}{B}\right) - B\left[\bm{\tau}{\bf \cdot}\nabla\left(\frac{p_\Delta}{B}\right)\right] \bm{\tau}
& = \nabla p_\Delta - \left(\bm{\tau}{\bf \cdot}\nabla p_\Delta\right) \bm{\tau}
+p_\Delta\left[ -\bm{\tau}{\bf \cdot}\nabla\bm{\tau} +\left(\nabla \bm{\tau}\right) {\bf \cdot}\bm{\tau}
-\nabla\ln B  + \left(\bm{\tau}{\bf \cdot}\nabla\ln B\right) \bm{\tau}\right] . 
\end{aligned}
\label{eq:term1.expand}
\end{equation}
In this expression, 
the second and third terms combine into 
$-\bm{\tau}{\bf \cdot}\nabla( p_\Delta \bm{\tau})$;
the fourth term vanishes
$ \left(\nabla \bm{\tau}\right) {\bf \cdot}\bm{\tau} 
= \nabla\left(\frac{1}{2}\bm{\tau}{\bf \cdot}\bm{\tau}\right) =0$
since $\bm{\tau}$ is a unit vector; 
and with the Gauss' law equation written as $\nabla{\bf\cdot}\bm{\tau} = -\bm{\tau}{\bf\cdot}\nabla\ln B$, 
the last term can be expressed as
$-p_\Delta (\nabla{\bf\cdot}\bm{\tau})\bm{\tau}$. 
Thus, equation (\ref{eq:term1.expand}) reduces to:
\begin{equation}
\begin{aligned}
{\bf B}\times(\nabla\times{\bm\Omega}) 
& = \nabla p_\Delta -p_\Delta \nabla\ln B  -\nabla{\bf\cdot}\left(p_\Delta \bm{\tau}\bm{\tau}\right) .
\end{aligned}
\label{eq:term1}
\end{equation}
Next, we expand the remaining term in (\ref{eq:2.ch2.rhs})
by using the Pfaffian differential equation (\ref{eq:2.16i}) in the gradient form:
\begin{equation}
\nabla e = T\nabla S +\frac{p_\parallel}{\rho}\nabla\ln \rho -\frac{p_{\Delta}}{\rho}\nabla\ln B . 
\end{equation}
This yields 
\begin{equation}
\begin{aligned}
\rho(T\nabla S -\nabla h) 
& = \rho\left( \nabla(e-h) -\frac{p_\parallel}{\rho}\nabla\ln \rho +\frac{p_{\Delta}}{\rho}\nabla\ln B\right)\\
& =-\nabla p_\parallel +p_{\Delta}\nabla\ln B . 
\end{aligned}
\label{eq:term2}
\end{equation}
by using expressions (\ref{eq:2.ch4}) for enthalpy and (\ref{eq:2.16j}) for internal energy. 
Finally, we combine the terms (\ref{eq:term2}) and (\ref{eq:term1}), 
which gives
\begin{equation}
{\bf B}\times(\nabla\times{\bm\Omega}) + \rho(T\nabla S -\nabla h) 
= 
-\nabla p_\perp -\nabla{\bf\cdot}\left(p_\Delta \bm{\tau}\bm{\tau}\right)
=-\nabla{\bf\cdot}{\sf p} 
\label{eq:C10}
\end{equation}
since the pressure tensor (\ref{eq:defn.p}) can be written in terms of $p_\Delta$ as:
\begin{equation}
{\sf p}=p_\parallel \bm{\tau}\bm{\tau} + p_\perp ({\sf I} -\bm{\tau}\bm{\tau})
=p_\perp {\sf I}+p_\Delta \bm{\tau}\bm{\tau} . 
\label{eq:p.alt}
\end{equation}
This yields the first form of the pressure divergence equation (\ref{eq:2.ch2}). 

We derive the second form of the pressure divergence equation (\ref{eq:C14}) 
by starting from the terms on its lefthand side:
\begin{equation}
\nabla{\bf\cdot}{\sf p} -\rho\nabla h . 
\label{eq:C14.lhs}
\end{equation}
Expanding the first term in (\ref{eq:C14.lhs}) 
by use of the gyrotropic expression (\ref{eq:p.alt}), we obtain 
\begin{equation}
\nabla{\bf\cdot}{\sf p} = \nabla p_\perp + \bm{\tau}\bm{\tau}{\bf\cdot}\nabla p_\Delta + p_\Delta \left(\bm{\tau}\nabla{\bf\cdot}\bm{\tau} + \bm{\tau}{\bf\cdot}\nabla\bm{\tau} \right)
\end{equation}
The scalar product of this expression with $\bm{\tau}$ yields
\begin{equation}
\bm{\tau}{\bf \cdot}\left( \nabla{\bf\cdot}{\sf p} \right) 
= \bm{\tau}{\bf \cdot}\left( \nabla p_\parallel  -p_\Delta \nabla\ln B \right)
\label{eq:term1'}
\end{equation}
through the Gauss' law equation in the form $\nabla{\bf\cdot}\bm{\tau} = -\bm{\tau}{\bf\cdot}\nabla\ln B$. 
The second term in (\ref{eq:C14.lhs}) can be expanded using the enthalpy (\ref{eq:2.ch4}).
After taking the scalar product with $\bm{\tau}$, 
this gives
\begin{equation}
-\rho\bm{\tau}{\bf\cdot}\nabla h 
=-\bm{\tau}{\bf\cdot}\nabla\left( \frac{3}{2} p_\parallel + p_\perp\right) 
+ \left( \frac{3}{2} p_\parallel + p_\perp\right) \bm{\tau}{\bf\cdot}\nabla\ln\rho . 
\label{eq:term2'}
\end{equation}
The combined terms (\ref{eq:term1'}) and (\ref{eq:term2'}) yield 
\begin{equation}
\bm{\tau}{\bf\cdot}\left( \nabla{\bf\cdot}{\sf p} -\rho\nabla h  \right) 
= -\bm{\tau}{\bf\cdot}\nabla\left( \frac{1}{2} p_\parallel + p_\perp\right) 
-p_\Delta\bm{\tau}{\bf\cdot}\nabla\ln B
+ \left( \frac{3}{2} p_\parallel + p_\perp\right) \bm{\tau}{\bf\cdot}\nabla\ln\rho . 
\label{eq:terms'}
\end{equation}
Now, we substitute the gradient of the expressions for $p_\parallel$ and $p_\perp$ 
in terms of adiabatic integrals (\ref{eq:S.eqns}): 
\begin{equation}
\begin{aligned}
\nabla p_\parallel & = p_\parallel \left( \nabla\ln c_\parallel(S)  + 3\nabla\ln \rho -2\nabla\ln B\right), \\
\nabla p_\perp & = p_\perp \left( \nabla\ln c_\perp(S)  -\nabla\ln \rho - \nabla\ln B\right) .
\end{aligned}
\end{equation}
Then equation (\ref{eq:terms'}) reduces to the form 
\begin{equation}
\bm{\tau}{\bf\cdot}\left( \nabla{\bf\cdot}{\sf p} -\rho\nabla h  \right) 
= -\frac{1}{2} p_\parallel\bm{\tau}{\bf\cdot}\nabla\ln c_\parallel(S) 
-p_\perp\bm{\tau}{\bf\cdot}\nabla\ln c_\perp(S) , 
\end{equation}
which gives the pressure divergence equation (\ref{eq:C14})  
after we use the relations (\ref{eq:2.9b}) for the adiabatic integrals 
in terms of ${\bar S}_\parallel$ and ${\bar S}_\perp$.

\appendix
\section*{Appendix C}
\setcounter{section}{3}
\setcounter{equation}{0}

In this appendix, 
we show how the Eulerian momentum equation (\ref{eq:2.2}) is obtained from 
the Euler-Lagrange equation (\ref{eq:3.20}) of the Lagrangian (\ref{eq:3.19}). 
We follow the approach in \cite{Newcomb62}. 

From expression (\ref{eq:3.19}), we have
\begin{equation}
\begin{aligned}
& \deriv{L_0}{X^i} = -\rho_0 \deriv{\Phi}{X^i},
\quad
\deriv{L_0}{\dot X^i} = \rho_0 \dot X^i, 
\\
& \deriv{L_0}{X_{ij}} = \left(\frac{p_{\parallel 0}}{\zeta^4} -\frac{p_{\perp 0}}{\zeta J} \right)
X_{ik}\tau_0^k \tau_0^j
+ \left(\frac{p_{\perp 0}\zeta}{J^2} + \frac{\zeta^2 B_0^2}{2\mu_0 J^2}\right) A_{ij}
-\frac{X_{ik} B_0^k B_0^j}{\mu_0 J} , 
\end{aligned}
\label{eq:D1}
\end{equation}
which uses expression (\ref{eq:3.9}) for $\zeta^2$,
%$\deriv{\zeta^2}{X_{ij}} = 2 X_{ik}\tau_0^k \tau_0^j$, 
and the derivative expression in (\ref{eq:3.5}) for $J$. 
Then, substituting the derivatives (\ref{eq:D1}) 
into the Euler-Lagrange equation (\ref{eq:3.20}), 
we obtain 
\begin{equation}
\begin{aligned}
E_{X^i}(L_0) 
& = -\rho_0 \deriv{\Phi}{X^i} 
-\deriv{}{t}\left(\rho_0 \dot X^i\right)
-\derv{x_0^j}\left(
\left(\frac{p_{\perp 0}\zeta}{J^2} + \frac{\zeta^2 B_0^2}{2\mu_0 J^2}\right) A_{ij}
\right)
\\&\qquad
-\derv{x_0^j}\left(
 \left( \frac{p_{\parallel 0}}{\zeta^4} -\frac{p_{\perp 0}}{\zeta J} -\frac{B_0^2}{\mu_0 J} \right) X_{ik}\tau_0^k \tau_0^j 
\right) . 
\end{aligned}
\label{eq:D2}
\end{equation}
To proceed, we will need the derivative identity:
\begin{equation}
A_{ij}\derv{x_0^j} f   = \derv{x_0^j}\left( A_{ij} f \right) 
= J\deriv{f}{X^i} = J\nabla_i f\big|_{{\bf x}={\bf X}}
\label{eq:D5}
\end{equation}
holding for any function $f$
(see \cite{Newcomb62}). 
This identity arises from the middle property in (\ref{eq:3.5}) as follows. 
Differentiating with respect to $x_0^k$ 
gives $\partial J/\partial x_0^j = X_{ij} \partial A_{ik}/\partial x_0^k + A_{ik} \partial X_{ij}/\partial x_0^k$. 
By using commutativity of partial derivatives 
$\partial X_{ij}/\partial x_0^k=\partial X_{ik}/\partial x_0^k$
and substituting Jacobi's formula for derivative of a determinant
$\partial J/\partial x_0^i = A_{jk} \partial X_{jk}/\partial x_0^i$, 
we find $\partial A_{ik}/\partial x_0^k =0$. 
This leads to the identity (\ref{eq:D5}),
after applying the chain rule and again using the middle property in (\ref{eq:3.5}). 

There are two main steps for simplifying (\ref{eq:D2}). 
First, 
through relations (\ref{eq:3.9}), (\ref{eq:3.11b}), (\ref{eq:3.13}),
followed by use of the identity (\ref{eq:D5}), 
we can express the two divergence terms in (\ref{eq:D2}) in the form:
\begin{align}
& \begin{aligned}
\derv{x_0^j}\left( 
\left( p_{\perp} + \frac{B^2}{2\mu_0}\right) A_{ji}
\right)
& = J \derv{X^i}\left( 
\left( p_{\perp} + \frac{B^2}{2\mu_0}\right) 
\right) , 
\end{aligned}
\label{eq:D4a}
\\
& \begin{aligned}
\derv{x_0^j}\left( 
\frac{J}{\zeta^2}\left(p_\Delta -\frac{B^2}{\mu_0}\right)\zeta \tau_i \tau_0^j 
\right)
& = \derv{x_0^j}\left( 
\left(p_\Delta -\frac{B^2}{\mu_0}\right) \tau_i \tau^k A_{kj} 
\right) 
\\
& = \derv{X^k}\left( 
\left(p_\Delta -\frac{B^2}{\mu_0}\right) \tau_i \tau^k 
\right) . 
\end{aligned}
\label{eq:D4b} 
\end{align}
Note these terms imply that 
\begin{equation}
\deriv{L_0}{X_{ij}} = A_{kj}\left( 
\delta_{ik}\left(p_\perp+\frac{B^2}{2\mu_0}\right)
+\left(p_\Delta -\frac{B^2}{\mu_0}\right) \tau_i \tau_k
\right) . 
\label{eq:D1b}
\end{equation}

Second, 
from the density relation (\ref{eq:3.2}) and the fluid element flow equation (\ref{eq:3.0a}), 
we see that in (\ref{eq:D2}) the first term is simply 
\begin{equation}
-J\rho \deriv{\Phi}{X^i} , 
\label{eq:D3a}
\end{equation}
while the second term can be expressed as
\begin{equation}
\begin{aligned}
-\left( \dot J\rho u^i + J \frac{d}{dt}(\rho u^i) \right) 
& = 
-\rho u^i A_{kj}\deriv{\dot X^k}{x_0^j} 
- J\left( (\rho u^i)_t + \dot X^j \nabla_j (\rho u^i) \right) 
\\
& = 
- J\left( (\rho u^i)_t + \nabla_j (\rho u^i u^j) \right)\big|_{{\bf x}={\bf X}}
\end{aligned}
\label{eq:D3b}
\end{equation}
by use of relation (\ref{eq:3.4}). 

Thus, the Euler-Lagrange equation (\ref{eq:D2}) becomes:
\begin{equation}
\begin{aligned}
E_{X^i}(L_0) 
= -J \left( 
\rho \nabla_i \Phi +(\rho u^i)_t  
+\nabla_i\left( p_{\perp} + \frac{B^2}{2\mu_0} \right) 
+\nabla_j\left( \left(p_\Delta -\frac{B^2}{\mu_0}\right) \tau_i \tau^j \right) 
\right)\bigg|_{{\bf x}={\bf X}} .
\end{aligned}
\label{eq:D6}
\end{equation}
The stationary points of the action principle are given by the equation 
$E_{X^i}(L_0) =0$. 
From expression (\ref{eq:D6}), the resulting equation coincides with the 
Eulerian momentum equation (\ref{eq:2.2}). 

Note that, in terms of the Lagrangian variables, 
the Euler-Lagrange equation (\ref{eq:D2}) is 
a nonlinear wave system for $X^i(x_0^j,t)$,
where $X_{ij} = \partial X^i/\partial x_0^j$ 
and $A_{ij} = J^{-1} \big( \partial X^i/\partial x_0^j \big)^{-1}$. 
(See also \cite{Golovin11}; \cite{Webb05b}; \cite{Webb19} for the MHD case).

\appendix
\section*{Appendix D}
\setcounter{section}{4}
\setcounter{equation}{0}

%The use of variational principles in fluid dynamics has a long history 
%stretching back in time to the works of Euler and Lagrange.

\cite{Eckart63} uses the Lagrangian map
and  Jacobians to describe fluids. \cite{Lundgren63} 
uses Eulerian and Lagrangian variations in MHD 
and in plasma physics (see also \cite{Newcomb62}). In \cite{Lundgren63}
a small parameter $\epsilon$ is used to describe the variations, 
where the physical quantity $\psi$ has the functional form: 
$\psi=\psi({\bf x},{\bf x}_0,\epsilon)$ 
in which ${\bf x}={\bf X}({\bf x}_0,t)$ is the Lagrangian map. 
Eulerian ($\delta\psi$) and Lagrangian ($\Delta\psi$) variations of $\psi$ 
are defined as:
\beqn
\delta\psi=\lim_{\epsilon\to 0}\left(\deriv{\psi}{\epsilon}
\right)_{\bf x},\quad 
\Delta\psi=\lim_{\epsilon\to 0}\left(\deriv{\psi}{\epsilon}\right)_{{\bf x}_0}.
\label{eq:L1}
\eeqn
Thus, for an Eulerian variation, $\delta\psi$ is evaluated with ${\bf x}$ 
held constant, whereas for a Lagrangian variation $\Delta\psi$, 
${\bf x}_0$ is held constant. Thus $\delta{\bf x}=0$ and $\Delta{\bf x}_0=0$.

Using the chain rule for differentiation, it follows that:
\begin{align}
\delta\psi=&\Delta\psi+\delta{\bf x}_0{\bf\cdot}\nabla_0\psi, \nonumber\\
\Delta\psi=& \delta\psi+\Delta{\bf x}{\bf\cdot}\nabla\psi. \label{eq:L2}
\end{align}
\cite{Dewar70} used a variational principle for linear WKB MHD waves 
in a non-uniform background plasma flow.  
 \cite{Webb05a} used a variational principle to 
describe non-WKB waves in a non-uniform background flow.
\cite{Webb05b} used similar ideas to describe nonlinear waves 
in a non-uniform flow by variational methods. 
 
In this appendix we describe the use of Eulerian and Lagrangian variations
in defining the Poisson bracket for CGL plasmas. 

We derive the CGL plasma non-canonical bracket (\ref{eq:5.3a})
starting from the canonical Poisson bracket (\ref{eq:5.5}). 
This canonical bracket is properly defined in a Lagrangian frame, i.e. 
a physical reference frame moving with the fluid flow, 
where all quantities are functions of the fluid element labels $x_0^i$ and time $t$. 
Non-scalar quantities (e.g. vectors, differential forms, tensors) 
are expressed in terms of their components with respect to the  Cartesian basis vectors 
of this frame (corresponding to the coordinates $x_0^i$, $i=1,2,3$).
These components will be designated by a subscript $0$.  

The canonical bracket (\ref{eq:5.5}) has the component form 
\begin{equation}
\left\{\F,\G\right\}=\int\left(\F_{q^i}\G_{p^i} -\F_{p^i}\G_{q^i}\right)\, d^3x_0, 
\label{eq:E1}
\end{equation}
where $q^i=x^i(x_0,t)$ and $p^i=\pi^i(x_0,t)$ comprise the canonical coordinates and momenta, 
with $\pi^i = \rho_0 \dot{x}^i$, 
as used in formulating Hamilton's equations (\ref{eq:3.33}). 
Note that the fluid element motion is expressed through the relation 
\begin{equation}
\dot{x}^i(x_0,t) = u^i(x(x_0,t),t) . 
\label{eq:E2}
\end{equation}
Here $\F$ and $\G$ are functionals which depend on $q^i$, $p^i$, as well as 
a set of advected quantities ${\sf a}(x_0^j,t)$ which are used in describing the dynamics. 

For a CGL plasma, the basic advected quantities are listed in (\ref{eq:4.8}). 
%We will also include the advected scalars $\bar S_{\parallel 0}$ and $\bar S_{\perp 0}$ 
%for the reasons explained in section~\ref{sec:Poissonbracket}. 
We will take 
\begin{equation}
{\sf a} = (S_0,%\bar S_{\parallel 0},\bar S_{\perp 0},
{\bf B}_0/\rho_0,\rho_0\,d^3x_0) . 
\label{eq:E3}
\end{equation}
Note we asume $S_\parallel$ and $S_\perp$ are functionals of $s$. 
Accordingly, functionals will be expressed as 
\begin{equation}
\F = \int F_0({\sf Z}_0)\,d^3x_0 
\label{eq:E4a}
\end{equation}
in terms of the Lagrangian variables 
\begin{equation}
{\sf Z}_0 = ({\bf q},{\bf p},{\sf a}) . 
\label{eq:E5}
\end{equation}

The non-canonical Poisson bracket (\ref{eq:5.3a}) employs the Eulerian variables (\ref{eq:5.7}). 
The transformation from Lagrangian variables (\ref{eq:E5}) to these Eulerian variables 
is effected in the following four steps. 

Firstly, the Eulerian form of a functional (\ref{eq:E4a}) is given by 
\begin{equation}
\F = \int F({\sf Z})\,d^3x 
\label{eq:E4b}
\end{equation}
with 
${\sf Z}=\big(\rho,\sigma,{\bf B},{\bf M}\big)$. 
The vector variables here will be expanded in components 
with respect to the Eulerian basis vectors 
corresponding to $x^i$ viewed as coordinates in an Eulerian frame. 
It will be convenient to take the basis vectors to be derivative operators 
(via the standard correspondence between vectors and directional derivatives
e.g. \cite{Schutz80}):
\begin{equation}
{\bf x} = x^i \partial_{x^i},
\quad
{\bf x}_0 = x_0^i \partial_{x_0^i} . 
\label{eq:E6}
\end{equation}

\def\Lvar{\Delta}
\def\Evar{\delta}

Secondly, using the notation in \cite{Newcomb62},
we define $\Lvar x^i$ to represent a variation of $x^i(x_0,t)$ 
in which $x_0^i$ and $t$ are held fixed: $\Lvar x_0^i=0$ and $\Lvar t=0$. 
Note that $\Lvar x^i$ will itself be some function of $x_0^i$ and $t$; we will write it 
in the Eulerian form: 
\begin{equation}
\Lvar x^i(x_0,t) = \epsilon^i(x,t)
\label{eq:E7a}
\end{equation}
where, on the righthand side, $x^i$ is regarded as a function of $x_0^i$ and $t$. 
Likewise, $\Lvar \pi^i = \Lvar (\rho_0 \dot{x}^i) = \rho_0 \delta \dot{x}^i$ is a corresponding variation of $\pi^i(x_0,t)$, 
where $\rho_0$ is unchanged because it is a function of only $x_0^i$ and $t$. 
In Eulerian form: 
\begin{equation}
\Lvar \pi^i(x_0,t) = J(x) \rho(x) \Lvar u^i(x,t) ,
\label{eq:E7b}
\end{equation}
using the relation (\ref{eq:E2}) and the advection result (\ref{eq:3.2}),
where $J$ is the determinant of the Jacobian matrix of partial derivatives of $x^i(x_0,t)$
(cf (\ref{eq:3.3})). 

The third step is to derive formulae for the variation of the Eulerian variables ${\sf Z}$. 
These formulae depend on the specific tensorial nature of each variable
and will be obtained through the variation of the advected variables (\ref{eq:E3}) 
in component form. 
Hereafter we will suppress the $t$ dependence in all variables and quantities 
whenever it is convenient.

We start with the scalar field $S_0$. 
Since it is advected (i.e. frozen in), this implies $S(x) = S_0(x_0)$. 
Applying a variation (\ref{eq:E7a}), we consequently see that
\begin{equation}
\Lvar S(x)= \Lvar S_0(x_0) = 0 . 
\label{eq:E12a}
\end{equation}
%Likewise, we have
%\begin{equation} 
%\Lvar S_\parallel(x) = 0, 
%\quad
%\Lvar S_\perp(x) = 0 . 
%\label{eq:E12b}
%\end{equation} 

Next we consider $\rho(x)\, d^3x$, which is properly viewed as an advected differential 3-form
(see e.g. \cite{Schutz80}). 
Its advection property is expressed by (\ref{eq:3.2}). 
Since $x_0$ and $t$ are held fixed, a variation (\ref{eq:E7a}) yields 
\begin{equation} 
\Lvar(\rho(x)J(x))= J(x)\Lvar\rho(x) +\rho(x)\Lvar J(x) =0 . 
\label{eq:E13}
\end{equation}
Now we use the variation of the determinant relation $d^3x = J(x)d^3x_0$,
which is given by 
$\Lvar\, d^3x = \partial_{x^i}(\Lvar x^i)\, d^3x = (\Lvar J(x)) d^3x_0$. 
This yields the result
\begin{equation}
\Lvar J(x) = J(x)\partial_{x^i} \epsilon^i(x). 
\label{eq:E14}
\end{equation} 
Substituting this variation into (\ref{eq:E13}), we obtain
\begin{equation} 
\Lvar\rho(x) = -\rho(x)\partial_{x^i}\epsilon^i(x) . 
\label{eq:E15}
\end{equation}

Last we consider the vector field ${\bf b}(x)\equiv {\bf B}(x)/\rho(x)$, 
which has the advection property (\ref{eq:3.7}). 
We use this property in component form:
$b^i(x)\partial_{x^i} = b_0^i(x_0)\partial_{x_0^i}$. 
Again, since all of the quantities on the righthand side are held fixed in a variation (\ref{eq:E7a}), we see that
\begin{equation}
\Lvar (b^i(x)\partial_{x^i}) = 
(\Lvar b^i(x)) \partial_{x^i} + b^i(x) \Lvar\partial_{x^i} . 
\label{eq:E16}
\end{equation}
To determine $\Lvar\partial_{x^i}$, we use the coordinate basis relation 
$\partial_{x^i} \hook dx^j=\delta_i^j$ 
where $\delta_i^j$ is the Kronecker symbol. 
Taking the variation gives
$\Lvar \partial_{x^i} \hook dx^j= - \partial_{x^i} \hook \Lvar dx^j$,
where 
\begin{equation}
\Lvar dx^j = d\Lvar x^j = d\epsilon^j(x) = dx^k \partial_{x^k}\epsilon^j(x) .
\end{equation}
Since $\Lvar \partial_{x^i}$ must have the form of a linear transformation, 
say $\phi_i{}^k$, applied to $\partial_{x^k}$, 
we find $\phi_i{}^k \partial_{x^k} \hook dx^j = \phi_i{}^j = - \partial_{x^i} \hook dx^k \partial_{x^k}\epsilon^j(x)= -\partial_{x^i}\epsilon^j(x)$. 
Thus, we have 
\begin{equation}
\Lvar \partial_{x^i} =-\partial_{x^i}\epsilon^j(x) \partial_{x^j} . 
\end{equation}
Substituting this formula into (\ref{eq:E16}) yields 
\begin{equation}
\Lvar b^i(x) = b^j(x)\partial_{x^j} \epsilon^i(x) . 
\label{eq:E17}
\end{equation}

By applying the variations (\ref{eq:E12a}), (\ref{eq:E15}), and (\ref{eq:E17})
to the quantities $\sigma(x) = \rho(x) S(x)$, 
%$\sigma_\parallel(x) = \rho(x) S_\parallel(x)$,  
%$\sigma_\perp(x) = \rho(x) S_\perp(x)$, 
and $B^i(x)=\rho(x) b^i(x)$, 
we readily obtain
\begin{equation}
%\begin{aligned}
\Lvar \sigma(x)  = -\sigma(x)\partial_{x^i}\epsilon^i(x) , 
%\\
%\Lvar \sigma_\parallel(x) & = -\sigma_\parallel(x)\partial_{x^i}\epsilon^i(x) , 
%\\
%\Lvar \sigma_\perp(x) & = -\sigma_\perp(x)\partial_{x^i}\epsilon^i(x) , 
%\end{aligned}
\label{eq:E18}
\end{equation}
and
\begin{equation}
\Lvar B^i(x) = - (\partial_{x^j}\epsilon^j(x)) B^i(x) +B^j(x)\partial_{x^j} \epsilon^i(x) . 
\label{eq:E19}
\end{equation}

To complete the derivation of $\delta{\sf Z}$, 
we will also need to find the variation of $M^i(x) = \rho(x) u^i(x)$,
which can be obtained directly from the relation (\ref{eq:E2}). 
Taking the variation of this relation yields
\begin{equation}
\Lvar u^i(x) = \partial_t \Lvar x^i(x_0) = d\epsilon^i(x)/dt . 
\label{eq:E20}
\end{equation}
This result implies that
$\Lvar M^i(x) = (\Lvar \rho(x)) u^i(x) + \rho(x)(\Lvar u^i(x)) = 
\rho(x) \big[ d\epsilon^i(x)/dt  -(\partial_{x^j} \epsilon^j(x))u^i(x) \big]$,
and thus we obtain 
\begin{equation}
\Lvar M^i(x) = \rho(x) \frac{d}{dt}\epsilon^i(x) - M^i(x) \partial_{x^j} \epsilon^j(x)
= \rho(x)\partial_t \epsilon^i(x) + M^j(x)\partial_{x^j}\epsilon^i(x) - M^i(x) \partial_{x^j} \epsilon^j(x) .
\label{eq:E21}
\end{equation}

Now, the fourth step consists of transforming the variational derivatives 
$\F_{q^i}\equiv \delta\F/\delta x^i$ and $\F_{p^i}\equiv \delta\F/\delta \pi^i$ 
into an equivalent form with respect to the Eulerian variables that comprise ${\sf Z}$. 
Consider a variation of a functional $\F$. 
From (\ref{eq:E4a}), we obtain 
\begin{equation}
\Lvar \F 
=\int \left(F_{0\,q^i} \Lvar x^i +F_{0\,p^i} \Lvar \pi^i\right)\, d^3x_0
= \int \left( J^{-1}F_{0\,q^i} \epsilon^i +F_{0\,p^i} \rho \frac{d}{dt}\epsilon^i \right)\, d^3x
\label{eq:E22a}
\end{equation}
where we have used (\ref{eq:E7a})--(\ref{eq:E7b}) and (\ref{eq:E20}), 
along with $\Lvar{\sf a}=0$ 
which holds because the quantities (\ref{eq:E3}) comprising ${\sf a}$ are advected. 
Similarly, from (\ref{eq:E4b}),
we get
\begin{equation}
\begin{aligned}
\Lvar \F 
=&\int \Big(
\left( F_\rho\Lvar\rho +F_\sigma \Lvar\sigma 
%+F_{\sigma_\parallel} \Lvar\sigma_\parallel  
%+F_{\sigma_\perp} \Lvar\sigma_\perp
+F_{B^i}\Lvar B^i +F_{M^i}\Lvar M^i \right) J 
+ F\delta J 
\Big)\,d^3x_0 
\\
=&\int \Big(
\left( F - F_\rho \rho -F_\sigma \sigma 
%- F_{\sigma_\parallel} \sigma_\parallel -F_{\sigma_\perp} \sigma_\perp
- F_{B^i} B^i - F_{M^i} M^i \right) \partial_{x^j} \epsilon^j 
\\&\qquad 
+ F_{B^i} B^j\partial_{x^j} \epsilon^i + F_{M^i} \rho \frac{d}{dt}\epsilon^i  
\Big)\, d^3x 
\\
=&\int \Big(
\big( \rho \partial_{x^j} F_\rho 
+\sigma \partial_{x^j} F_\sigma 
%+\sigma_\parallel \partial_{x^j}F_{\sigma_\parallel} 
%+ \sigma_\perp \partial_{x^j} F_{\sigma_\perp} 
+ M^i \partial_{x^j} F_{M^i} 
\\&\qquad
+ B^i \partial_{x^j}F_{B^i} 
-\partial_{x^i} (F_{B^j} B^i) \big) \epsilon^j 
+ F_{M^i} \rho \frac{d}{dt}\epsilon^i 
\Big)\, d^3x 
\end{aligned}
\label{eq:E22b}
\end{equation}
where we have substituted (\ref{eq:E15}), (\ref{eq:E18}), (\ref{eq:E19}), (\ref{eq:E21}),
integrated by parts, 
and then used the cancellation
\begin{equation}
F_\rho \partial_{x^j}\rho +F_\sigma \partial_{x^j}\sigma 
%+F_{\sigma_\parallel} \partial_{x^j}\sigma_\parallel 
%+ F_{\sigma_\perp} \partial_{x^j}\sigma_\perp 
+ F_{B^i} \partial_{x^j}B^i + F_{M^i} \partial_{x^j} M^i -\partial_{x^j} F  =0
\label{eq:E22c}
\end{equation}
which holds by chain rule. 
Finally, from the two expressions (\ref{eq:E22a}) and (\ref{eq:E22b}), 
we equate the coefficients of $\epsilon^i$, and likewise the coefficients of $d\epsilon^i/dt$, 
since $\epsilon^i$ and $d\epsilon^i/dt$ are arbitrary functions of $x$. 
This yields the key result: 
\begin{equation}
\begin{aligned}
F_{0\,q^i} & = 
\big( \rho \partial_{x^i} F_\rho 
+\sigma \partial_{x^i} F_\sigma 
%+\sigma_\parallel \partial_{x^i}F_{\sigma_\parallel} 
%+ \sigma_\perp \partial_{x^i} F_{\sigma_\perp} 
+ M^j \partial_{x^i} F_{M^j} 
\\&\qquad
+ B^j \partial_{x^i}F_{B^ji} 
-\partial_{x^j} (F_{B^i} B^j) \big)J ,
\\
F_{0\,p^i} & = F_{M^i} ,
\end{aligned}
\label{eq:E23}
\end{equation}
which are the transformation formulae for the variational derivatives. 

Substitution of these formulae (\ref{eq:E23}) into the canonical bracket (\ref{eq:E1}) 
gives
\begin{equation}
\begin{aligned}
\left\{\F,\G\right\}
=\int \Big( &
G_{M^j} \big( \rho \partial_{x^j} F_\rho 
+\sigma \partial_{x^j} F_\sigma 
%+\sigma_\parallel \partial_{x^j}F_{\sigma_\parallel} 
%+ \sigma_\perp \partial_{x^j} F_{\sigma_\perp} 
+ M^i \partial_{x^j} F_{M^i} 
\\&
+ B^i \partial_{x^j}F_{B^i} 
-\partial_{x^i} (F_{B^j} B^i) \big)
- F_{M^j} \big( \rho \partial_{x^j} G_\rho 
+\sigma \partial_{x^j} G_\sigma 
%+\sigma_\parallel \partial_{x^j}G_{\sigma_\parallel} 
\\&
%+ \sigma_\perp \partial_{x^j} G_{\sigma_\perp} 
+ M^i \partial_{x^j} G_{M^i} 
+ B^i \partial_{x^j}G_{B^i} 
-\partial_{x^i} (G_{B^j} B^i) \big)
\Big)\, d^3x
\\
=\int \Big( &
\rho (G_{M^j}\partial_{x^j} F_\rho -F_{M^j}\partial_{x^j} G_\rho)
+\sigma (G_{M^j}\partial_{x^j} F_\sigma -F_{M^j}\partial_{x^j} G_\sigma)
\\&
%+\sigma_\parallel (G_{M^j}\partial_{x^j} F_{\sigma_\parallel} -F_{M^j}\partial_{x^j} G_{\sigma_\parallel})
%+\sigma_\perp (G_{M^j}\partial_{x^j} F_{\sigma_\perp} -F_{M^j}\partial_{x^j} G_{\sigma_\perp})
%\\&
+ M^i(G_{M_j} \partial_{x^j} F_{M^i} - F_{M_j} \partial_{x^j} G_{M^i})
+ B^i(G_{B_j} \partial_{x^j} F_{B^i} - F_{B_j} \partial_{x^j} G_{B^i})
\\&
+ B^i (F_{B^j} \partial_{x^i} G_{M^j} - G_{B^j} \partial_{x^i} F_{M^j})
\Big)\, d^3x
\end{aligned}
\label{eq:E24}
\end{equation}
after integration by parts to remove derivatives off of $B^i$. 
This completes the derivation of the CGL Poisson bracket, 
which is the counterpart of the MHD bracket \cite{MorrisonGreene82}. 
Note that, for using the bracket (\ref{eq:E24}), 
we can convert all terms into vector notation as shown in (\ref{eq:5.3a}).  

Equations (\ref{eq:E2}) for $\dot{x}^i(x_0,t)$ 
and (\ref{eq:E7a}) for the variation $\delta x^i(x_0,t)$
can be set up in an alternative way by means of 
the Lagrangian map (\ref{eq:4.0}), 
using diffeomorphisms $g(t)$ on Euclidean space
similarly to the set up for the Euler-Poincar\'e action principle in section~\ref{sec:EulerPoincare}. 
However, here the variations are not the same as the ones used for the Euler-Poincar\'e action principle, 
since those variations took place in the Eulerian frame and involved fixing $x^i$, 
with $x_0^i$ being a function of $x$ and $t$. 
To proceed,
we view $g(t)$, at any fixed time $t$, as an element in the group of diffeomorphisms $G\equiv {\rm Diff}({\mathbb R}^3)$ 
acting on Euclidean space in terms of the Cartesian coordinates $x_0^i$. 
The Lagrangian map (\ref{eq:4.0}) then can be expressed in component form:
\begin{equation}
x^i(x_0,t) = g(t) x_0^i . 
\label{eq:E30}
\end{equation}
(Concretely, $g$ can be thought of as a matrix in the fundamental representation of the group $G$.)
Thus, we can write 
\begin{align}
& \epsilon^i(x,t) = \delta (g(t)x_0^i) = (\delta g(t) g^{-1}(t))x^i = (\delta g(t) g^{-1}(t))^i,
\label{eq:E31a}
\\
& u^i(x,t) = \partial_t (g(t)x_0^i) = (g_t(t)g^{-1}(t))x^i = (g_t(t)g^{-1}(t))^i, 
\label{eq:E31b}
\end{align}
where 
\begin{equation}
\delta g(t) g^{-1}(t) \equiv (\delta g(t) g^{-1}(t))^i \partial_{x^i}, 
\quad
g_t(t) g^{-1}(t) \equiv (g_t(t) g^{-1}(t))^i \partial_{x^i} 
\label{eq:E32}
\end{equation}
represent right invariant vector fields on the group $G$,
which are identified with Eulerian vector fields (directional derivatives) 
at the point $x^i$ in Euclidean space. 
Note the property of right-invariance means that, for any fixed element $h$ in $G$, 
$g\to gh$ implies 
$\delta g g^{-1}\to \delta (gh) (gh)^{-1} = \delta g h h^{-1}g^{-1} = \delta g g^{-1}$
and 
$g_t g^{-1}\to (gh)_t (gh)^{-1} = g_t h h^{-1}g^{-1} = g_t g^{-1}$, 
due to $\partial_t h= \delta h\equiv 0$. 
The variation of the Eulerian variables ${\sf Z}$ can be shown to arise
from using the push-forward action of $g(t)$, and the pull-back action of $g(t)^{-1}$, on scalar fields, vector fields, differential forms, and tensor fields in the Lagrangian frame. 

A closer connection to the variations used for the Euler-Poincar\'e action principle
arises if the variations derived in 
(\ref{eq:E12a}), (\ref{eq:E15}), (\ref{eq:E18}), (\ref{eq:E19}),
and  (\ref{eq:E21}),
are reformulated in the following way. 
A variation of any one of these quantities, $\delta {\sf Z}(x)$, 
can be expressed as the sum of 
a contribution from varying only its dependence on $x^i$, 
in addition to a contribution from varying ${\sf Z}$ with $x^i$ being unchanged. 
For example, consider 
$\delta S(x)= (\delta S)(x) + (\delta x^i)\partial_{x^i} S(x)$. 
The pointwise contribution is given by 
\begin{equation}
(\delta S)(x) = \delta S(x) - \epsilon^i(x)\partial_{x^i} S(x)
= -\epsilon^i(x)\partial_{x^i} S(x)
\equiv \Delta S(x) 
\label{eq:E40}
\end{equation} 
from the total variation (\ref{eq:E12a}). 
Likewise, 
$\delta \rho(x) = (\delta\rho)(x) + (\delta x^i)\partial_{x^i}\rho(x)$ leads to 
\begin{equation} 
(\delta\rho)(x) = - \partial_{x^i}(\rho(x)\epsilon^i(x)) 
\equiv \Delta \rho(x) 
\label{eq:E41}
\end{equation}
from the total variation (\ref{eq:E15}). 
As another example, 
$\delta B^i(x)= \delta(B^i)(x) + (\delta x^j)\partial_{x^j} B^i(x)$ gives
\begin{equation}
(\delta B^i)(x) = \delta(B^i(x)) - \epsilon^j(x)\partial_{x^j} B^i(x) 
= - \partial_{x^j}(\epsilon^j(x) B^i(x)) + B^j(x)\partial_{x^j} \epsilon^i(x) 
\equiv \Delta B^i(x) 
\label{eq:E42}
\end{equation}
using the total variation (\ref{eq:E19}). 
When the analogous pointwise variations are considered for the components of the advected quantities (\ref{eq:E3}), 
we see that they take the form of a Lie derivative:
\begin{equation}
(\delta S)(x) = -\lieder{\bm{\epsilon}} S(x),
\quad
(\delta b^i)(x) = -\lieder{\bm{\epsilon}} b^i(x),
\quad
(\delta \rho e_{ijk})(x) = -\lieder{\bm{\epsilon}} (\rho(x) e_{ijk}) . 
\label{eq:E43}
\end{equation}
Note that here we have expressed the volume element as a 3-form: 
$d^3x = e_{ijk} dx^i\wedge dx^j\wedge dx^k$, 
where $e_{ijk}$ is the Levi-Civita symbol.
The Lie derivative in the variations (\ref{eq:E43}) denotes the standard Lie derivative formula in component form. 
These formulae can be written more properly by including the basis vectors: e.g. 
$(\delta b^i(x))\partial_{x^i} = -\lieder{\bm{\epsilon}} (b^i(x)\partial_{x^i})$,
where the Lie derivative then acts in the standard geometrical way.
This result coincides with the general formulation in \cite{Holm83c}. 
In addition, it shows that the variations of the non-advected Eulerian variables 
given by (\ref{eq:E41}) and (\ref{eq:E42}) 
can be expressed in terms of Lie derivatives:
\begin{equation}
\Delta \rho(x) = -(\partial_i\epsilon^i)\rho(x) -\lieder{\bm{\epsilon}} \rho(x) , 
\quad
\Delta B^i(x) = -(\partial_{x^j}\epsilon^j)B^i(x) -\lieder{\bm{\epsilon}} B^i(x). 
\label{eq:E44}
\end{equation}
These formulae are used in the classical work of \cite{Newcomb62}. 
They can be used to derive the transformation formula (\ref{eq:E23}) 
for the variational derivatives by taking a pointwise variation of a functional:
\begin{equation}
\Delta \F 
=\int \Big(
F_\rho\Delta\rho +F_\sigma \Delta\sigma 
%+F_{\sigma_\parallel} \Delta\sigma_\parallel  +F_{\sigma_\perp} \Delta\sigma_\perp
+F_{B^i}\Delta B^i +F_{M^i}\Delta M^i 
\Big)\,d^3x , 
\label{eq:E45}
\end{equation}
which differs from (\ref{eq:E23}) by lacking a contribution from varying the volume element. 
Nevertheless, this leads to the same expression (\ref{eq:E24}) for the bracket
(cf the cancellation of terms (\ref{eq:E22c})).

\appendix
\section*{Appendix E}
\setcounter{section}{5}
\setcounter{equation}{0}

In this appendix, 
we outline the general approach of \cite{Holm98} 
on the Euler Poincar\'e variational principles, 
with application to CGL plasmas.
The action is of the form:
\begin{equation}
{\mathcal J}=\int L [u,a]\,d^3x\, dt. 
\label{eq:G1}
\end{equation}
The stationary points of ${\mathcal A}$ are given by $\delta{\mathcal J}=0$,
where the variables $a$ are advected quantities (\ref{eq:advect})
subject to the variations:
\begin{equation}
\delta a=-{\cal L}_{\bm\eta}(a). 
\label{eq:G2}
\end{equation}
Here ${\bm\eta}$ is an arbitrary, sufficiently differentiable vector field.
Alternatively, ${\bm\eta}$ arises from a Lagrangian map as discussed 
in section~\ref{sec:EulerPoincare}. 

For the CGL plasma case, 
the Lagrangian density $L[u,a]$ is given by (\ref{eq:3.16}),
where the $a$ are the advected quantities $S$, $\rho\,d^3x$, 
${\bf B}/\rho\equiv{\bf b}$ 
(cf (\ref{eq:4.8})). The magnetic flux 2-form 
$B_x dy\wedge dz+B_y dz\wedge dx+B_z dx\wedge dy$ is a Lie dragged
invariant of the flow (\cite{Webb14a}).  
(Note $p_\parallel$, $p_\perp$ and $e$ are defined in terms of 
$\rho$, $S$ and $B$ via the equation of state for $e(\rho,S, B)$). 
The expression for the action $\mathcal J$ is the same as in (\ref{eq:EPaction}). 
From (\ref{eq:G2}) and (\ref{eq:4.8}),  
the variation $\delta a$ is obtained:
\begin{equation}
\begin{aligned}
\delta S & = -{\cal L}_{\bm\eta} S 
= -{\bm\eta}{\bf\cdot}\nabla S, 
\\
\delta \rho\, d^3x & = -{\cal L}_{\bm\eta} (\rho\, d^3x) 
= -\nabla{\bf\cdot}(\rho {\bm\eta})\,d^3x,
\\
\delta{\bf B}&=\nabla\times({\bm\eta}\times{\bf B})
-{\bm\eta}(\nabla{\bf\cdot}{\bf B}). 
%\delta {\bf b} & = -{\cal L}_{\bm\eta} {\bf b}
%= {\bf b}{\bf\cdot}\nabla {\bm \eta} - {\bm \eta} {\bf\cdot}\nabla {\bf b}
%\\&\qquad\qquad
%= \nabla\times\left( {\bm \eta} \times {\bf b} \right)
%+ \left(\nabla{\bf\cdot}{\bf b}\right) {\bm \eta} 
%- \left(\nabla{\bf\cdot}{\bm \eta} \right) {\bf b} , 
\end{aligned}
\label{eq:G4}
\end{equation}
%which uses $\delta x^i=0$ 
%(cf (\ref{eq:4.9a})--(\ref{eq:4.9c})). 

The variational equation $\delta{\mathcal J}=0$ is expressed as:
\begin{equation}
\delta {\mathcal J}=\int \left\langle {\bm\eta},{\bf F}\right\rangle \, d^3x\,dt =0, 
\label{eq:G5}
\end{equation}
which can be shown to imply for arbitrary ${\bm\eta}$ that
\begin{equation}
{\bf F}=
\derv{t}\left(\frac{\delta L}{\delta {\bf u}}\right)
+ad^*_{\bf u}\left(\frac{\delta L}{\delta{\bf u}}\right)_R
- \frac{\delta L}{\delta a}\diamond a=0 
\label{eq:G6}
\end{equation}
(see \cite{Holm98} for details). 
This is the general form of the Euler Poincar\'e (EP) equation. 
The diamond operator $\diamond$ is defined by the property (\ref{eq:4.13}). 

For the CGL plasma action principle, 
the EP equation (\ref{eq:G6}) has the form:
\begin{equation}
\derv{t}(\rho {\bf u}) +\nabla{\bf\cdot}(\rho {\bf u}\otimes {\bf u})
+\rho\nabla\left(\frac{1}{2} u^2\right)=
\frac{\delta L}{\delta a}\diamond a. 
\label{eq:G7}
\end{equation}
The term in (\ref{eq:G7}) involving the diamond operator 
can be determined from the variation:
\begin{equation}
\begin{aligned}
\int \frac{\delta L}{\delta a} {\delta a} \,d^3x
=& \int \left( \frac{\delta L}{\delta\rho} \delta\rho
+ \frac{\delta L}{\delta S} \delta S
+\frac{\delta L}{\delta{\bf B}}{\bf\cdot} \delta {\bf B} \right)\,d^3x
\\
=& \int \bigg(\frac{\delta L} {\delta \rho} 
\left[-\nabla{\bf\cdot}(\rho{\bm\eta}) \right]
+\frac{\delta L}{\delta S}\left[-{\bm\eta}{\bf\cdot}\nabla S\right]
\\&\qquad
+\frac{\delta L}{\delta {\bf B}}{\bf\cdot}
\left[\nabla\times\left({\bm\eta}\times {\bf B}\right) 
-{\bm\eta}(\nabla{\bf\cdot}{\bf B})\right]
%+{\bf b}(\nabla{\bf\cdot}{\bm\eta})\right] 
\bigg)\,d^3x
\\
= & \int {\bm\eta}{\bf\cdot}\left[
\rho \nabla\frac{\delta L}{\delta \rho}
-\frac{\delta L}{\delta S}\nabla S
+{\bf B}\times\left[\left(\nabla\times 
\frac{\delta L}{\delta {\bf B}}\right)
-\frac{\delta L}{\delta {\bf B}}
(\nabla{\bf\cdot}{\bf B})\right]
%-\nabla\left(\frac{\delta L}{\delta {\bf b}}{\bf\cdot}{\bf b}\right)
\right]
 \,d^3x , 
\end{aligned}
\label{eq:G8}
\end{equation}
where boundary integral terms have been discarded. 
From (\ref{eq:G8}) and (\ref{eq:G5}), 
we identify the diamond operator term 
$(\delta L/\delta a)\diamond a$ as being the term in the square brackets.
Substituting the variational derivatives of $L$, which are given by (\ref{eq:4.15}), 
we obtain
\begin{equation}
\begin{aligned}
\frac{\delta L}{\delta a}\diamond a =& 
\rho\left( T\nabla S-\nabla h +\nabla\left(\frac{1}{2}u^2-\Phi\right)\right) 
+{\bf J}\times {\bf B}
\\&
+\frac{\bf B}{\mu_0} (\nabla{\bf\cdot}{\bf B})
+{\bf B}\times \left(\nabla\times {\bm\Omega}\right)
-{\bm\Omega}(\nabla{\bf\cdot}{\bf B}),  
\end{aligned}
\label{eq:G9}
\end{equation}
where
\begin{equation}
{\bf J}=\frac{\nabla\times{\bf B}}{\mu_0}, \quad {\bm\Omega}
=\frac{p_\Delta}{B} {\bm\tau}, \quad {\bm\tau}=\frac{\bf B}{B}. 
\label{eq:G10}
\end{equation}

The auxiliary calculations
\begin{equation}
\begin{aligned}
{\bf B}\times(\nabla\times{\bm\Omega})=&
-p_\Delta{\bm\tau}{\bf\cdot}\nabla {\bm\tau}
-p_\Delta \left({\sf I}- {\bm\tau}{\bm\tau}\right){\bf\cdot}\nabla(\ln B) 
+\left({\sf I}- {\bm\tau}{\bm\tau}\right){\bf\cdot}\nabla p_\Delta, 
\\
\nabla{\bf\cdot}{\sf p}=&
\nabla p_\perp 
+p_\Delta\left(( \nabla{\bf\cdot}{\bm\tau}){\bm\tau} +{\bm\tau}{\bf\cdot}\nabla {\bm\tau}\right)
+{\bm\tau}{\bm\tau}{\bf\cdot}\nabla p_\Delta, 
\\
\nabla{\bf\cdot}{\bf B}=&
B\left(\nabla{\bf\cdot}{\bm\tau}+{\bm\tau}{\bf\cdot}\nabla\ln B\right), 
\end{aligned}
\label{eq:G11}
\end{equation}
and the thermodynamic relation
\begin{equation}
\rho\left(T\nabla S-\nabla h\right)=p_\Delta\nabla (\ln B)-\nabla p_\parallel, 
\label{eq:G12}
\end{equation}
can be combined to give the identity:
\begin{equation}
{\bf B}\times(\nabla\times{\bm\Omega})-{\bm\Omega}(\nabla{\bf\cdot}{\bf B})
+\nabla{\bf\cdot}{\sf p}+\rho\left(T\nabla S-\nabla h\right)=0. \label{eq:G13}
\end{equation}
This equation was derived in (\ref{eq:C10}). 
%It arises here in a fairly straightforward way, 
%by comparing the terms in $\nabla{\bf\cdot}{\sf p}$ 
%in the momentum equation with the other terms in 
%$\delta L/\delta a\diamond a$ in (\ref{eq:G9}) that arise in the EP analysis.
%It reduces to the thermodynamic form of the pressure gradient in MHD and in gas
%dynamics in terms of the entropy gradient $\nabla S$ and enthalpy gradient 
%$\nabla h$ in the limit as ${\bm\Omega}\to 0$ or as $p_\Delta\to 0$. 

Substituting the identity (\ref{eq:G13}) into (\ref{eq:G9}) gives:
\begin{equation}
\frac{\delta L}{\delta a}\diamond a=\rho\nabla\left(\frac{1}{2} u^2\right)
-\rho\nabla\Phi +{\bf J}\times{\bf B}
+\frac{\bf B}{\mu_0} \nabla{\bf\cdot}{\bf B} -\nabla{\bf\cdot}{\sf p}, 
\label{eq:G14}
\end{equation}
which, when used in the EP equation (\ref{eq:G7}) yields the Eulerian momentum 
equation (\ref{eq:2.2}).

\appendix
\section*{Appendix F}
\setcounter{section}{6}
\setcounter{equation}{0}

In this appendix, 
we summarize the complete Lie point symmetry group 
for the CGL system (\ref{eq:2.1})--(\ref{eq:2.8}). 
We will restrict attention to the case where there is no gravity 
(i.e.\ $\Phi=0$ in (\ref{eq:2.2})). 
(For the general theory of Lie point symmetries, 
see \cite{Ovsjannikov78, Ovsjannikov94}, \cite{Golovin09}, \cite{Bluman02, Bluman10}, \cite{Ibragimov85}, \cite{Olver93}).

The independent and dependent variables in equations (\ref{eq:2.1})--(\ref{eq:2.8})
consist of $t$, $x^i$, $\rho$, $u^i$, $B^i$, $S$, $p_\parallel$, $p_\perp$. 
Thus, every Lie point symmetry arises from a generator of the form 
\beqn
\X= 
\xi^t\partial_t +\xi^i\partial_{x^i}
+\xi^\rho\partial_\rho +\xi^{u^i}\partial_{u^i} +\xi^{B^i}\partial_{B^i}+\xi^{S}\partial_{S}
+\xi^{p_\parallel}\partial_{p_\parallel} +\xi^{p_\perp}\partial_{p_\perp}
\label{eq:F.1}
\eeqn
under which the equations (\ref{eq:2.1})--(\ref{eq:2.8}) are invariant 
on the space of solutions. 
Here 
\beqn
\xi^t,\ 
\xi^i,\ 
\xi^\rho,\
\xi^{u^i},\
\xi^{B^i},\
\xi^{S},\
\xi^{p_\parallel},\
\xi^{p_\perp}
\label{eq:F.2}
\eeqn
are functions of the independent and dependent variables. 
For computational purposes, 
it is easiest to work with the characteristic form of the generator 
in which only the dependent variables undergo a transformation:
\beqn
\hat\X= 
\hat\xi^\rho\partial_\rho +\hat\xi^{u^i}\partial_{u^i} +\hat\xi^{B^i}\partial_{B^i}+\hat\xi^{S}\partial_{S}
+\hat\xi^{p_\parallel}\partial_{p_\parallel} +\hat\xi^{p_\perp}\partial_{p_\perp}
\label{eq:F.3a}
\eeqn
where
\beqn
\hat\xi^v = \xi^v - \xi^t v_t - \xi^i v_{x^i}
\label{eq:F.3b}
\eeqn
for each dependent variable $v=(\rho,u^i,B^i,S,p_\parallel,p_\perp)$. 
The generator $\hat\X$ has the convenient property that it commutes with total derivatives with respect to $t,x^i$,
and so the prolongation of $\hat\X$ acting on derivatives of $v$ is easy to compute. 

The determining equations for Lie point symmetries are given by 
applying $\pr\hat\X$ to each equation (\ref{eq:2.1})--(\ref{eq:2.8})
and evaluating the resulting system on the solution space of equations (\ref{eq:2.1})--(\ref{eq:2.8}) which is carried out by putting the equations (\ref{eq:2.1})--(\ref{eq:2.8})
into a solved form with respect to a set of leading derivatives. 
(See \cite{Olver93} for a general discussion.)
The resulting system of determining equations then splits with  
respect to all derivative variables that appear in the system. 
This yields an overdetermined linear system of equations 
for the functions (\ref{eq:F.2}). 
The system can be solved straightforwardly by use of computer algebra
(e.g. Maple or Mathematica). 

The Lie point symmetry generators are found to be given by 
the 10 Galilean transformation generators (\ref{eq:6.6a})
and the following 3 scaling generators
\begin{equation}
\begin{aligned}
S_1 & = t\partial_t+x^i\partial_{x^i} ,
\\
S_2 &  = t\partial_t +2 \rho\partial_\rho - u^i\partial_{u^i} ,
\\
S_3 & = 2\rho\partial_\rho+2p_\perp\partial_{p_\perp} +2 p_\parallel\partial_{p_\parallel} +B^i\partial_{B^i} . 
\end{aligned} 
\label{eq:F.4}
\end{equation}
In terms of the quantities 
$p=(p_\parallel+2 p_\perp)/3$ and $p_{\Delta}=p_\parallel-p_\perp$, 
the third scaling symmetry has the equivalent form:
\begin{equation}
S_3 =2\rho\partial_\rho+2p\partial_{p} +2 p_{\Delta}\partial_{p_{\Delta}} +B^i\partial_{B^i} . 
\end{equation}

For comparison, 
the Lie point symmetries for MHD (\cite{Fuchs91} and \cite{RogersAmes89})
are given by the 10 Galilean transformation generators (\ref{eq:6.6a}) 
and the 2 scaling generators $S_1$ and $S_2$ plus a scaling generator
$S_{3'} =2\rho\partial_\rho+2p\partial_{p} +B^i\partial_{B^i}$
which differs from $S_3$ by omitting the term involving $p_\Delta$. 

A full study of subalgebras of the Galilean Lie algebra was given in 
\cite{Ovsjannikov78,Ovsjannikov94, Ovsjannikov99}) and \cite{Grundland95}).

%\appendix
\section*{Appendix G}
\setcounter{section}{7}
\setcounter{equation}{0}
Here we derive the Lie invariance condition (\ref{eq:6.23a})
for the action, which follows from (\ref{eq:6.1b}). 
% ${\hat X}(L_0)$ in 
We start from the appropriate expansion for $\pr\hat{\X}L_0$: 
\beqn
\pr{\hat \X}(L_0)=\hat{\xi}^i \deriv{L_0}{x^i}
+D_t\left({\hat\xi}^i\right)\deriv{L_0}{\dot{x}^i}
+D_{x_0^j}\left({\hat\xi}^i\right)\deriv{L_0}{x_{ij}}. \label{eq:I1}
\eeqn
The derivatives of $L_0$ in (\ref{eq:I1}) are given in Appendix C, namely:
\begin{equation}
\deriv{L_0}{X^i}=-\rho_0\deriv{\Phi}{x^i},\quad 
\deriv{L_0}{\dot{X}^i}=\rho_0 u^i, \quad
\deriv{L_0}{X^{ij}}=\left({\sf p}+\sf{M}_B\right)^{ik}A_{kj}. \label{eq:I2}
\end{equation}
Using (\ref{eq:I2}) in (\ref{eq:I1}) gives 
\begin{align}
\pr{\hat \X}(L_0)=&-\rho_0{\bm\xi}{\bf\cdot} \nabla\Phi 
+D_t\left(\hat{\bm\xi}\right){\bf\cdot}\rho_0{\bf u}
+D_{x_0^j}\left(\hat{\xi}^i\right)\left({\sf p}+{\sf M}_B\right)^{ik} 
A_{kj}\nonumber\\
\equiv& \rho J u^i \left[\frac{d\hat{\xi}^i}{dt}
- \hat{\bm\xi}{\bf\cdot} \nabla u^i\right]+R_1, \label{eq:I3}
\end{align}
where
\begin{align}
R_1=&-\rho_0\hat{\bm\xi}{\bf\cdot}\nabla\Phi
+\rho_0\hat{\bm\xi}{\bf\cdot}\nabla\left(\frac{1}{2} u^2\right) 
+D_{x_0^j}\left(\hat{\xi}^i\right)
\left({\sf p}+{\sf M}_B\right)^{ik} A_{kj}\nonumber\\
&\equiv J\left\{\nabla{\bf\cdot}\left(\rho{\hat{\bm\xi}}\right)
\left(\Phi-\frac{1}{2} u^2\right)
-\nabla{\bf\cdot}\left[\rho\hat{\bm\xi}
\left(\Phi-\frac{1}{2}u^2\right)
\right]\right\}
+D_{x_0^j}\left(\hat{\xi}^i\right)
\left({\sf p}+{\sf M}_B\right)^{ik} A_{kj}. 
\label{eq:I4}
\end{align}

%which can be reduced to the form:
%\beqn
%{\hat X}(L_0)=\rho J{\bf u}{\bf\cdot}
%\left(\frac{d{\hat{\bm\xi}}}{dt}
%-{\hat{\bm\xi}}{\bf\cdot}\nabla{\bf u}\right) +R_1, \label{eq:I2}
%\eeqn
%where
%\beqn
%R_1=J\left\{ \nabla{\bf\cdot}\left(\rho {\hat{\bm\xi}}\right)\left[\Phi
%-\frac{1}{2}u^2\right]
% -\nabla{\bf\cdot}\left[\rho {\hat{\bm\xi}}
%\left(\Phi-\frac{1}{2}u^2\right)\right]\right\}
%+D_{x_0^j}\left({\hat\xi}^i\right)\left(p^{ik}+M_B^{ik}\right)A_{kj}.
%\label{eq:I3}
%\eeqn

By noting that:
\beqn
\nabla_k\left(p^{ik}\right)
=-\left\{{\bf B}\times\left(\nabla\times{\bm\Omega}\right)
-{\bm\Omega}\nabla{\bf\cdot}{\bf B}
+\rho\left(T\nabla S-\nabla h\right)\right\}^i, \label{eq:I5}
\eeqn
 and
\beqn
\nabla_k\left(M^{ik}_B\right)
=-\left\{{\bf J}\times{\bf B}+{\bf B} \frac{\nabla{\bf\cdot}{\bf B}}{\mu_0}
\right\}^i, \label{eq:I6}
\eeqn
We obtain:
\begin{align}
A_{kj} D_{x_0^j}\left(\hat{\xi}^i\right) \left({\sf p}+{\sf M}_B\right)^{ik}
=&J\biggl\{D_{x^k}\left[\hat{\xi}^i\left({\sf p}+{\sf M}_B\right)^{ik}\right]
+\rho \hat{\bm\xi}{\bf\cdot}\left(T\nabla S-\nabla h\right)\nonumber\\
&+\hat{\bm\xi}{\bf\cdot}\left[{\bf B}\times (\nabla\times{\bm\Omega})
-{\bm\Omega} \nabla{\bf\cdot}{\bf B}\right]
+\hat{\bm\xi}{\bf\cdot}\left[{\bf J}\times{\bf B}
+{\bf B}\frac{\nabla{\bf\cdot}{\bf B}}{\mu_0}\right]\biggr\}. 
\label{eq:I7}
\end{align}
Noting
\begin{equation}
\tilde{\bf J}={\bf J}-\nabla\times{\bm\Omega} 
=\frac{\nabla\times\tilde{\bf B}}{\mu_0},
 \quad \tilde{\bf B}={\bf B}\left(1-\frac{\mu_0 p_\Delta}{B^2}\right), 
\label{eq:I8}
\end{equation}
we obtain the equations:
\begin{align}
&{\hat{\bm\xi}}{\bf\cdot}
\left[{\bf B}\times (\nabla\times {\bm\Omega})
-{\bm\Omega} \nabla {\bf\cdot B}\right]
+{\hat{\bm\xi}}{\bf\cdot}
\left[{\bf J}\times{\bf B}
+{\bf B}(\nabla{\bf\cdot}{\bf B})/\mu_0\right]\nonumber\\
=&{\tilde{\bf J}}{\bf\cdot}\left({\bf B}\times {\hat{\bm\xi}}\right) 
+({\hat{\bm\xi}}{\bf\cdot}{\tilde{\bf B}}) \nabla{\bf\cdot B}/\mu_0
\nonumber\\
=&\frac{1}{\mu_0}
\biggl\{\nabla\times \left[\left(\hat{\bm\xi}\times {\bf B}\right)
\times\tilde{\bf B}\right]-{\tilde{\bf B}}
{\bf\cdot}\left[\nabla\times(\hat{\bm\xi}\times{\bf B})
-\hat{\bm\xi} \nabla{\bf\cdot B}\right]\biggr\}. \label{eq:I9}
\end{align}
In the derivation of (\ref{eq:I9}) we used the identity:
\begin{equation}
\nabla{\bf\cdot}\left({\bf A}\times{\bf C}\right)=
(\nabla\times{\bf A}){\bf\cdot C}
-(\nabla\times{\bf C}){\bf\cdot A}, \label{eq:I9a}
\end{equation}
with ${\bf A}=\hat{\bm\xi}\times{\bf B}$ and ${\bf C}=\tilde{\bf B}$.
%and using the vector identity:
%\beqn
%\nabla{\bf\cdot}\left({\bf A}\times{\bf C}\right)
%=(\nabla\times{\bf A}){\bf\cdot}{\bf C}
%-(\nabla\times{\bf C}){\bf\cdot}{\bf A}, \label{eq:I6}
%\eeqn

Use (\ref{eq:I9}) in (\ref{eq:I7}) gives:
\begin{align}
A_{kj}D_{x_0^j}\left({\hat\xi}^i\right)
\left(p^{ik}+M_B^{ik}\right)
=&J\biggl\{D_{x^k}\left[{\hat\xi}^i
\left(p^{ik}+M_B^{ik}\right)\right]
+\rho{\bm{\hat\xi}} {\bf\cdot} \left[T\nabla S
-\nabla h\right]\nonumber\\
&+\nabla{\bf\cdot}
\left[\frac{({\hat{\bm\xi}}\times{\bf B})\times{\tilde{\bf B}}}{\mu_0}
\right]
-\frac{\tilde{\bf B}}{\mu_0} {\bf\cdot}
\left[\nabla\times({\hat{\bm\xi}}\times{\bf B})-{\hat{\bm\xi}} 
\nabla{\bf\cdot}{\bf B}\right]\biggr\}. \label{eq:I10}
\end{align}
Using (\ref{eq:I10}) in (\ref{eq:I4}) gives a simplification of $R_1$, 
which can in turn be used to obtain $\pr\hat{\X}(L_0)$ in (\ref{eq:I1}).
Substitution of the resultant $\pr\hat{\X}(L_0)$ in (\ref{eq:6.1b})
gives (\ref{eq:6.23a}) as the condition for Lie invariance 
of the action.

\appendix
{\bf{\section*{Appendix H}}}\label{sec:noether}
\setcounter{section}{8}
\setcounter{equation}{0}

In this appendix, we provide a derivation of Noether's first theorem, 
using the approach of \cite{Bluman89}. This 
analysis should be useful for readers not acquainted with 
the classical approach to Noether's first theorem.  
These ideas are used in Section 6.2 to describe Noether's theorem.

Consider a system of differential equations in the dependent 
variables $u^\alpha$ ($1\leq\alpha\leq m$) and independent variables 
$x^i$ ($1\leq i\leq n$) of the form:
\begin{equation}
R^s(x_i,u^\alpha,u^\alpha_i, u^\alpha_{ij},\ldots )=0,\quad 1\leq s\leq m,
\label{eq:N0}
\end{equation}
(the subscripts in $u^\alpha_i$, $u^\alpha_{ij},\ldots$, denote partial 
derivatives with the respect to the independent variables $x^i$ 
$(1\leq i\leq n)$, 
 which  arises from 
extremal variations of the action 
\begin{equation}
J[u]=\int_\Omega L\left({\bf x},u^\alpha_i, u^\alpha_{ij}, \ldots\right)
\ d{\bf x} ,
\label{eq:N1}
\end{equation}
which remain invariant under infinitesimal Lie transformations of the form:
\begin{equation}
x'{}^i=x^i+\epsilon \xi^i, \quad u'{}^\alpha= u^\alpha+\epsilon \eta^\alpha, 
\quad L'=L+\nabla_i\Lambda^i. \label{eq:N2}
\end{equation}
The variation of $J[{\bf u}]$ is defined as:
\begin{equation}
\delta J=\int_{\Omega'} L'\left({\bf x}', {\bf u}', u'{}^\alpha_i, 
u'{}^\alpha_{ij}, \ldots\right)\ d{\bf x}'
-\int_{\Omega} L\left({\bf x}, {\bf u}, u{}^\alpha_i,  
u{}^\alpha_{ij}, \ldots\right)\ d{\bf x}, \label{eq:N3}
\end{equation}
where $\Omega$ is the region of integration. The variation $\delta J$ to 
$O(\epsilon)$ in (\ref{eq:N3})
reduces to:
\begin{equation}
\delta J=\epsilon\int_{\Omega} \left(\pr\X L+LD_i\xi^i
+D_i\Lambda^i\right)\ d{\bf x} +O(\epsilon^2). \label{eq:N4}
\end{equation}
%The term $\pr\X L$ in (\ref{eq:N4}) represents changes in 
%$J[{\bf u}]$ due to the transformations of ${\bf x}'$ 
%and ${\bf u}'$, in which 
%the functional form of $L$ remains unchanged.  
The term $LD_i\xi^i$ in (\ref{eq:N4}) 
represents  changes in the volume element:
\begin{equation}
d{\bf x}'=\left[1+\epsilon D_i\xi^i+O(\epsilon^2)\right]\ d{\bf x}. 
\label{eq:N5} 
\end{equation}
The Lie derivative term:
\begin{equation}
\pr\X L=\left(\xi^i\derv{x^i}+\eta^\alpha\derv{u^\alpha}+\eta^\alpha_i 
\derv{u^\alpha_i} 
+\eta^\alpha_{ij} \derv{u^\alpha_{ij}}+\ldots \right) L, 
\label{eq:N6}
\end{equation}
describes the changes in $L({\bf x},u^\alpha,u^\alpha_i,u^\alpha_{ij},\cdots)$
due to the Lie transformations of $x^i$ and $u^\alpha$ in (\ref{eq:N2}), 
in which the form of $L$ does not change. The term $D_i\Lambda^i$ 
describes changes of $J$ due to changes in the form of $L$ due to 
a divergence transformation (under such a transformation the action
remains invariant).
 From (\ref{eq:N4}) the action $J[{\bf u}]$ remains invariant under 
the Lie transformations (\ref{eq:N2}) to $O(\epsilon)$ if:
\begin{equation}
\pr\X L+L D_i\xi^i+D_i\Lambda^i=0. \label{eq:N7}
\end{equation}

It turns out that there is an equivalent extended Lie symmetry operator, 
$\pr (\hat{X})$ of the form:
\begin{equation}
\pr (\hat{X})=\hat{\eta}^\alpha\derv{u^\alpha}
+D_i\left(\hat{\eta}^\alpha\right) \derv{u^\alpha_i}+
D_iD_j\left(\hat{\eta}^\alpha\right)\derv{u^\alpha_{ij}}+\ldots, 
\label{eq:N8}
\end{equation}
called the evolutionary operator (e.g. \cite{Olver93}) 
which describes  Lie transformations:
\begin{equation}
x'{}^i=x^i,\quad u'{}^\alpha=u^\alpha+\epsilon 
\hat{\eta}^\alpha, \label{eq:N9}
\end{equation}
where
\begin{equation}
\hat{\eta}^\alpha=\eta^\alpha-\xi^j D_j u^\alpha. \label{eq:N10}
\end{equation}
The operator $\pr(\hat{\X})$  
 is related to $\pr(\X)$ by the 
formula:
\begin{equation}
\pr (\X)=\pr (\hat{\X})+\xi^j D_j, \label{eq:N11}
\end{equation}
where $D_j=d/dx^j$ is the total, partial derivative with respect 
to $x^j$. From (\ref{eq:N8}),  
we obtain the formulae:
\begin{equation}
u'{}^\alpha_i=u{}^\alpha_i+\epsilon D_i\left(\hat{\eta}^\alpha\right), 
\quad u'{}^\alpha_{ij}=u{}^\alpha_{ij}
+\epsilon D_iD_j\left(\hat{\eta}^\alpha\right), \label{eq:N12}
\end{equation}
for the transformation of partial derivatives under  
 ${\rm pr}(\hat{X})$. These transformations are 
different from the transformations of derivatives formulae 
under the canonical prolonged symmetry operator $\pr\X$, namely
\begin{equation}
\eta^\alpha_i\equiv \pr (\X) u^\alpha_i=\pr \hat{\X} u^\alpha_i
+\xi^j D_ju^\alpha_i
=D_i\left(\hat{\eta}^\alpha\right)+\xi^ju^\alpha_{ji}, \label{eq:N13}
\end{equation}
and similarly for transformations of the higher order 
derivatives of $u^\alpha$. 

Evaluation of $\delta J[{\bf u}]$ using $\pr(\hat{\X})$ gives 
the variational equation:
\begin{equation}
\delta J=\epsilon\int\left(\pr\hat{\X}L\right)\ d{\bf x}
=\epsilon\int\left[D_i W^i[{\bf u},\hat{\bm\eta}]+\hat{\eta}^\gamma 
E_\gamma (L)\right] 
\ d{\bf x}, \label{eq:N14}
\end{equation}
from which it follows that:
\begin{equation}
\pr(\hat{\X})L=D_iW^i[{\bf u},\hat{\bm\eta}]+\hat{\eta}^\gamma E_\gamma(L), 
\label{eq:N15}
\end{equation}
where the $W^i[{\bf u},\hat{\bm\eta}]$ are surface terms given by:
\begin{equation}
W^i[{\bf u},\hat{\bm\eta}]=\hat{\eta}^\gamma \frac{\delta L}{\delta u^\gamma}
+\hat{\eta}^\gamma_j \frac{\delta L}{\delta u^\gamma_{ji}}
+\hat{\eta}^\gamma_{jk} \frac{\delta L}{u^\gamma_{jki}}+\ldots, \label{eq:N16}
\end{equation}
and $\delta L/\delta\psi$ is given by:
\begin{equation}
\frac{\delta L}{\delta\psi}\equiv E_\psi (L)=\deriv{L}{\psi}
-D_i\left(\deriv{L}{\psi_i}\right)+D_iD_j
\left(\deriv{L}{\psi_{ij}}\right)-\ldots, 
\label{eq:N17}
\end{equation}
Here $E_\psi$ denotes the Euler operator or variational 
derivative with respect to $\psi$ used in the Calculus of variations.
In particular, the equations:
\begin{equation}
E_\gamma(L)=\deriv{L}{u^\gamma}
-D_i\left(\deriv{L}{u^\gamma_i}\right)
+D_iD_j\left(\deriv{L}{u^\gamma_{ij}}\right)-\ldots =0, \label{eq:N18}
\end{equation}
are the Euler Lagrange equations for the variational principle
$\delta J=0$.

Using $\pr(\X)L$ from (\ref{eq:N11}), 
the Lie invariance condition (\ref{eq:N7}) for the action reduces to:
\begin{equation}
\pr(\hat{\X})L+D_i\left(L\xi^i +\Lambda^i\right)=0. \label{eq:N19}
\end{equation}
Using (\ref{eq:N15}) for $\pr(\hat{\X})L$ in (\ref{eq:N19}) we obtain 
the Noether theorem identity:
\begin{equation}
\hat{\eta}^\gamma E_\gamma(L)
+ D_i \left[W^i[{\bf u},\hat{\bm\eta}]+L\xi^i+\Lambda^i\right]=0. 
\label{eq:N20}
\end{equation}
For the case of a finite point Lie group of symmetries (i.e. for a finite 
number of point symmetries $\hat{\eta}^{\gamma}$), for which the 
Euler-Lagrange equations $E_\gamma(L)=0$ are satisfied, and 
(\ref{eq:N20}) reduces to the conservation law:
\begin{equation}
D_i \left[W^i[{\bf u},\hat{\bm\eta}]+L\xi^i+\Lambda^i\right]=0,
\label{eq:N21}
\end{equation}
associated with Noether's first theorem. 

If the symmetries $\hat{\eta}^\gamma$ depend on continuous functions 
$\{\phi^k({\bf x}: 1\leq k\leq N\}$ then the Lie pseudo algebra 
of symmetries is infinite
 dimensional. In this case Noether's second theorem implies that the 
Euler-Lagrange equations are not all independent and that 
there exists differential relations between the Euler-Lagrange equations 
(see e.g. \cite{Olver93} and \cite{Hydon11} for details). 

%\appendix

\centerline{\bf Competing Interests}
The authors: G.M. Webb, S.C. Anco, S.V. Meleshko, and G.P. Zank declare none.

\end{document}